\documentclass[fleqn,usenatbib]{mnras}

\usepackage{newtxtext,newtxmath,lscape}

\usepackage[T1]{fontenc}
\usepackage{ae,aecompl}

\pdfoutput=1

\usepackage{graphicx}	
\usepackage{amsmath}	
\usepackage{amssymb}	
\usepackage[caption=false]{subfig}
\usepackage{xcolor}
\usepackage{breakurl}

\defcitealias{lag17}{L17}
\defcitealias{die18}{D18}
\defcitealias{kaw18}{K18}

\title[A370 MUSE Mosaic]{Probing 3D Structure with a Large MUSE Mosaic: Extending the Mass Model of Frontier Field Abell 370}

\author[D. J. Lagattuta et al.]{
David J. Lagattuta,$^{1}$\thanks{E-mail: david-james.lagattuta@univ-lyon1.fr}
Johan Richard,$^{1}$
Franz E. Bauer,$^{2,3,4}$
Benjamin Cl\'{e}ment,$^{1}$
\newauthor
Guillaume Mahler,$^{1,5}$
Genevi\`{e}ve Soucail,$^{6}$
David Carton,$^{1}$
Jean-Paul Kneib,$^{7,8}$
\newauthor
Nicolas Laporte,$^{9}$
Johany Martinez,$^{1}$
Vera Patr\'{i}cio,$^{1,10}$
Anna V. Payne,$^{1,11}$
\newauthor
Roser Pell\'{o},$^{6}$
Kasper B. Schmidt,$^{12}$
and Geoffroy de la Vieuville$^{6}$
\\
$^{1}$Univ Lyon, Univ Lyon1, Ens de Lyon, CNRS, Centre de Recherche Astrophysique de Lyon UMR5574, F-69230, Saint-Genis-Laval, France\\
$^{2}$Instituto de Astrof{\'{\i}}sica and Centro de Astroingenier{\'{\i}}a, Facultad de F{\'{i}}sica, Pontificia Universidad Cat{\'{o}}lica de Chile,\\ Casilla 306, Santiago 22, Chile\\
$^{3}$Millennium Institute of Astrophysics (MAS), Nuncio Monse{\~{n}}or S{\'{o}}tero Sanz 100, Providencia, Santiago, Chile\\
$^{4}$Space Science Institute, 4750 Walnut Street, Suite 205, Boulder, Colorado 80301, USA\\
$^{5}$Department of Astronomy, University of Michigan, 1085 S. University Ave., Ann Arbor, MI 48109, USA\\
$^{6}$Institut de Recherche en Astrophysique et Plan\'{e}tologie (IRAP), Universit\'{e} de Toulouse, CNRS, UPS, F-31400 Toulouse, France\\
$^{7}$Laboratoire d'Astrophysique, Ecole Polytechnique F\'{e}d\'{e}rale de Lausanne (EPFL), Observatoire de Sauverny, CH-1290 Versoix, Switzerland\\
$^{8}$Aix Marseille Univ, CNRS, LAM, Laboratoire d'Astrophysique de Marseille, Marseille, France\\
$^{9}$Department of Physics and Astronomy, University College London, Gower Street, London WC1E 6BT, UK\\
$^{10}$Dark Cosmology Centre, Niels Bohr Institute, University of Copenhagen, Juliane Maries Vej 30, 2100 Copenhagen, Denmark\\
$^{11}$Institute for Astronomy, University of Hawaii, 2680 Woodlawn Drive, Honolulu, Hawaii 96822, USA\\
$^{12}$Leibniz-Institut f\"{u}r Astrophysik Potsdam (AIP), An der Sternwarte 16, D-14482 Potsdam, Germany
}

\date{Accepted 2019 February 28. Received 2019 February 27; in original form 2018 November 06}

\pubyear{2019}

\begin{document}
\label{firstpage}
\pagerange{\pageref{firstpage}--\pageref{lastpage}}
\maketitle

\begin{abstract}
We present an updated strong-lensing analysis of the massive cluster Abell 370 (A370), continuing the work first presented in \citet{lag17}.  In this new analysis, we take advantage of the deeper imaging data from the \textit{Hubble Space Telescope (HST)} Frontier Fields program, as well as a large spectroscopic mosaic obtained with the Multi-Unit Spectroscopic Explorer (MUSE).  Thanks to the extended coverage of this mosaic, we probe the full 3D distribution of galaxies in the field, giving us a unique picture of the extended structure of
the cluster and its surroundings.  Our final catalog contains 584 redshifts, representing the largest spectroscopic catalog of A370 to date.  Constructing the model, we measure a total mass distribution that is quantitatively similar to our previous work -- though to ensure a low rms error in the model fit, we invoke a significantly large external shear term.  Using the redshift catalog, we search for other bound groups of galaxies, which may give rise to a more physical interpretation of this shear.  We identify three structures in narrow redshift ranges along the line of sight, highlighting possible infalling substructures into the main cluster halo.  We also discover additional substructure candidates in low-resolution imaging at larger projected radii.  More spectroscopic coverage of these regions (pushing close to the A370 virial radius) and more extended, high-resolution imaging will be required to investigate this possibility, further advancing the analysis of these interesting developments.
\end{abstract}




\section{Introduction}
Probing the formation and build-up of matter from large to small scales is a key ingredient in understanding the Universe.  With an accurate picture of structure formation, we can gain critical insight into a variety of astrophysical topics, such as galaxy evolution \citep[e.g., ][]{koo06,fra08,wet13,con14}, cosmology \citep[e.g., ][]{per09,jul10,bla14}, and even the nature of dark matter itself \citep[e.g., ][]{nie13,li16,boz16}.  Clusters of galaxies -- gravitationally bound collections of tens to hundreds of individual galaxies -- act as ideal laboratories for this phenomenon, as they provide information about structure formation at several physical scales simultaneously.  Forming at the intersections of long filaments of the cosmic web \citep[e.g., ][]{bon96,spr06,kra12}, clusters trace the large-scale structure of the universe, while the accretion of smaller haloes on to the primary mass (either via these filaments or independently) showcases more compact distributions in the so-called ``non-linear'' regime of the $\Lambda$CDM cosmological model \citep{bul17}.  At the same time, mergers between two (or more) pre-existing clusters can highlight structure build-up at intermediate scales, and individual mass configurations within and around the cluster lead to localized pockets of substructure which represents another important aspect of mass accumulation.

While there are many ways to study the distribution of mass within clusters, including X-ray gas \citep[e.g., ][]{sta06,ett13,mer15}, the Sunyaev-Zeldovich effect \citep[e.g., ][]{mar12,shi16,lop17}, and density maps of cluster light \citep[e.g., ][]{gav04,bah14,seb16}, one of the most robust methods is gravitational lensing.  Unlike other techniques, lensing does not rely on kinematic effects and makes no assumptions about the dynamic state of the cluster.  Lensing is also achromatic, producing the same signal at all frequencies and making it trivially easy to combine information from multi-wavelength data sets.  

Such multi-band imaging is a key feature of the Frontier Fields (FFs) program, a campaign designed to obtain deep {\it Hubble} (\emph{HST}) and {\it Spitzer} Space Telescope imaging of six massive lensing clusters \citep{lot17} in order to better observe extremely faint and distant ($z > 5$) galaxies being magnified by the cluster, pushing the boundaries of the observable universe.  Among these faint objects are hundreds of strongly-lensed, multiply-imaged background galaxies, which are used to model the mass distributions of these clusters with unprecedented accuracy.  Since the release of the FF images, analysis of the data in conjunction with other ancillary products has provided numerous insights into cluster masses \citep[e.g.,][]{die15,wan15,ogr16,hoa16,men17}, including a confirmation that merging lensing clusters -- often the most efficient gravitational lenses \citep{won12,won13} -- have exceedingly complex mass distributions.  Specifically, many studies have shown significant substructure populations within cluster fields \citep[e.g.,][]{lim16,jau16,nat17,wil18} and an elevated number of ``jellyfish'' galaxies: gas-rich, infalling spiral galaxies being stripped down by the intra-cluster medium \citep[e.g.,][]{ebe14,mcp16}.

The presence of complex substructure points to an evolving dynamical mass state, and has significant impact on the accuracy of model cluster masses: failure to account for mass concentrations at large cluster-centric radii can bias lens models by up to 20\% \citep{ace17}. The FF cluster Abell 2744 is a prime example, where the weak-lensing analysis of \citet{jau16} revealed several massive subhaloes at the cluster redshift, located at projected radii of 0.5 -- 1 Mpc from the strong-lensing core. These substructures strongly influenced past central cluster lens modeling \citep{mah18}, where the inclusion of these haloes mimicked the effects of a previously important ad-hoc ``external shear'' term and provided a more physical interpretation of the observations. Results from Abell 2744 and other merging clusters \citep[e.g.,][]{gir15} highlight the fact that substructures are often located at considerable distance from the cluster core, and can be missed in narrowly-targeted imaging and spectroscopic campaigns. This is an important motivator for the Beyond Ultra-deep Frontier Fields And Legacy Observations (BUFFALO) project\footnote{\url{http://buffalo.ipac.caltech.edu/}}, a new survey designed to expand the {\it HST} FF (HFF) imaging data by as much as four times the area of the current pointings. This wider area will improve our ability to trace the overall mass profile and substructure characteristics of the dark and luminous components of the FF clusters to $\sim 75\%$ of the cluster virial radius ($R_{vir}$), using a combination of both strong- and weak-lensing techniques.  This will provide critical insight into each cluster's central assembly history, which can be compared to theoretical predictions from standard $\Lambda$CDM \citep{vog14,sch15,sch17,jau18} and alternative models \citep[e.g.,][]{kra12,koy16}.

While the high-resolution {\it HST} data sets provided by the HFF and BUFFALO projects are useful, imaging alone is not sufficient for this analysis.  Indeed, the only robust way to derive physical values from a detected lensing signal is through a combination of imaging {\it and} spectroscopy.  Historically, lensing clusters have lacked comprehensive spectroscopic coverage due to the inefficient and often time-consuming observing process utilized by traditional instruments. However, this paradigm is changing. Thanks to the arrival of integral field unit spectrographs (IFUs) such as the Multi-Unit Spectroscopic Explorer (MUSE; \citealt{bac10}), we can now obtain hundreds of high-quality spectra in a given field with only a few hours of integration \citep[e.g.,][]{bac15}.  In particular, the wide ($1 \times 1$ arcmin) field of view and high throughput at optical wavelengths (4800--9300 \AA) makes MUSE an efficient tool for capturing redshifts of cluster galaxies, nearby infalling objects, and distant multiply-imaged background systems all at once.  At the same time, by studying the redshift distribution of other ``interloper'' galaxies in the field, it is possible to detect compact groups of objects in front of or behind the cluster (i.e., line-of-sight substructures) that are not often obvious in imaging.  Over the past few years, several lensing studies have taken advantage of MUSE spectroscopy \citep[e.g.,][]{ric15,kar15,bin16,pat16,gri16,cam17b,gri18,mah18} to great effect.

In this work, we study the merging cluster Abell 370 (A370) using a combination of HFF imaging and MUSE spectroscopy.  Centred at $z = 0.375$ \citep{mel88}, A370 is a highly elongated and efficient lens, with an Einstein radius of 39\arcsec\ at $z \approx 2$, a mass of $M_{< 250 \rm kpc} = 3.8 \times 10^{14} M_{\odot}$ \citep{ric10}, and an X-ray luminosity $L_{\rm X} = 1.1 \times 10^{45}$ erg s$^{-1}$ \citep{mor07}.  A370 is notable for hosting the first confirmed giant gravitational arc, at $z = 0.725$ \citep{sou87,ham87,pat18} and for having a high magnification area ($\mu > 5 - 10$) roughly twice as large as any other FF cluster \citep{ric14}.

This project follows up our previous investigation into A370 (\citealt{lag17}; hereafter L17) which modeled the total mass distribution with the help of a two-hour MUSE pointing in the very core of the cluster.  That analysis revealed a number of interesting properties of the system, including a largely flat central mass profile, boxy-shaped mass contours, and the possible existence of a third large-scale halo in the northeast. However, our conclusions regarding structure at extended radii were limited, due to a lack of spectroscopy beyond the central core. Here we expand the original MUSE data with a $2\arcmin\ \times 2\arcmin$ mosaic that covers nearly the entire region where strong-lensing constraints are expected to be found, as well as a large fraction of the deep, multi-band imaging region of the HFF data set.  This allows us to investigate our initial claims more rigorously, and also gives us an opportunity to search directly for substructures or other complex mass distributions.  Overall, this provides a more complete picture of the structure of A370 and serves as a pathfinder for other upcoming programs with {\it HST} (e.g., BUFFALO) and the James Webb Space Telescope (\emph{JWST}).

This paper is organized as follows: in Section \ref{sec:data} we describe the data used in our analysis and briefly discuss the reduction techniques.  In Section \ref{sec:redshifts} we explain our spectral extraction process and present an updated catalog of redshifts in the A370 field.  We use these redshifts to generate an initial set of mass models, which we present in Section \ref{sec:modeling}.  Next, we identify possible substructure candidates in Section \ref{sec:substruc} and modify our models to include the effects of the additional mass.  We also discuss our results in this section, paying special attention to the physical interpretation of external shear.  Finally we briefly conclude in Section \ref{sec:end}.  Throughout this work we assume a standard cosmological model of $\Omega_M = 0.3$, $\Omega_{\Lambda} = 0.7$ and $H_0 = 70$ km s$^{-1}$ Mpc$^{-1}$.  In this framework, a span of 1 arcsecond corresponds to a physical distance of 5.162 kpc at the cluster redshift of $z = 0.375$.  Finally, unless otherwise stated all magnitudes are measured using the AB system.
 
\section{Data}
\label{sec:data}

To improve our understanding of the cluster we use a combination of imaging and spectroscopic data in this work, both to construct a model of the A370 mass distribution and to identify overdensities in the redshift distribution which can be indicative of line-of-sight substructure.

\subsection{HST}
\label{subsec:HST}

As in \citetalias{lag17}, we again use the publicly-available HFF image stacks of A370\footnote{\url{https://archive.stsci.edu/prepds/frontier/abell370.html}} to identify cluster members and multiple-image candidates, and as a basis for spectral extraction in MUSE data (Section \ref{sec:redshifts}).  The complete HFF data set covers seven broadband filters ($F435W$, $F606W$, $F814W$, $F105W$, $F125W$, $F140W$, and $F160W$), forming the deepest optical and near-IR imaging ever taken of A370.  For the three optical bands we use the final Epoch 1 v1.0 stacks (consisting of 20, 10, and 52 {\it HST} orbits, respectively), while for the near-IR bands we use the Epoch 2 v1.0 stacks (consisting of 25, 12, 13, and 28 {\it HST} orbits) which were not available in our previous work.

The addition of deeper near-IR data is especially useful for identifying high-redshift dropout candidates which can be lensed more efficiently by the cluster.  Multi-band photometry also enhances our analysis by providing important colour information about each object.  In particular, we use an optical colour-colour selection criterion to identify cluster members that otherwise lack secure spectroscopic redshifts (see Section \ref{subsec:construction}), allowing us to build a more complete and robust lens model.

To better resolve small, point-source-like objects in the field we use the images with 30 mas resolution, and to ensure the best possible data quality we use the ``selfcal'' charge transfer inefficiency (CTI)-corrected data for the ACS images and the ``bkgdcor'' data for WFC3, which applies corrections for persistence effects and detector artefacts, as well as an improved handling of time-variable sky emission.  The full details of data reduction and cleaning are described in the HFF data archive for A370\footnote{\url{https://archive.stsci.edu/pub/hlsp/frontier/abell370/images/hst/v1.0-epoch2/hlsp_frontier_hst_acs-00_abell370_v1.0-epoch2_readme.pdf}} and in \citet{lot17}. 

\subsection{MUSE}
\label{subsec:MUSE}

\begin{figure*}
\includegraphics[width=\textwidth]{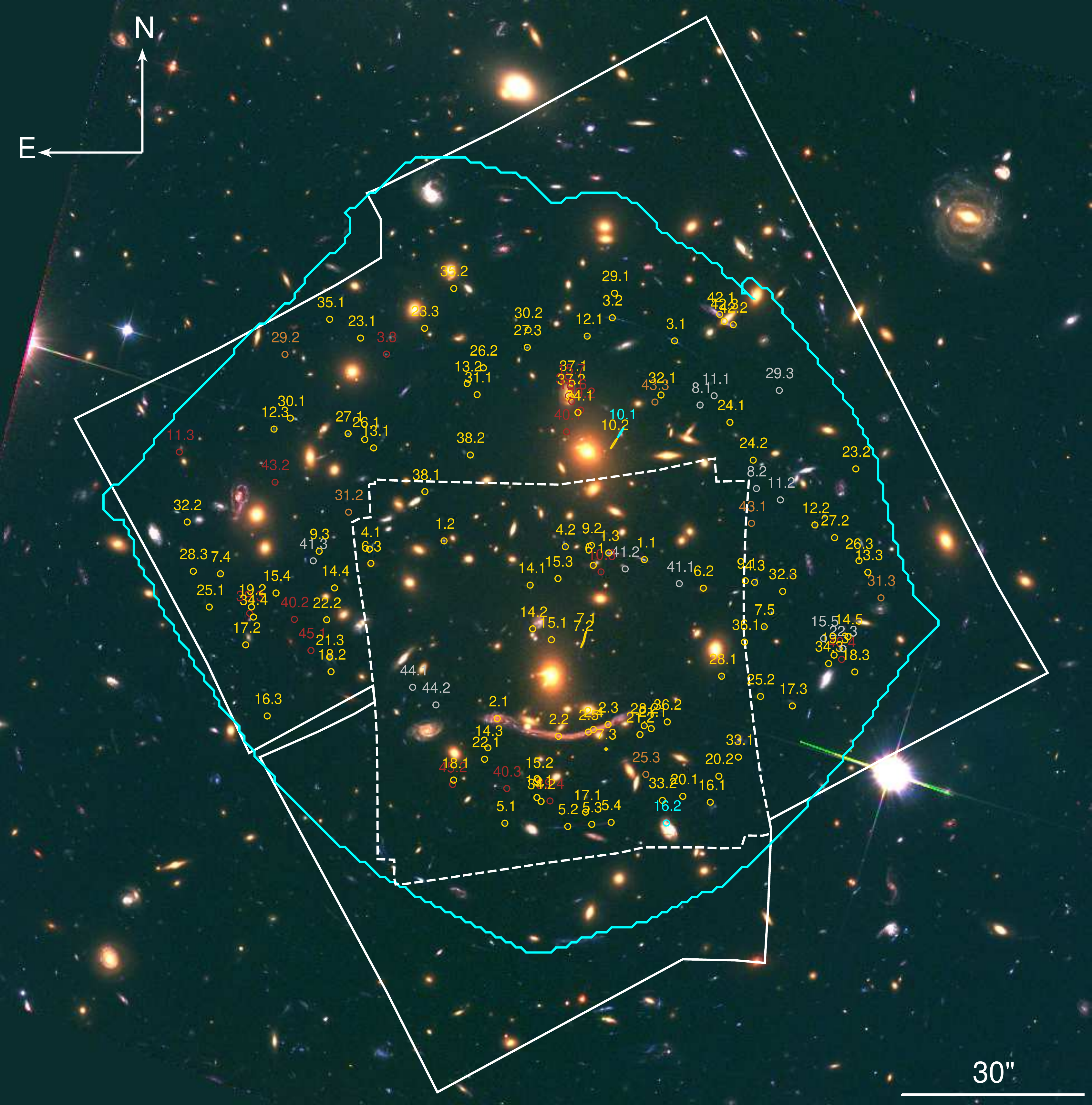}
\caption{Colour image of the A370 field, using the Epoch 1.0 $F435W$, $F606W$, and $F814W$ data sets from the Hubble Frontier Fields program.  The region covered by the MUSE mosaic is represented by the thick white line; for reference, we show the original GTO MUSE footprint used in \citetalias{lag17} as a dashed white line.  In addition, all multiply-imaged galaxy constraints used in the lens modeling are shown as coloured circles.  The colour of each constraint (gold, silver, bronze, and copper) is based on its ranking according to the HFF public modeling challenge (see Section \ref{subsec:construction} and Table \ref{tbl:Multi-Images} for complete details.)  The copper-coloured constraints appear darker than their bronze counterparts.  Cyan coloured circles are items that are predicted but not used in the modeling.  The cyan contour traces out the ``multiple-image zone'', the region where all multiply-imaged objects are expected to fall out to high redshift ($z = 10$).} 
\label{fig:colorImg}
\end{figure*}

For spectroscopic information we turn to MUSE, observing the A370 field using a large mosaic covering $\sim 4$ arcmin$^2$.  This mosaic, program 096.A-0710(A) (PI: Bauer) is an expansion of an initial Guaranteed Time Observing (GTO) program 094.A-0115(A) (PI: Richard) which focused on the central cluster core.  This full mosaic covers a $2'$$\times$$2'$ area centred on the cluster, providing nearly complete coverage of the ``multiple-image zone'': the area where multiple-images of all background galaxies out to high redshift ($z$$=$10) are expected to fall (Fig. \ref{fig:colorImg}).

\begin{figure}
\includegraphics[width=0.5\textwidth]{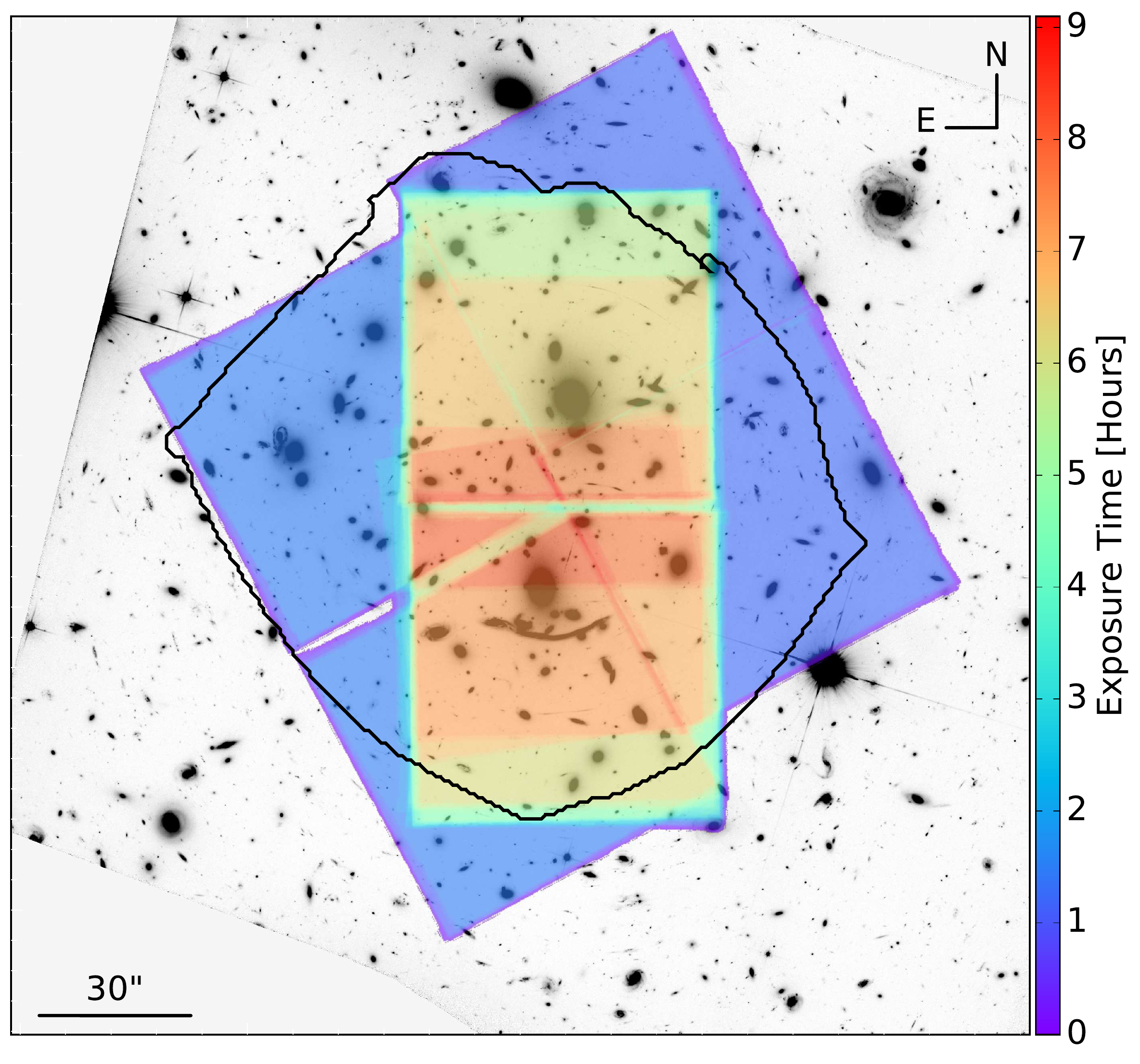}
\caption{Exposure-time map of the MUSE mosaic covering A370.  The deeper central region is designed to overcome noise due to strong intracluster light while the outer region covers the remaining section of the cluster core, all with relatively uniform line sensitivity.  The thick black line again shows the multiple-image zone and highlights the nearly complete coverage of this area, greatly improving our ability to spectroscopically detect lens-model constraints (section \ref{subsec:catMulti}).} 
\label{fig:ExpMap}
\end{figure}

The mosaic was designed to have a factor of $\sim$3--4 higher exposure in the central portion (highest lensing area) of A370 in order to achieve relatively uniform line sensitivity in the face of strong intracluster light, which raises the background for line-detection by up to a factor of $\sim$2 at the centre. To this end, we obtained 2 hours on-source exposure rotated by 28$\degr$ to cover the bulk of the {\it HST} WFC $\sim$5 arcmin$^{2}$ footprint (excluding a bright star), while a further 4 hours on-source exposure were devoted to two pointings centred on the N-S high magnification region. We incorporated the archival 2 hours on-source exposure to deepen the exposure to 8 hours in the very centre. In total, the mosaic comprises 18 hours of on-source exposure. The full MUSE exposure map can be seen in Fig. \ref{fig:ExpMap}.

MUSE was employed in non-AO mode for all observations, and individual exposures were limited to $<$15--30 minutes to minimize sky variation and cosmic ray effects. The source exposures for program 094.A-0115(A) (4$\times$1800\,s) were acquired on November 20, 2014, while exposures for program 096.A-0710(A) (32$\times$930s, 3$\times$953s, 37$\times$962s, 2$\times$963s) were obtained in queue-mode between Oct 8, 2015 and Sept 28, 2016 in 15 separate observing blocks. Each observing block was ideally comprised of four equal exposures, rotated by 90$\degr$ from one another and dithered within a 1\arcsec\ box.  Observations were taken in both 'clear' and 'photometric' conditions ('clear' requested), with DIMM seeing values between 0.37--1.09$\arcsec$ (0.68$\arcsec$ median; $<$0.8$\arcsec$ requested), airmasses between 1.09-1.42 (1.17 median; $<$1.6 requested) and $<$$\pm$7 days from full Moon.

Data were reduced largely following the standard procedures of the \texttt{esorex} pipeline (muse-kit-2.4.1; \citealt{wei16}). Basic calibration files were used to perform bias subtraction and flat fielding, using illumination and twilight exposures taken closest to the date of the source exposure. Flux calibration and telluric correction were performed using the standard star taken closest to the date of the source exposure (generally but not always the same night). Spectral response curves were visually inspected for problems; all appeared reasonable and were applied to all individual cubes.  Autocalibration was performed to improve IFU-to-IFU and slice-to-slice flux variations, using masks generated via Source Extractor (hereafter referred to as SExtractor; \citealt{ber96}).

Individual datacubes were aligned to the {\it HST} $F814W$ reference frame using a combination of SExtractor to identify bright sources in the individual datacubes and {\it HST} image, and custom software to calculate the offsets between them.  \texttt{ZAP} \citep{sot16} was applied to individual cubes to mitigate sky-line residuals. Individual datacubes were then renormalized based on photometric offset compared to PSF-matched {\it HST} $F606W$ and $F814W$ images; flux offsets ranged from 0.95--1.08  with a median of 1.02. Finally, individual exposure-weighted cubes were stacked on a common grid to create the final mosaic.

An image highlighting the coverage of the A370 MUSE mosaic can be seen in Fig. \ref{fig:colorImg}.

\section{Spectroscopy}
\label{sec:redshifts}

\subsection{Spectral Extraction}
We obtain spectral information for objects in the MUSE mosaic using two complementary techniques: a targeted extraction based on {\it HST} imaging, and a ``blind'' identification of emission lines.  The full procedure is described in \citet{mah18}, but we briefly describe the process here.  

For the targeted search, we identify objects by running SExtractor on an inverse-variance weighted stacked image of all seven broadband HFF filters.  Prior to stacking the data, we first remove bright intra-cluster light from each image by subtracting the median value of a 21$\times$21 pixel box surrounding each point.  This significantly improves the contrast between bright and faint objects, resulting in cleaner spectral extraction.  The actual extraction is based on the SExtractor segmentation map for each object, convolved with the MUSE PSF and resampled to match its pixel scale.  All MUSE spaxels that fall in this broadened mask region are combined, weighted by the signal-to-noise ratio, and collapsed from 3D to 1D, which we take as the final spectrum of each object.

Conversely, the blind search operates on the MUSE cube directly, using \texttt{MUSELET}\footnote{\url{http://mpdaf.readthedocs.io/en/latest/muselet.html}} to detect emissions lines without regard to any {\it HST} imaging.  To do this, the program creates a series of pseudo narrow-band images over the full MUSE wavelength range, summing the flux from five spectral bins around a given wavelength slice (spanning 6.25 \AA) and subtracting the nearby continuum: the average of the 25 \AA\ immediately redward and blueward of the averaged narrow band.  The program then runs an additional iteration of SExtractor on these images and notes the positions of any bright flux peaks.  If multiple emission lines appear at the same spatial location, \texttt{MUSELET} attempts to fit a redshift for the object by matching it with a library of known spectral features.  Otherwise, it simply reports the coordinates and wavelength of the emission line(s), which can then be inspected manually.  

Since \texttt{MUSELET} operates independently from the \textit{HST} data, many of these lines can also be seen in spectra extracted from the targeted search.  Therefore, to avoid duplicate efforts while measuring redshifts (section \ref{subsec:zMeasure}) we first match \texttt{MUSELET} detections to any \textit{HST} object that falls within a 1\arcsec\ radius of its centroid.  In cases where multiple \textit{HST} objects fall within this radius, we simply take the closest object to be the match.  The average separation between the matched \texttt{MUSELET} and \emph{HST} centroids is less than 0.1\arcsec, resulting in a high likelihood that matched entries identify the same object.  Nevertheless, we still compare the two values during redshift measurement, to make sure this is the case.  While this accounts for a large fraction of lines, there are still 21 ``orphan'' \texttt{MUSELET} detections that do not have a counterpart in the \textit{HST} SExtractor catalog. These are typically extremely faint (or continuum-free) galaxies at high redshift ($z > 3$) displaying strong Lyman $\alpha$ emission \citep[e.g.,][]{mas18}.  Since the targeted extraction method does not automatically generate a spectrum for these systems, we instead create one by hand by combining all spaxels in the MUSE data within a 1\arcsec\ radius from the \texttt{MUSELET}-measured centroid.

\subsection{Redshift Measurements}
\label{subsec:zMeasure}

Once all spectra are extracted and coincident \emph{HST}- and \texttt{MUSELET}-based entries are  merged together, we analyse the data to measure redshifts.  As an initial guess, we run the automated software \texttt{AutoZ} \citep{bal14} on all extracted spectra, without regard to object size or brightness.  The software attempts to find a best-fitting redshift for each spectrum using cross-correlation with user-supplied redshift templates; for templates we use a series of previously-classified MUSE spectra (transformed to $z = 0$) with prominent optical emission and absorption features, following \citet{ina17}.  The correlation coefficient (cc) output by \texttt{AutoZ} provides an objective measure to determine the quality of the template match, where a value cc $> 5$ is considered a ``good'' fit to the data.

After running the automatic method, we review the results using a customized \texttt{Python}-based redshifting tool that allows a user to manually refine redshifts up to $\delta z = 0.0001$ precision, independent of the redshift.  We also compare these measurements to any \texttt{MUSELET} value associated with the object.  We inspect all objects down to $m_{F814W} = 26.5$ in the deeper, central region of the MUSE mosaic, and $m_{F814W} = 25.0$ in the shallower outer regions.  In both cases this marks the point where continuum- or absorption-based features become too faint to be detected, and also corresponds to a steep decline in objects with an \texttt{AutoZ} cc > 5.  We stress however that any objects with a \texttt{MUSELET} detection or an \texttt{AutoZ} cc > 5 are included in this subsample, regardless of the magnitude cutoff.  

\subsection{Redshift Catalog}
\begin{figure*}
\begin{center}
\includegraphics[width=0.70\textwidth]{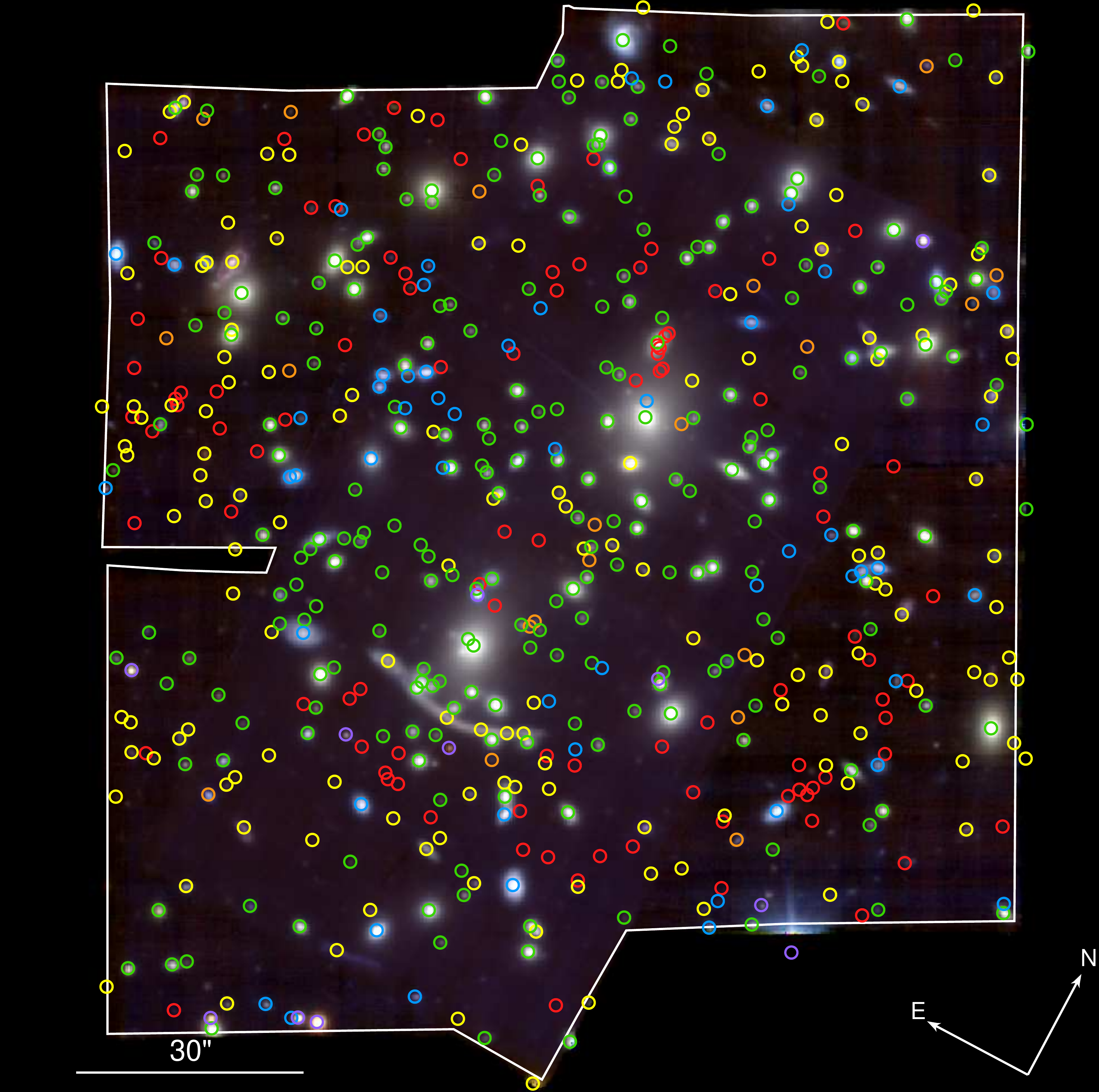}
\includegraphics[width=\textwidth]{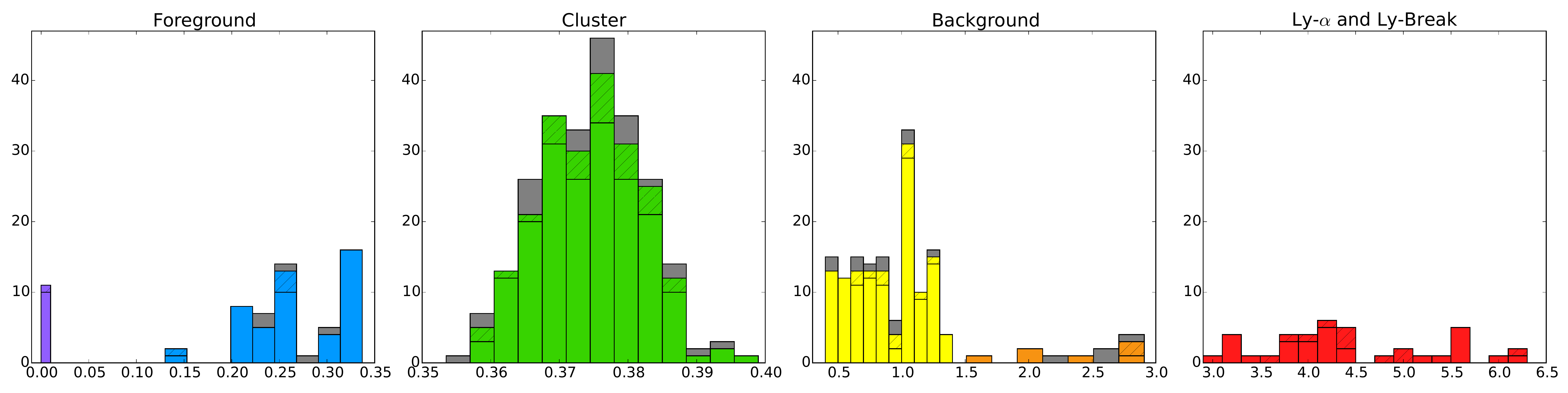}
\end{center}

\caption{Top: Spatial distribution of all objects identified in the redshift catalog.  Objects are colour-coded based on their redshift range and match the colours used in the lower panels.  The RGB colour image is generated from the MUSE data itself.  Bottom: Redshift distribution of all objects in the catalog (for clarity, all members of a multiply-imaged object are combined into a single entry).  Colours are as follows: purple -- stars ($z=0$); blue -- foreground galaxies ($0 < z <0.35$); green -- cluster galaxies ($0.35 \leq z \leq 0.4$); yellow -- ``near'' background galaxies ($0.4 < z < 1.5$); orange -- MUSE redshift desert ($1.5 < z < 2.9$); red -- ``far'' background galaxies ($z > 2.9$).  Colours and shading of the histogram bars are based on the spectral confidence value, as follows:  plain bar -- confidence 3 (high confidence); hashed bar -- confidence 2 (medium confidence); gray bar -- confidence 1 (low confidence).  For complete spectral information, see Table \ref{tbl:Redshifts} and the online catalog supplement.  Details of the spectral classification (including the confidence measurement) can be found in Appendix \ref{app:Redshifts}.}
\label{fig:spectra}
\end{figure*}

The final master catalog contains 584 objects, though there are only 506 unique systems when accounting for multiple-imaging of background galaxies. We identify and refine multiply-imaged objects using an iterative process based on the current lens model.  These steps are described in the next subsection and in Section \ref{sec:modeling}.  The full catalog spans a redshift range $0 \leq z \leq 6.2855$ and is made up of stars, cluster members, multiply-imaged galaxies, and other foreground and background interlopers.  The spectral distribution and spatial location of all objects with measured redshifts can be seen in Fig. \ref{fig:spectra}.  In addition, we also provide a sample of the first few entries of the catalog in Table \ref{tbl:Redshifts}. The full catalog is available as an online supplement to the electronic version of this manuscript.

We now describe portions of the redshift sample that are important for our subsequent lensing analysis, as well as a few additional objects that are interesting in their own right.  We also compare our catalog to the Grism Lens-Amplified Survey from Space (GLASS; \citealt{sch14,tre15}), a program providing catalogs of lower resolution, near-IR spectroscopy of all six FF clusters.  Given the higher sensitivity and resolution of MUSE relative to the \emph{HST} grism, the number of objects with secure, high-confidence redshifts is significantly higher in our catalog compared to that of GLASS, though we still find several objects in common between the two.  While in most cases the GLASS catalog agrees (within measurement uncertainty) with our updated MUSE results, there are some instances where MUSE is able to correct mis-identified features in the GLASS data and provide a more accurate redshift for the object.  Likewise, there are also instances where the presence of emission lines in the observed NIR from GLASS improves the confidence of a redshift initially based on low-S/N features in the MUSE spectra.  A summary of these comparisons can be seen in Appendix \ref{app:GLASS} and Table \ref{tbl:GlassCompare}.

\subsubsection{Multiply-Imaged Systems}
\label{subsec:catMulti}

By covering nearly the entire multiple-image zone of A370 (Fig. \ref{fig:colorImg}) the new redshift catalog contains a substantial number of lensing constraints that can significantly improve the overall mass model.  Thanks to the deeper data in the cluster core, we are able to revise redshifts of previously-known systems, while the wider coverage at larger distances allows us to confirm (or modify) counterimages missed in the GTO cube and identify entirely new systems located in different parts of the cluster. 

We begin by revisiting the lensing constraints used to construct the mass model presented in  \citetalias{lag17} (hereafter referred to as the GTO model).  Of the 22 multiple-image systems used in that work, only four (Systems 1, 2, 5, and 6) were completely contained within the GTO data, leading to secure spectroscopic redshifts for all images in the set.  All other systems had at least one ``missed'' counterimage lying beyond the MUSE footprint (identified only by lens model predictions and HFF imaging), including six (Systems 3, 8, 10, 11, 12, and 13) which fell entirely outside of the cube.  With our larger catalog, we are able to examine the missed counterimages of the partial GTO systems, remarkably finding that (without exception) our model predictions are correct: we measure the expected redshifts in the mosaic at the same positions of the HFF imaging candidates.  This showcases the predictive power of our initial lens model, and highlights the usefulness of even partial spectroscopic coverage.  We note, however, that without the additional high-resolution imaging, our initial guesses would likely not have been as accurate. 

\begin{figure*}
\includegraphics[width=\textwidth]{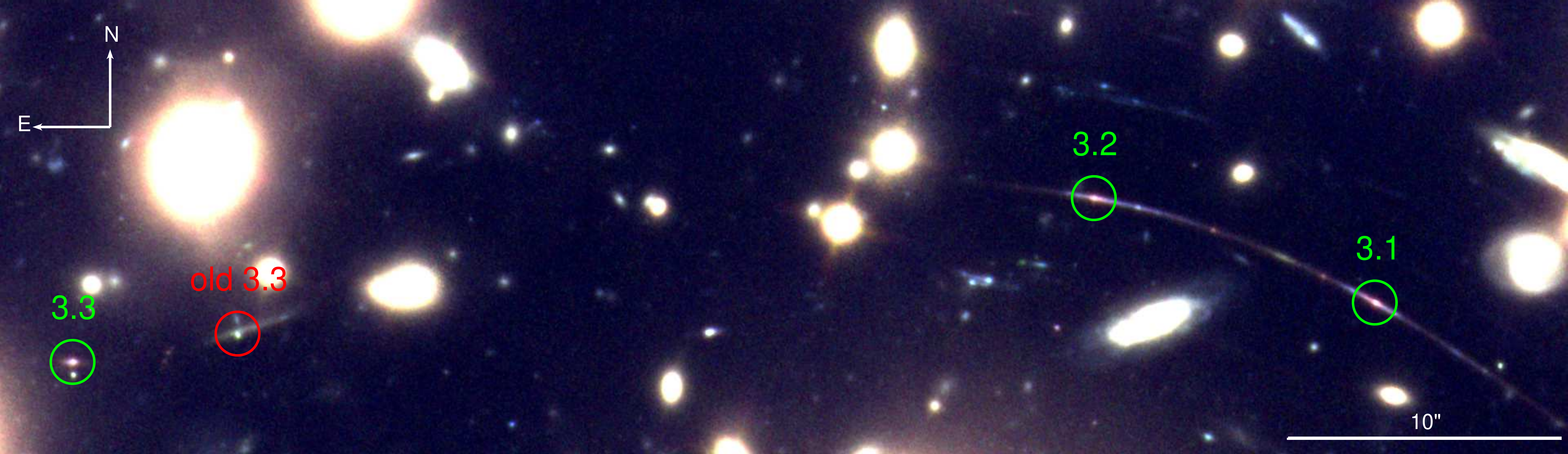}
\caption{Updated configuration of System 3 ($z = 1.9553$), including the new location of Image 3.3.  We correct the position of Image 3.3 thanks to new spectroscopic and visual evidence: the \ion{C}{III}] emission seen in the core of Images 3.1 and 3.2 is also found in the new counterimage (see Fig. \ref{fig:SpecFigs}), but absent in the old candidate (red circle).  Additionally, the deeper near-IR HFF data shows that the new counterimage has a pink colour (using an $F814W$, $F105W$, and $F160W$ colour scheme) which better matches the core of the Image 3.1/3.2 pair.}
\label{fig:Sys3.3}
\end{figure*}

Additionally, we provide new redshifts for many of the systems outside of the GTO cube, and adjust the redshift of one GTO system that was too faint to be accurately measured.  These updates are as follows:

\begin{itemize}
\item System 3:  While this system was too far north to be seen in the GTO cube, Images 3.1 and 3.2 did have low-resolution grism spectroscopy from GLASS (\citealt{die18}; hereafter D18), and we used this redshift ($z_{\rm GLASS} = 1.95$) in our earlier mass model.   In the mosaic catalog, we detect \ion{C}{III}] emission in the central regions of these two images, giving rise to an updated redshift $z = 1.9553$.  More importantly, we modify the position of the third counterimage (Image 3.3), shifting it from the original galaxy identified in \citet{ric10} to a nearby object slightly to the east (Fig. \ref{fig:Sys3.3}) based on the presence of faint \ion{C}{III}] emission in the new counterimage that does not appear in the original candidate.  We note that this new result agrees with the prediction presented in \citet[][hereafter K18]{kaw18}, made without the use of other spectroscopic data.\\

\item System 9: Although Image 9.1 fell inside the original GTO data set, the spectrum was too faint to measure any definitive lines.  Therefore, like System 3, we relied on GLASS spectroscopy ($z_{\rm GLASS} = 1.52$) for our previous mass model.
With deeper mosaic data in the core, we are able to detect faint \ion{C}{III}] emission, leading to an improved redshift $z = 1.5182$\\

\item System 10: We identify faint \ion{C}{III}] in Images 10.1 and 10.2, leading to a redshift $z = 2.7512$.  We also detect possible (though extremely weak; S/N $\sim$ 1.6) traces of \ion{C}{III}] in a separate counterimage (10.3; also predicted in \citetalias{die18}) at the same redshift.  That this redshift is exactly the same as the one measured for System 7 brings up the intriguing possibility that both systems are in fact images of the same galaxy.  We address this possibility in Section \ref{sec:modeling}.\\

\item System 12: All images in this system show strong, broad Lyman $\alpha$ absorption, along with other, narrower UV absorption features such as \ion{O}{I}[$\lambda$1301] and \ion{Si}{II}[$\lambda$1303].  Image 12.1 (which lies in the deeper central region of the mosaic) also shows faint \ion{C}{III}] emission, which we use to set the systemic velocity of the system ($z = 3.4809$).\\

\item System 13: Both original images in this system (Images 13.1 and 13.2) show strong Lyman $\alpha$ emission, resulting in a redshift $z = 4.2480$.  Furthermore, we also discover a faint counterimage (Image 13.3) in the southwest of the cluster with the same redshift, which has similar \emph{HST} colours.  While this new image is far from the prediction for 13.3 made with the GTO model, the lack of a secure redshift for the system in that model made an accurate guess more difficult.\\

\item Systems 8 and 11: Even with the new mosaic we are still unable to detect any features in spectra of Systems 8 and 11, though given their observed colours this is not entirely surprising.  Photometric estimates for System 8 place it in the MUSE ``redshift desert'' ($z \sim 1.5 - 3$) where bright optical lines are too red to be observed, while Lyman $\alpha$ emission is still too blue.  Although \ion{C}{III}] can be used to measure redshifts in this range, we do not detect \ion{C}{III}] emission in either of the known counterimages of this system.  Conversely, the photometric redshift of System 11 (an $F814W$ dropout galaxy) is $z \sim 7.8$, suggesting it is too distant to be measured by MUSE.  We do note, however, that thanks to the enhanced near-IR HFF coverage, we are able to identify a new counterimage with matching \emph{HST} colours on the far eastern side of the multiple-image region, which we use in subsequent mass modeling.
\end{itemize}

After making adjustments to the known constraints, we search for new multiply-imaged objects in the field.  While the larger mosaic makes finding these systems significantly easier than before -- we simply identify groups of images that have the same redshift inside the multiple-image zone -- we still use the GTO mass model as a guide.  Specifically, we check that the observed image configuration of a candidate system is at least broadly consistent with its model prediction, to ensure that the targets are not simply a chance alignment of individual galaxies.  While in principle, we allow for a large separation (up to 10\arcsec) between prediction and observation before rejecting the image -- to avoid potentially biasing the new model -- we note that none of the actual candidates deviates from its initial prediction by more than 3\arcsec.  At the same time, model predictions can also reveal additional counterimages missed by our spectroscopic campaign, either because they are too faint to be detected automatically, or because they are highly contaminated by a nearby bright source.  

\begin{figure}
\includegraphics[width=0.5\textwidth]{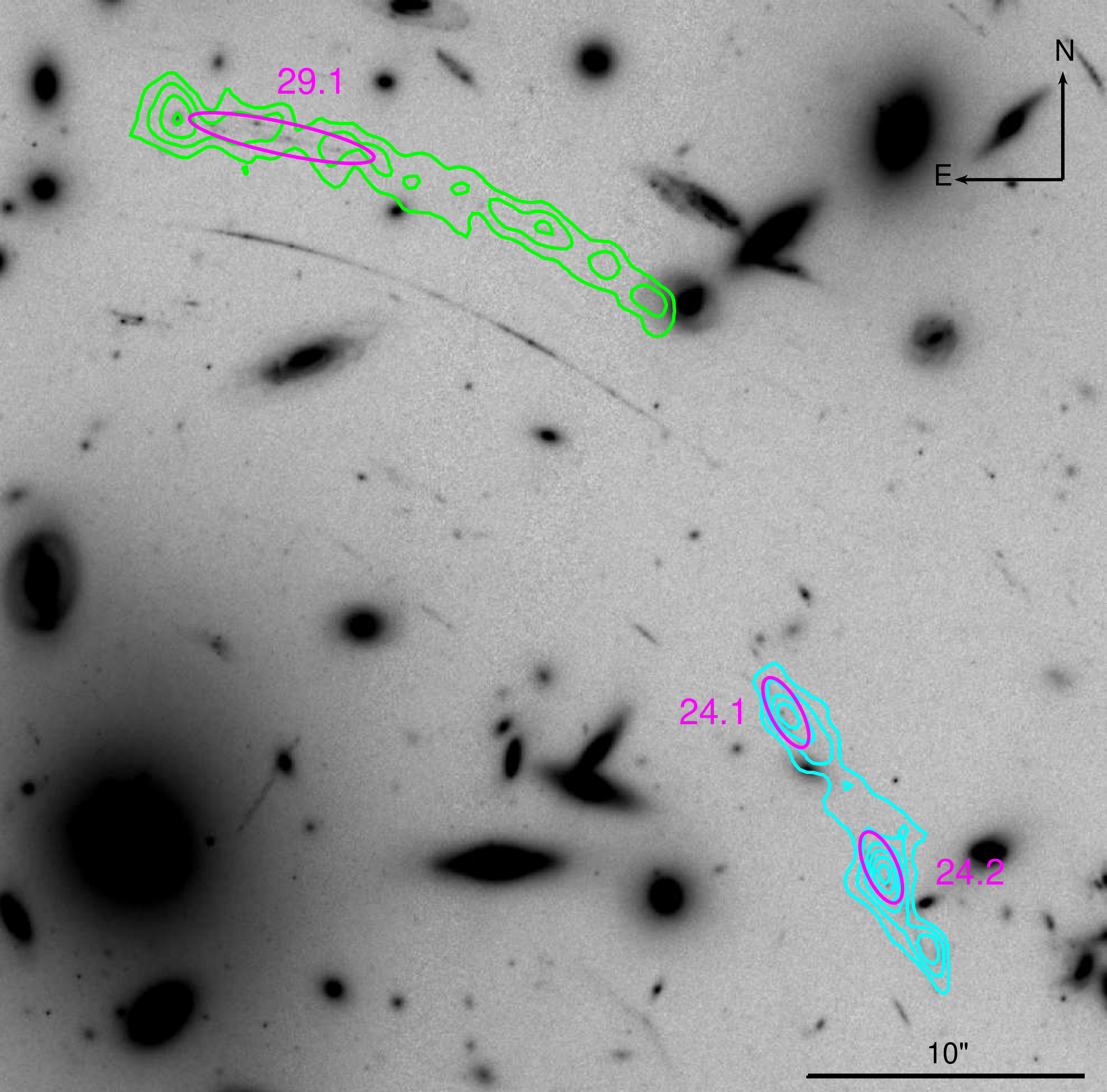}
\caption{Two examples of extended Lyman $\alpha$ arcs found in the A370 MUSE mosaic: the merging pair System 24 ($z = 4.9160$; cyan contours) and the giant arc System 29 ($z = 4.4897$; green contours).  In both cases the narrow-band Lyman $\alpha$ flux is significantly larger than the observed continuum counterpart, highlighted by the magenta ellipses.}
\label{fig:Arcs}
\end{figure}

Overall, we identify 18 new systems with secure spectroscopic redshifts (labelled as Systems 23 through 40) in the MUSE data, with redshifts between $z = 2.9$ and $z = 6.3$.  This more than doubles the number of spectroscopically confirmed multiply-imaged galaxies in A370, and considerably increases the total number of robust model constraints.  Given their higher redshifts, many of these new images are located further from the cluster centre than the GTO systems, providing better information about the A370 mass distribution out to larger radii.  Most systems are identified by strong Lyman $\alpha$ emission (except System 38 which is instead a Lyman break galaxy, identified by a prominent trough in the UV continuum), and while many are point-like, some do present features of extended emission -- which is expected based on previous MUSE-based Lyman $\alpha$ studies \citep[e.g.,][]{wis16,wis18,lec17}  This is especially apparent in the merging pair of System 24, and the large extended arc seen in System 29 (Fig. \ref{fig:Arcs}).  Furthermore, the two most-distant lensed galaxies are located beyond $z = 6$ (System 35, $z = 6.1735$; System 36, $z = 6.2855$) -- placing them close to the end of the epoch of reionization, possibly making them interesting objects for future study (e.g., \citealt{her17}.)

Of the 18 systems, three are independently presented in other works: \citetalias{die18} identifies System 24 (labelled System 20 in that paper), while \citetalias{kaw18} detects System 26 (labelled System 47) and System 29 (split into two systems, 34 and 35).  These studies also partially identify five other systems, but without additional spectroscopic information they miss (or mis-identify) counterimages that we detect in MUSE.  These are summarized as follows:

\begin{itemize}
\item System 25 (\citetalias{die18} system 16): Images 25.1 (D16.2) and 25.2 (D16.1) are identified though slightly offset from the bright Lyman $\alpha$ emission, while Image 25.3 is missed.  Though it lies in the central part of the cluster, this image is significantly fainter than the others in MUSE, and does not have an obvious {\it HST} counterpart.  Its long extended Lyman $\alpha$ halo and matching spectral line profile, along with its prediction in the GTO model makes us confident it is a real counterimage.\\

\item System 27 (\citetalias{die18} system 27): Images 27.1 (D27.2) and 27.3 (D27.1) are identified.  The third image in \citetalias{die18} (D27.3) is a mis-identified object at a different redshift ($z = 1.2754$).  The correct counterimage (Image 27.2) is slightly further southwest than the candidate presented in that work and is significantly fainter.\\

\item System 28 (\citetalias{die18} system 21): Images 28.1 (D21.1) and 28.2 (D21.2) are identified.  Image 28.3, a faint companion at the far eastern side of the cluster is missed.\\

\item System 38 (\citetalias{die18} system 30):  Image 38.1 (D30.1) is correctly identified, while Image 38.2 (D30.2) is missed.  However, this may simply be a transcription error as the coordinates of D30.2 point to a patch of blank sky, and the published declination of D30.1 and D30.2 are identical. \\

\item System 40 (\citetalias{kaw18} system 42): \citetalias{kaw18} identifies the three southern images in this system: Images 40.2, 40.3, and 40.4 (K42.1, K42.2, and K42.3, respectively) but misses the northernmost Image 40.1, which lies very close to the northern BCG.  Though heavily contaminated by the BCG, we still see faint traces of Lyman $\alpha$ emission in narrow-band MUSE imaging.  
\end{itemize}

To be as complete as possible we do attempt to measure redshifts from all other multiple-image candidates presented in \citetalias{die18} and \citetalias{kaw18}, though we are largely unable to do so: many of these systems are quite faint, and their extracted spectra contain only noise.  At the same time, much like Systems 8 and 11, many have photometric redshift estimates that are not optimal for a MUSE detection.  One major exception is \citetalias{die18} system 15, a bright spiral galaxy at $z = 1.0315$ being galaxy-galaxy lensed by a single cluster member in the northwest corner. Conversely, we are able to reject \citetalias{die18} system 24 as being not multiply-imaged, but rather a chance alignment of three faint cluster members (D24.1, $z = 0.3788$; D24.2, $z = 0.3721$; D24.3, $z = 0.3749$).  We also measure a tentative redshift for \citetalias{die18} image D14.1 ($z = 1.2777$) based on a faint [\ion{O}{II}] line, though it is at low confidence and we do not see this line in the two counterimages D14.2 and D14.3 (located in the shallower outskirts of the mosaic.)  Without a strong confirmation, we do not use this redshift during the lens modeling process (Section \ref{sec:modeling}), opting instead to fit the system's redshift as a free parameter.  We note, however, that the best-fitting redshift in our final model ($z = 1.272 \pm 0.016$) is remarkably close to the tentative spectroscopic value.

\begin{figure}
\includegraphics[width=0.5\textwidth]{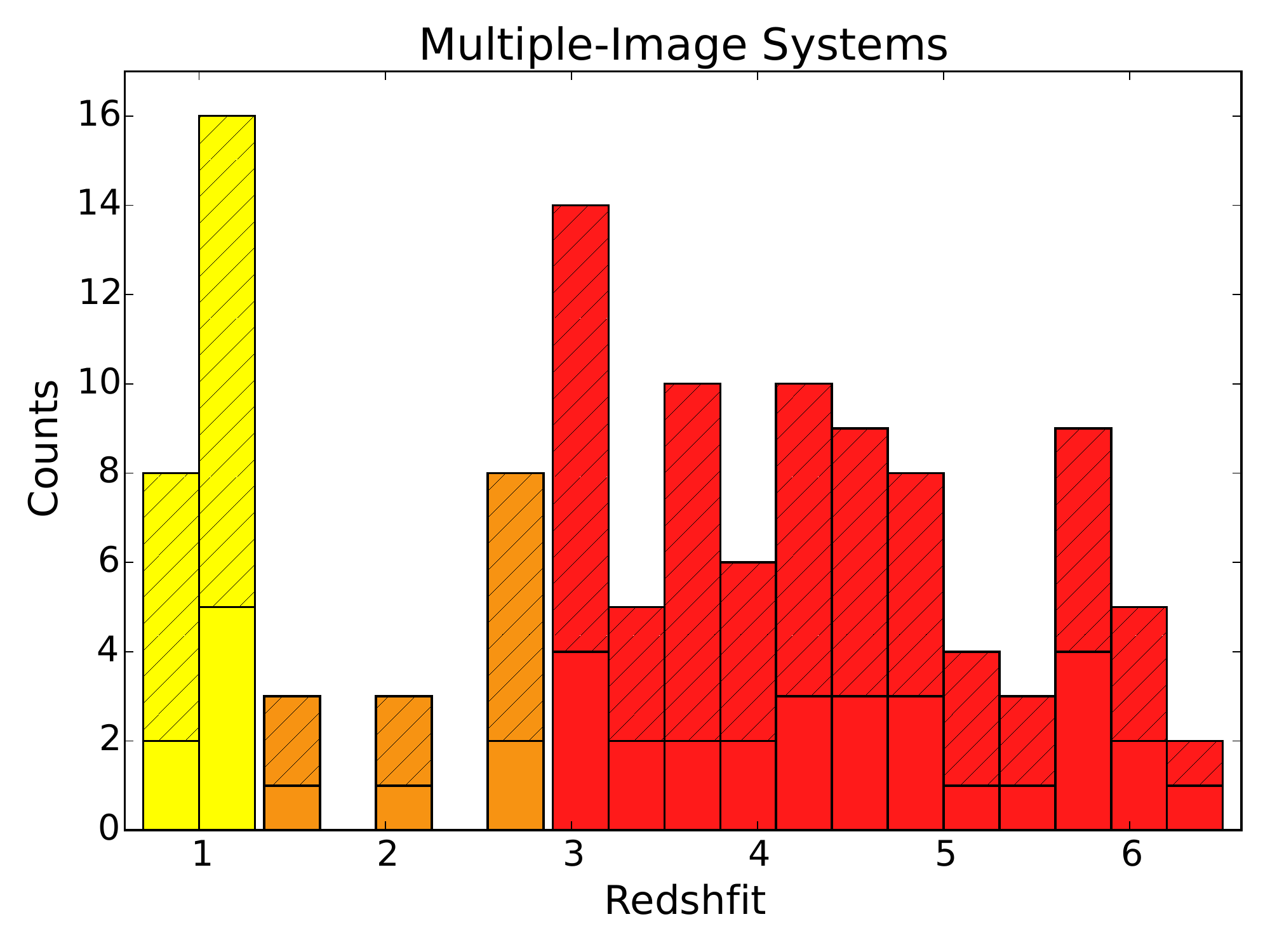}
\caption{Redshift histogram of multiply-imaged galaxies in the A370 field.  Solid-colour bars represent individual systems, while the hashed bars include all counterimages that make up these systems.  The colour scheme here mimics the redshift ranges used in Fig. \ref{fig:spectra}}
\label{fig:multiHist}
\end{figure}

Following discussions held during the HFF Public Modeling Challenge, a collaborative project bringing together several lens-modeling teams to work with common data sets as a way to better characterize systematic differences between modeling techniques,\footnote{\url{http://archive.stsci.edu/prepds/frontier/lensmodels/}} we include D14 and D15 as constraints to the model (as systems 41 and 42, respectively), along with three others (Systems 43 to 45) without any redshift measurement.  We will discuss this selection more in Section \ref{subsec:construction}.  All constraints (with or without a measured redshift) are presented in Table \ref{tbl:Multi-Images}, and their spatial distribution in A370 is shown in Figure \ref{fig:colorImg}.  The redshift distribution of spectroscopic multiple-images can be seen in Figure \ref{fig:multiHist}.

\begin{table}
  \centering
  \caption{Multiply-Imaged Systems}
  \label{tbl:Multi-Images}
  \begin{tabular}[t]{lllcc}
    \hline
    ID & RA & Dec & $z^{\rm a}$  & class$^{\rm b}$\\
    \hline
    1.1    & 39.967047   &  -1.5769172  &  0.8041     &   gold     \\  
    1.2    & 39.976273   &  -1.5760558  &  0.8041     &   gold     \\  
    1.3    & 39.968691   &  -1.5766113  &  0.8041     &   gold     \\  
    2.1    & 39.973825   &  -1.5842290  &  0.7251     &   gold     \\  
    2.2    & 39.971003   &  -1.5850422  &  0.7251     &   gold     \\  
    2.3    & 39.968722   &  -1.5845058  &  0.7251     &   gold     \\  
    2.4    & 39.969394   &  -1.5847328  &  0.7251     &   gold     \\  
    2.5    & 39.969630   &  -1.5848508  &  0.7251     &   gold     \\  
    3.1    & 39.965658   &  -1.5668560  &  1.9553     &   gold     \\  
    3.2    & 39.968526   &  -1.5657906  &  1.9553     &   gold     \\  
    3.3    & 39.978925   &  -1.5674624  &  1.9553     &   bronze   \\  
    4.1    & 39.979704   &  -1.5764364  &  1.2728     &   gold     \\  
    4.2    & 39.970688   &  -1.5763221  &  1.2728     &   gold     \\  
    4.3    & 39.961971   &  -1.5779671  &  1.2728     &   gold     \\  
    5.1    & 39.973473   &  -1.5890463  &  1.2775     &   gold     \\  
    5.2    & 39.970576   &  -1.5891946  &  1.2775     &   gold     \\  
    5.3    & 39.969472   &  -1.5890961  &  1.2775     &   gold     \\  
    5.4    & 39.968580   &  -1.5890045  &  1.2775     &   gold     \\  
    6.1    & 39.969405   &  -1.5771811  &  1.0633     &   gold     \\  
    6.2    & 39.964334   &  -1.5782307  &  1.0633     &   gold     \\  
    6.3    & 39.979641   &  -1.5770904  &  1.0633     &   gold     \\  
    7.1    & 39.969788   &  -1.5804299  &  2.7512     &   gold     \\  
    7.2    & 39.969882   &  -1.5807608  &  2.7512     &   gold     \\  
    7.3    & 39.968815   &  -1.5856313  &  2.7512     &   gold     \\  
    7.4    & 39.986567   &  -1.5775688  &  2.7512     &   gold     \\  
    7.5    & 39.961533   &  -1.5800028  &  2.7512     &   gold     \\  
    8.1    & 39.964485   &  -1.5698065  &  \{2.884 $\pm$ 0.084\} &   silver   \\  
    8.2    & 39.961889   &  -1.5736473  &  \{2.884 $\pm$ 0.084\} &   silver   \\  
    9.1    & 39.962402   &  -1.5778911  &  1.5182     &   gold     \\  
    9.2    & 39.969486   &  -1.5762654  &  1.5182     &   gold     \\  
    9.3    & 39.982022   &  -1.5765337  &  1.5182     &   gold     \\  
    10.1*,**& 39.968142   &  -1.5710778  &  2.7512     &   --       \\  
    10.2** & 39.968454   &  -1.5715778  &  2.7512     &   gold     \\  
    10.3** & 39.969046   &  -1.5774833  &  2.7512     &   copper   \\  
    11.1   & 39.963839   &  -1.5693802  &  \{7.040 $\pm$ 0.181\} &   silver   \\  
    11.2   & 39.960789   &  -1.5741702  &  \{7.040 $\pm$ 0.181\} &   silver   \\  
    11.3   & 39.988460   &  -1.5719676  &  \{7.040 $\pm$ 0.181\} &   copper   \\  
    12.1   & 39.969682   &  -1.5666360  &  3.4809     &   gold     \\  
    12.2   & 39.959198   &  -1.5753221  &  3.4809     &   gold     \\  
    12.3   & 39.984100   &  -1.5709127  &  3.4809     &   gold     \\  
    13.1   & 39.979513   &  -1.5717782  &  4.2480     &   gold     \\    
    13.2   & 39.975210   &  -1.5688203  &  4.2480     &   gold     \\  
    13.3   & 39.956759   &  -1.5775032  &  4.2480     &   gold     \\  
    14.1   & 39.972309   &  -1.5780910  &  3.1309     &   gold     \\  
    14.2   & 39.972192   &  -1.5801027  &  3.1309     &   gold     \\  
    14.3   & 39.974254   &  -1.5855770  &  3.1309     &   gold     \\  
    14.4   & 39.981313   &  -1.5782202  &  3.1309     &   gold     \\  
    14.5   & 39.957673   &  -1.5804590  &  3.1309     &   gold     \\  
    15.1   & 39.971328   &  -1.5806040  &  3.7084     &   gold     \\  
    15.2   & 39.971935   &  -1.5870512  &  3.7084     &   gold     \\  
    15.3   & 39.971027   &  -1.5777907  &  3.7084     &   gold     \\   
    15.4   & 39.984008   &  -1.5784556  &  3.7084     &   gold     \\   
    15.5   & 39.958795   &  -1.5805488  &  3.7084     &   silver   \\  
    15.6   & 39.970391   &  -1.5696387  &  3.7084     &   copper   \\  
    15.7   & 39.970450   &  -1.5689437  &  3.7084     &   copper   \\  
    16.1   & 39.964016   &  -1.5880782  &  3.7743     &   gold     \\  
    16.2*  & 39.966037   &  -1.5890355  &  3.7743     &   --       \\  
    16.3   & 39.984414   &  -1.5841111  &  3.7743     &   gold     \\  
    17.1   & 39.969758   &  -1.5885333  &  4.2567     &   gold     \\  
    17.2   & 39.985403   &  -1.5808406  &  4.2567     &   gold     \\  
    17.3   & 39.960235   &  -1.5836508  &  4.2567     &   gold     \\  
    18.1   & 39.975830   &  -1.5870613  &  4.4296     &   gold     \\  
    18.2   & 39.981476   &  -1.5820728  &  4.4296     &   gold     \\  
    18.3   & 39.957362   &  -1.5820861  &  4.4296     &   gold     \\
    \hline
    \end{tabular}
\end{table}
\begin{table}\ContinuedFloat
\centering
\caption{(continued) Multiply-Imaged Systems}
  	\begin{tabular}[t]{lllcc}
    \hline
    ID & RA & Dec & $z^{\rm a}$  & class$^{\rm b}$\\
    \hline
    19.1   & 39.971996   &  -1.5878654  &  5.6493     &   gold     \\  
    19.2   & 39.985142   &  -1.5790944  &  5.6493     &   gold     \\  
    19.3   & 39.958316   &  -1.5813093  &  5.6493     &   gold     \\  
    20.1   & 39.965279   &  -1.5878055  &  5.7505     &   gold     \\  
    20.2   & 39.963619   &  -1.5868798  &  5.7505     &   gold     \\  
    21.1   & 39.966733   &  -1.5846943  &  1.2567     &   gold     \\  
    21.2   & 39.967252   &  -1.5849694  &  1.2567     &   gold     \\  
    21.3   & 39.981539   &  -1.5814028  &  1.2567     &   gold     \\  
    22.1   & 39.974406   &  -1.5861017  &  3.1309     &   gold     \\  
    22.2   & 39.981675   &  -1.5796852  &  3.1309     &   gold     \\  
    22.3   & 39.957906   &  -1.5810108  &  3.1309     &   silver   \\   
    23.1   & 39.9801126  &  -1.5667264  &  5.9386     &   gold     \\  
    23.2   & 39.9573149  &  -1.572744   &  5.9386     &   gold     \\  
    23.3   & 39.9771658  &  -1.5662748  &  5.9386     &   gold     \\  
    24.1   & 39.9631110  &  -1.5706030  &  4.916      &   gold     \\  
    24.2   & 39.9620400  &  -1.5723407  &  4.916      &   gold     \\  
    25.1   & 39.9870836  &  -1.5790992  &  3.8145     &   gold     \\  
    25.2   & 39.9617028  &  -1.5832126  &  3.8145     &   gold     \\  
    25.3   & 39.966982   &  -1.5867999  &  3.8145     &   bronze   \\   
    26.1   & 39.979924   &  -1.571393   &  3.9359     &   gold     \\  
    26.2   & 39.97446    &  -1.5680963  &  3.9359     &   gold     \\  
    26.3   & 39.95717    &  -1.5769717  &  3.9359     &   gold     \\  
    27.1   & 39.9806909  &  -1.5711198  &  3.0161     &   gold     \\  
    27.2   & 39.9582916  &  -1.5759068  &  3.0161     &   gold     \\  
    27.3   & 39.9724399  &  -1.5671511  &  3.0161     &   gold     \\  
    28.1   & 39.963492   &  -1.5822806  &  2.9101     &   gold     \\  
    28.2   & 39.967058   &  -1.5845583  &  2.9101     &   gold     \\  
    28.3   & 39.987817   &  -1.5774528  &  2.9101     &   gold     \\  
    29.1   & 39.968425   &  -1.5646657  &  4.4897     &   gold     \\  
    29.2   & 39.983596   &  -1.5674774  &  4.4897     &   bronze   \\   
    29.3   & 39.960838   &  -1.5691328  &  4.4897     &   silver   \\   
    30.1   & 39.9833507  &  -1.5704081  &  5.6459     &   gold     \\  
    30.2   & 39.972404   &  -1.5663533  &  5.6459     &   gold     \\  
    31.1   & 39.9747496  &  -1.5693301  &  5.4476     &   gold     \\  
    31.2   & 39.980667   &  -1.5747346  &  5.4476     &   bronze   \\   
    31.3   & 39.956156   &  -1.5786786  &  5.4476     &   bronze   \\   
    32.1   & 39.9662845  &  -1.5693446  &  4.4953     &   gold     \\  
    32.2   & 39.988097   &  -1.5751871  &  4.4953     &   gold     \\  
    32.3   & 39.960682   &  -1.5783795  &  4.4953     &   gold     \\  
    33.1   & 39.9627222  &  -1.5860036  &  4.882      &   gold     \\  
    33.2   & 39.9662152  &  -1.5879961  &  4.882      &   gold     \\  
    34.1   & 39.9701070  &  -1.5701499  &  5.2437     &   gold     \\  
    34.2   & 39.971805   &  -1.5880395  &  5.2437     &   gold     \\  
    34.3   & 39.9585664  &  -1.5817008  &  5.2437     &   gold     \\  
    34.4   & 39.985048   &  -1.579559   &  5.2437     &   gold     \\  
    35.1   & 39.981538   &  -1.5658624  &  6.1735     &   gold     \\  
    35.2   & 39.9758248  &  -1.5644423  &  6.1735     &   gold     \\  
    36.1   & 39.9624435  &  -1.5807098  &  6.2855     &   gold     \\  
    36.2   & 39.965996   &  -1.5843845  &  6.2855     &   gold     \\  
    37.1**   & 39.9703912  &  -1.5687943  &  5.6489     &   gold     \\  
    37.2**   & 39.970428   &  -1.5694203  &  5.6489     &   gold     \\  
    38.1   & 39.977154   &  -1.5737917  &  3.20     &   gold     \\  
    38.2   & 39.975063   &  -1.5721045  &  3.20     &   gold     \\  
    39.1   & 39.970546   &  -1.5693801  &  4.9441     &   copper   \\   
    39.2   & 39.969977   &  -1.5700367  &  4.9441     &   copper   \\   
    39.3   & 39.985223   &  -1.5793885  &  4.9441     &   copper   \\   
    39.4   & 39.971395   &  -1.5880200  &  4.9441     &   copper   \\       
    40.1   & 39.970632   &  -1.5710393  &  4.3381     &   copper   \\    
    40.2   & 39.983162   &  -1.5796664  &  4.3381     &   copper   \\    
    40.3   & 39.973383   &  -1.5874465  &  4.3381     &   copper   \\    
    40.4   & 39.957967   &  -1.5815081  &  4.3381     &   copper   \\   
    41.1   & 39.965442   &  -1.5780222  &  \{1.272 $\pm$ 0.016\} &   silver   \\  
    41.2   & 39.967933   &  -1.5773472  &  \{1.272 $\pm$ 0.016\} &   silver   \\  
    41.3   & 39.982296   &  -1.5769750  &  \{1.272 $\pm$ 0.016\} &   silver   \\  
    \hline
    \end{tabular}
\end{table}
\begin{table}\ContinuedFloat
\centering
\caption{(continued) Multiply-Imaged Systems}
  	\begin{tabular}[t]{lllcc}
    \hline
    ID & RA & Dec & $z^{\rm a}$  & class$^{\rm b}$\\
    \hline    
    42.1   & 39.963579   &  -1.5656333  &  1.0315     &   gold     \\  
    42.2   & 39.962958   &  -1.5661111  &  1.0315     &   gold     \\  
    42.3   & 39.963375   &  -1.5659528  &  1.0315     &   gold     \\  
    43.1   & 39.962117   &  -1.5752500  &  \{1.973 $\pm$ 0.034\} &   bronze   \\  
    43.2   & 39.984054   &  -1.5733556  &  \{1.973 $\pm$ 0.034\} &   copper   \\  
    43.3   & 39.966563   &  -1.5696694  &  \{1.973 $\pm$ 0.034\} &   bronze   \\  
    44.1   & 39.977717   &  -1.5827917  &  \{2.336 $\pm$ 0.027\} &   silver   \\  
    44.2   & 39.976646   &  -1.5836028  &  \{2.336 $\pm$ 0.027\} &   silver   \\  
    45.1   & 39.982400   &  -1.5811000  &  \{8.593 $\pm$ 0.432\} &   copper   \\  
    45.2   & 39.975875   &  -1.5872600  &  \{8.593 $\pm$ 0.432\} &   copper   \\  
    \hline
   \end{tabular}
  \medskip\\
  \begin{flushleft}  
  $^{\rm a}$ Unless otherwise specified (i.e., a measurement having fewer than 4 significant digits), spectroscopic redshifts have an uncertainty $\delta z = \pm 0.0001$.  Redshifts enclosed in braces are not spectroscopic, instead fit by the model as free parameters.\\[2pt]
  $^{\rm b}$ Objects are classified according to the FF Public Modeling Challenge convention (see Section \ref{subsec:construction}.)  \emph{Gold} systems are considered the most reliable constraints, while \emph{copper} systems are the most uncertain.\\[2pt]
  $^*$ These systems are not used as constraints in our mass model.  Our best-fitting solution merges Image 10.1 and 10.2 into a single constraint, while Image 16.2 is predicted but is not seen in either MUSE or \emph{HST} data (see details in \citetalias{lag17}). \\[2pt]
  $^{**}$ While originally identified as unique systems, updated modeling suggests they are in fact counterimages of other objects (Section \ref{subsubsec:complexGroups}).\\
  \end{flushleft}
\end{table}

\subsubsection{Cluster Members}
\label{subsec:catCluster}

Cluster members represent another significant component of the spectroscopic catalog, and as in \citetalias{lag17} they make up the majority of redshifts measured in the field.  With a larger sample of cluster members, we can better characterize small-scale mass components in the A370 lens model, giving us a more complete picture of the total mass distribution.  At the same time, an accurate representation of the (3-dimensional) positions of these galaxies provides information about the internal structure and dynamics of the cluster itself.  

To identify these galaxies from the catalog directly, we simply select all objects with a redshift close to the A370 systemic value ($z_{\rm sys} = 0.375$) and treat these systems as cluster members.  Although we do not initially apply a hard limit to this selection, we find that the range from $z = 0.35$ to $z = 0.4$ is considerably overdense compared to the rest of the catalog, and the redshift distribution of galaxies within this range are well-fit by a Gaussian with $\mu = 0.3745$ and $\sigma = 0.0077$, corresponding to a physical velocity dispersion $\sigma_{V} = 1680$ km s$^{-1}$, which is consistent with a massive lensing cluster.  We therefore take the $0.35 \leq z \leq 0.4$ limits to be the A370 cluster boundary, which is only slightly larger than a 3-$\sigma$ spread about the Gaussian mean.  This matches the redshift parameter space we use in the GTO catalog, and we note that a Gaussian redshift distribution is expected for a system that is gravitationally bound and dynamically relaxed \citep[e.g.,][]{wen13} -- though any deviation from a pure Gaussian can be a sign of substructure \citep[e.g.,][]{gir15}.  While we could in principle apply a radial cutoff to the selection criteria as well, the new MUSE mosaic is well within the virial radius of the cluster ($r_{\rm vir} \sim 1$ Mpc, corresponding to $\sim$195\arcsec\ at $z = 0.375$), so an assumption that all objects are cluster members is not especially unwarranted.  Applying these limits, we find a total of 244 cluster objects, representing a four-fold increase from the GTO catalog.  

We easily identify the two brightest cluster galaxies (BCGs) in the catalog, and note that each has a different redshift, with the northern galaxy (catalog ID 11917; $z = 0.3780$) slightly redder than $z_{\rm sys}$ and the southern galaxy (ID 8844; $z = 0.3733$) slightly bluer.  However, unlike other merging clusters with significant MUSE coverage, such as Abell 2744 \citep{mah18} or MACSJ0416 (\citealt{cam17a}; Richard et al., in prep) we do not see distinct sub-populations of cluster redshifts centred around each BCG.  Instead, the single-peaked Gaussian used to initially identify the cluster is a good fit to the entire population.  This may imply that the merger is occurring nearly face-on, or -- given the elongated X-ray contours \citep[see e.g.,][ \citetalias{lag17}]{ric10} and similarly-shaped centralized mass distribution (Section \ref{subsec:results}) -- that the merging components have already passed through each other once, allowing the two populations to mix.  

Aside from the BCGs, we also investigate the bright northern ``crown'' of galaxies presented in \citetalias{lag17}, finding that (as we expect) all have redshifts consistent with being cluster members.  Projecting these galaxies into 3-dimensional space, we find that they are not tightly grouped together but instead circle around the outskirts of the cluster, approximately creating a ring above the BCGs.  While this would suggest that the crown galaxies are not strictly associated with each other and instead a 2-D projection of several galaxies falling towards the central potential, we do not include peculiar velocity effects in our projections, which can create additional separation in redshift space.  We note, however, that the three bright galaxies in the eastern-most section of the crown do appear close together and have a small velocity separation ($\Delta V \sim 325$ km s$^{-1}$).  This coincides with the location of the ``crown clump'' of the GTO model, providing additional physical motivation for this mass component. 

\begin{figure}
\includegraphics[width=0.5\textwidth]{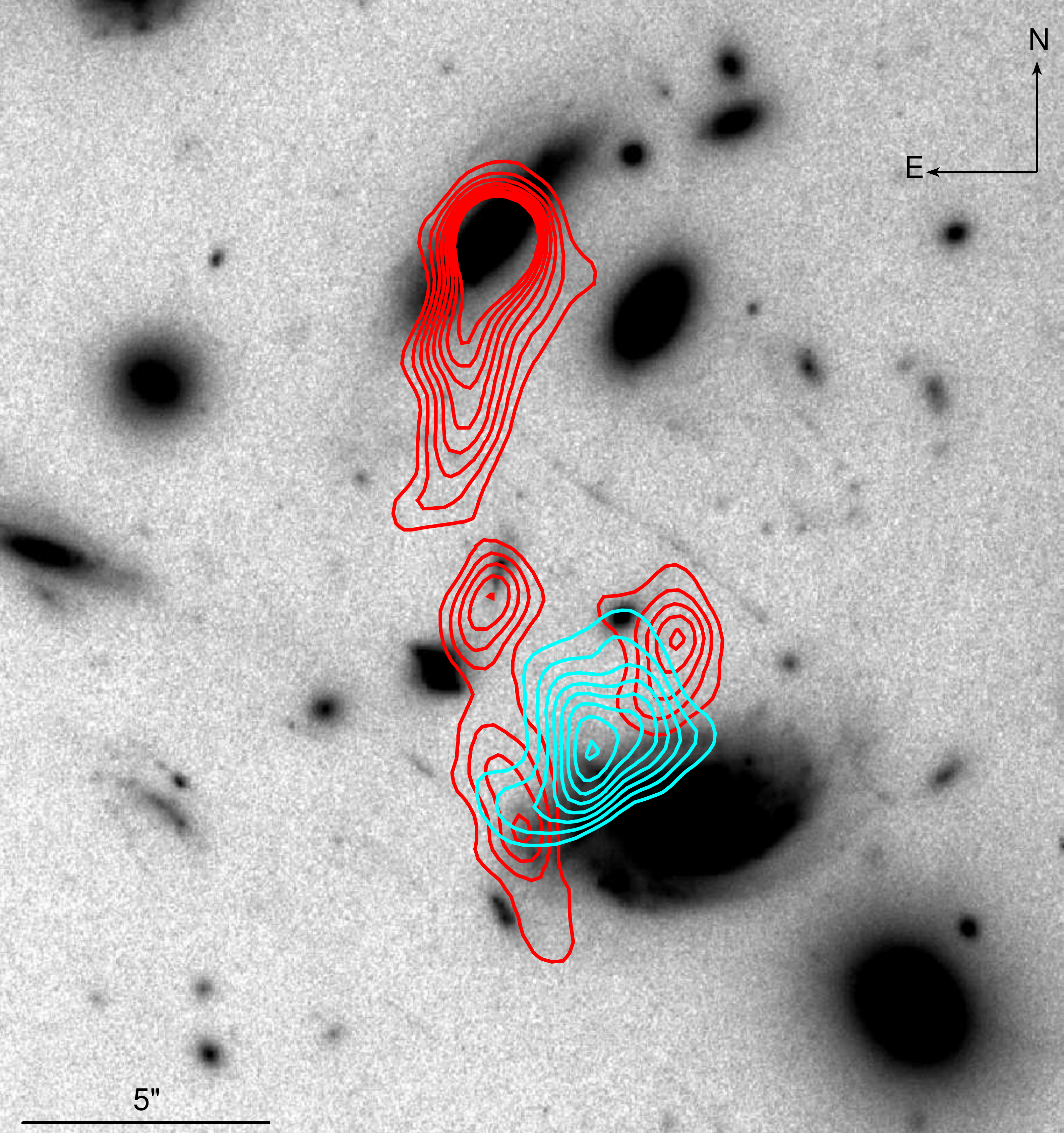}
\caption{[\ion{O}{II}] emission associated with the unusual object CL49 from \citetalias{lag17} (cyan contours; $\lambda = 5157$ \AA) and a nearby jellyfish galaxy (ID 8006; red contours; $\lambda = 5167$ \AA).  Stepping through the MUSE cube we find that the emission flows seamlessly from the cyan contours to the red, suggesting they are part of the same ``tail'' flowing out of the jellyfish galaxy at an offset velocity $\Delta V \sim 420$ km s$^{-1}$}
\label{fig:jellyfish}
\end{figure}

While the majority of cluster members have only absorption features in their spectra and appear as passive, early-type galaxies in \emph{HST} imaging, nearly 40 show [\ion{O}{II}] or H $\alpha$ emission (or both).  Most emission-line objects appear near the outskirts of the cluster and have redshifts close to the low- or high-end of the redshift distribution, suggesting that they are infalling or otherwise interacting with the intra-cluster medium (ICM).  The most prominent examples of this are four jellyfish galaxies which show evidence of extreme tidal stripping and have prominent ``tails'' of shocked gas in the MUSE cube.  Three of these galaxies (IDs 15715, 12808, and 16798) are located in the north of the cluster, near the crown.  The fourth (object 8006) is further to the south and positioned eastward of the southern BCG.   We explicitly mention this object (particularly its tail of shocked gas), as it helps to resolve a peculiar observation seen in the GTO cube.  Specifically, in our previous paper we discussed the unusual cluster member CL49, an object with a divergent velocity field and no {\it HST} counterpart.  Thanks to the larger mosaic data, we now see that this emission is simply the bottom section of the jellyfish gas stream (Figure \ref{fig:jellyfish}).  

\subsubsection{Foreground and Background Objects}
\label{subsec:catOthers}

The final subset of the redshift catalog contains all remaining objects: stars, foreground galaxies, and non-multiply imaged background galaxies.  These help to fill in the line-of-sight picture of A370, and (in some cases) also provide information about the lens model.

We identify 11 stars in the field, including the very bright star (ID 10011) southwest of the cluster at the extreme edge of the MUSE footprint.  Moving outside of the Milky Way, we detect 57 galaxies in front of A370, with redshifts between $z = 0.1$ and $z = 0.35$.  A large fraction of these objects are late-type galaxies with blue colours and prominent H $\alpha$ and [\ion{O}{II}] emission, though six show only absorption features in their spectra and have more elliptical morphologies.  While the foreground systems do not display any obvious spatial clustering in the HFF data -- galaxies are more or less evenly distributed throughout the mosaic footprint -- we do find slight spectral overdensities at $z \sim 0.26$ and $z \sim 0.32$ which could indicate the presence of coincident galaxy groups.  Given their relative paucity, it is unlikely that these candidate groups will significantly affect the overall mass model, but we investigate this more thoroughly in Section \ref{sec:substruc}.

There are considerably more background galaxies in the catalog, especially in the range from just behind the cluster to the start of the MUSE redshift desert.  In particular, we identify 134 (singly-imaged) galaxies between $z = 0.4$ and $z = 1.5$ which, like their foreground counterparts, largely appear blue and pink in optical (ACS) colour and show prominent emission lines -- though we do find eight purely passive early-type galaxies randomly spread throughout the mosaic.  The observed emission lines are predominantly [\ion{O}{II}] and [\ion{O}{III}], as H $\alpha$ redshifts out of MUSE spectra at $z = 0.425$.  While some objects appear to be well-defined late type galaxies, a larger fraction have clumpy, irregular morphologies, suggesting that many of these systems are actively forming stars.  One interesting example is Object 15491, located in the far north of the mosaic.  This galaxy has a strong blue power-law continuum, and several emission features with extremely high equivalent widths, including a strong \ion{Mg}{II}[$\lambda2797,2804$] doublet.  While weak extended emission can be seen in the HFF imaging, it has a bright point-like centre suggesting the object is an AGN. 

As in the foreground, galaxies in this redshift range are evenly spread throughout the mosaic and largely have a flat redshift distribution.  However, we do find one prominent redshift overdensity at $z \sim 1.05$, where the number of objects is more than double any other redshift bin.  This overdensity, which is also detected in \citet{die18}, consists of a significantly larger number of galaxies than either of the foreground groups, and may be indicative of additional substructure.  As background objects, these galaxies also have a much better chance of affecting the positions of observed lensing constraints, so we investigate their effects on the mass model in Section \ref{sec:substruc}.

Within the redshift desert, we find two objects (ID 10793; $z = 2.3830$ and ID 13028; $z = 2.8050$) with moderate \ion{C}{III}] emission in the northeast and northwest of the cluster, respectively, above the crown. We also identify a strong \ion{C}{III}] emitter in the southwest of the cluster (ID 4139; $z = 1.9655$) with a bright UV continuum and several other emission and absorption features, including \ion{Fe}{II}$^*$ [$\lambda\sim2400\AA$,$\lambda\sim2600\AA$] emission, which is useful for probing galactic winds \citep[see e.g.][]{fin17}. Additionally, we identify a fourth potential \ion{C}{III}] emitter (ID 11481; $z = 2.5607$) in the northeast, and three objects with weak UV absorption features (ID 6655; $z = 2.1240$, ID 13600; $z = 2.5442$, and ID 14822; $z = 2.8897$) at the western edges of the mosaic, but we stress that these identifications are made at low confidence (C = 1) and could simply be noise.

Finally, we detect 14 singly-imaged galaxies in the catalog at distances beyond the redshift desert ($z \ga 3$).  Given the high redshifts of these objects, they are largely located in the outskirts of the mosaic, outside of the A370 multiple-image zone.  In nearly all cases, galaxies are identified by strong Lyman $\alpha$ emission, though we do detect two Lyman-break systems with several other prominent UV-continuum absorption features (ID 12137; $z = 4.2510$ and ID 14650; $z= 4.2561$) in the northern region of the cluster.  With the exception of the very lowest redshift galaxies ($z \sim 2.9$) and the Lyman break objects, most systems are barely resolved in the HFF imaging with limited morphological information.  Likewise, the extracted spectra often show little to no continuum emission, relying solely on the Lyman $\alpha$ feature for redshift identification.  The typical line flux for the Lyman $\alpha$ emission is $2.5 \times 10^{-18}$ erg s$^{-1}$ cm$^{-2}$ \AA$^{-1}$ which, when coupled by the relatively low magnification values ($\mu \sim 2.3$) associated with these sources, indicates that the lines are intrinsically bright, especially when compared to their multiply-imaged counterparts at similar redshifts.

\section{Lens Modeling}
\label{sec:modeling}

\subsection{Method}

To model the total cluster mass we use the publicly-available software \texttt{LENSTOOL}\footnote{\url{https://projets.lam.fr/projects/lenstool/wiki}} \citep{kne96,jul07,jul09}.  Based on a ``parametric'' approach, \texttt{LENSTOOL} constructs an overall distribution using a series of individual components, each with an analytic profile. This is in contrast to ``non-parametric'' methods such as WSLAP+ \citep{die07}, which utilize a free-form grid of pixels without regard to a particular shape. The specific processes we employ in this work mirror those used in  \citetalias{lag17}, and a complete description of the procedure can be found there.  However, to aid the reader we briefly describe the general technique here.  

We fit the system with a set of elliptical mass potentials, including both large cluster-scale haloes and smaller galaxy-scale masses.  Each potential is constructed with a dual Pseudo-Isothermal Elliptical profile (dPIE; \citealt{eli07}), a distribution that matches empirical cluster data well and has enough flexibility to model variations at all scales.  The basic dPIE halo is described by seven parameters, including position ($\alpha$ and $\delta$), position angle ($\theta$), ellipticity ($\epsilon$), central velocity dispersion ($\sigma_0$), and two scale radii ($r_{\rm core}$ and $r_{\rm cut}$) which modify the inner and outer mass slopes of the profile, respectively.  To optimize a given model, we take known multiple image systems (Section \ref{subsec:catMulti}) as constraints, minimizing the rms distance between the model-predicted locations of these objects and their actual positions on the sky.  \texttt{LENSTOOL} generates a model prediction by first transforming constraints from the lens plane (observed positions) to the source plane (intrinsic, undeflected positions), grouping all counterimages of a given system together, and then calculating the barycentre coordinates of each group.  These barycentre positions are then transformed back to the lens plane, where they can be compared to the original observations.  Throughout this process, the code continually updates and tests new parameter configurations, using a Bayesian Markov Chain Monte Carlo (MCMC) sampling routine to search for the maximum likelihood region of parameter space.  Once this region is discovered,  \texttt{LENSTOOL} probes the area using a fixed number of MCMC realizations, treating the most probable configuration as the best-fitting model and averaging the others to determine parameter uncertainties.  The exact number of sample iterations is specified by the user, which in this work we set to 100. 

During model optimization we allow all parameters of the cluster-scale potentials to vary freely, with the exception of  $r_{\rm cut}$, which we fix at 800 kpc (155\arcsec) since the mosaic data still do not extend far enough to put a meaningful constraint on its value.  Conversely, we fix many of the parameters of the galaxy-scale haloes, since these components represent individual cluster-member galaxies and their values can be measured from the data directly.  In particular, we match the $\alpha$, $\delta$, $\epsilon$, and $\theta$ parameters for each galaxy to their observed values in the HFF $F814W$ image.  To further reduce the model parameter space, we do not fit the remaining galaxy-scale parameters independently.  Instead, we only model the parameters of a typical $L^*$ galaxy, then calibrate all other potentials with an empirical scaling relation, given by
\begin{equation}
\begin{split}
\sigma_{0,\rm gal} &= \sigma_0^* \left(\frac{L_{\rm gal}}{L^*}\right)^{1/4},\\
r_{\rm core, gal} &= r_{\rm core}^* \left(\frac{L_{\rm gal}}{L^*}\right)^{1/2},\\
r_{\rm cut, gal} &= r_{\rm cut}^* \left(\frac{L_{\rm gal}}{L^*}\right)^{1/2},
\end{split}
\label{eqn:scaling}
\end{equation}
where $\sigma_0^*$, $r_{\rm core}^*$, and $r_{\rm cut}^*$ are the optimized components of the $L^*$ galaxy.  We note, however that we fix the value of $r_{\rm core}^*$ to 15 kpc, as it is difficult to constrain a true core radius for small-scale mass distributions.  Additionally, we limit the variation of $r_{\rm cut}^*$ to be between 10 and 30 kpc, in order to account for the effects of tidal stripping within the cluster \citep[e.g.][]{hal07}, and because there is often a strong degeneracy between $r_{\rm cut}^*$ and $\sigma_0^*$.  For reference, $L^* = 3.19 \times 10^{10} L_\odot$ at $z = 0.375$, corresponding to an apparent magnitude $m_{F814W} = 19.78$.

\subsection{Model Construction}
\label{subsec:construction}

Choosing a robust set of model components (both potentials and multiple-image constraints) is crucial to this work, as it allows us to reproduce the true distribution of A370 as accurately as possible.  Since there is no direct evidence to constrain the number of large-scale dark matter haloes in the cluster, we begin by mirroring the GTO model.  Namely, we include four such mass components in our initial guess: two located around the BCGs, a third ``bridge'' clump connecting the BCG haloes and flattening the central mass profile, and finally the crown clump in the northeast of the cluster.  However, after the initial model run, we experiment with different numbers of cluster-scale potentials (Section \ref{subsec:results}) to see if an alternative parametrization is more appropriate.

When selecting galaxy-scale potentials, we first investigate sources in the catalog that fall in the cluster redshift range ($0.35 \leq z \leq 0.4$).  While we can in principle include all of these galaxies as potentials, doing so would make the model needlessly complex, as faint (i.e., low-mass) cluster members do not significantly contribute to the total mass budget.  From past experience we find that applying a nominal magnitude cutoff ($m_{F814W} < 22.6$) keeps the model computationally manageable without affecting its accuracy, so we maintain these ``bright'' cluster members as model components and disregard the rest.  To supplement these galaxies, we apply a colour-colour cut to the entire HFF field in order to select any potential cluster members without a spectroscopic redshift.  Using the confirmed cluster galaxies as a guide, we find that the dual colour-magnitude red-sequence given by
\begin{equation}
\begin{split}
(-0.11 \times m_{F814W}) + 4.06 \leq (m_{F435W} - m_{F606W}) \\
(-0.11 \times m_{F814W}) + 4.45 \geq (m_{F435W} - m_{F606W})
\end{split}
\end{equation}
and
\begin{equation}
\begin{split}
(-0.04 \times m_{F814W}) + 1.74 \leq (m_{F606W} - m_{F814W}) \\
(-0.04 \times m_{F814W}) + 1.93 \geq (m_{F606W} - m_{F814W})
\end{split}
\end{equation}
can accurately detect these objects.  While this selection function does identify some additional galaxies, the gain is small (15 galaxies) and limited to the far outskirts of the cluster, thanks to the large spectroscopic coverage of the MUSE mosaic.  Combining all systems, we have a total of 126 galaxy-scale components in the model.  Of these, nearly all are included in the $L^*$-based scaling relation, but we remove six special objects from the list (see Fig.~\ref{fig:MassProperties}) and model them separately.  In particular, we remove the two BCGs because we do not expect them to follow the same empirical relation \citep[e.g.,][]{ric10}, while an additional four galaxies (IDs 6023, 12084, 12305, and 12718; see Table \ref{tbl:ModParams}) lie close enough ($< \sim2\arcsec$) to a multiple image constraint to induce a galaxy-galaxy lensing signal.  Because of the high magnification induced by the galaxy-galaxy interaction, these objects are given an undue weight in the scaling relation, leading to a biased parameter fit.

For the multiply-imaged features, we include all of the systems identified in the spectroscopic catalog, as they provide the strongest constraints on the model.  However, we also add additional systems with less secure (or missing) redshifts, as these objects can still reveal information about the cluster mass distribution.  While we do not apply weights to any multiple-image constraint, we do classify images based on their overall reliability, following the convention of the HFF public modeling challenge.  As part of the challenge, any modeler can suggest an image to add to the constraint set (either a known multiple-image or a new candidate) and all groups then vote on the object's inclusion.  Based on the results of the vote, images can be classified in one of three categories: \emph{gold}, \emph{silver}, or \emph{bronze}, or remain uncategorized and not included in the challenge set.  

Images in the \emph{gold} category are considered the most reliable constraints, with a broad agreement between groups that they are real multiple-images.  These systems often have a distinct and recognizable morphology and a well-defined redshift -- in fact, based on the rules of the challenge a \emph{gold}-class constraint is \textit{required} to have a spectroscopic redshift.  Similarly, \emph{silver}-class objects are also considered reliable, but may be missing an essential component, such as a distinct morphological identification or a spectroscopic redshift.  While these images can generate some disagreement between modeling groups, they are still highly-ranked by a majority of teams.  Constraints in the \emph{bronze}-class are less certain than the other categories -- often, these images have only a tentative spectroscopic redshift and/or an unusual feature (such as a dissimilar shape or colour) that makes them appear different from their counterimages.  However, there is still enough of an agreement between groups to include them in the challenge.  Any remaining candidate systems (which generate even more disagreement between modeling teams) are dropped from consideration.

The A370 challenge considers a total of 150 candidate lens images, gathered from \citetalias{lag17}, \citetalias{die18}, and the new systems presented in this work.  Of all images considered, a vast majority (103 images) are rated as \emph{gold}-class constraints, along with a small minority of \emph{silver}-class (12 images) and \emph{bronze}-class (7 images) objects as well.  Although there are only three categories used in the public modeling challenge, our model of A370 includes a fourth category, which we label as \emph{copper} constraints.  While these objects are not formally considered as part of the modeling challenge, their exclusion can often be explained by special circumstances, and we believe they are all real multiply-imaged systems.  In particular, Systems 39 and 40, and Images 15.6 and 15.7  (all of which have spectroscopic redshifts) were only discovered after the modeling challenge had finished, while System 45 (a $z > 7$ dropout candidate identified by \citealt{ish18}) was suggested too late in the process to be voted on by all groups.  The remaining \emph{copper}-class constraints (such as Images 10.3 and 11.3) have been evaluated by the modeling teams, but did not receive enough favorable votes to make the final cut.  In spite of this, we feel there is enough additional evidence, such as shape and multi-band colour, to include these images regardless.  Overall, we include 15 \emph{copper}-class images in our model, which, when combined with all other constraints leads to a grand total of 137 multiple-image objects in the field.  The final classification of all ``good'' objects can be seen in Table \ref{tbl:Multi-Images}. 

We note that the additional systems identified by \citetalias{kaw18} were not available during the modeling challenge voting period.  Since these objects have no votes and also no spectroscopic data, even here, we do not include them as model constraints.  However, we use them after the fact as predictive images, in order to see if they are consistent with our new model (Section \ref{subsubsec:others}).  

\subsection{Results}
\label{subsec:results}

After selecting the appropriate set of model components and constraints, we input everything into \texttt{LENSTOOL} and proceed to measure a best-fitting mass distribution for A370.  Initially we only include \emph{gold}-class images as model constraints, as this avoids potentially biasing the fit with less reliable data.  However, we progressively add additional constraints as the model develops.  

\subsubsection{The \emph{gold} model}
In order to probe as broad an area of parameter space as possible, we place large uniform priors on all variables during optimization, centring the limits of each parameter's prior on its corresponding GTO model value.  Based on the results of this \emph{gold}-model fit, we find that the shape and magnitude of the mass distribution remain broadly similar to the GTO model, though some individual components have significantly shifted. In particular, relative to  \citetalias{lag17} the two cluster-scale mass clumps near the northern BCG rearrange themselves from a V-shaped configuration to a more linear shape, with the eastern clump moving slightly northward, and the western clump moving further south, creating a more obvious bridge between the northern and southern cluster haloes.  The relative PAs of these components remain the same, however, and still create boxy iso-mass contours that trace the shape of the X-ray gas.  We note that the GTO V-configuration is likely due to the limitations of using purely elliptical potentials, rather than a true configuration of the mass distribution, and that a linear configuration is likely closer to reality, given the orientation of the two merging clusters.  Conversely, the crown clump remains largely unchanged from its GTO orientation and mass, providing further evidence for its importance in the model.

We evaluate the quality of the fit by measuring the rms error of the constraint predictions, finding that -- despite using more than double the number of constraints -- the average rms separation only increases by a small fraction, from 0\farcs94 in the GTO model to 1\farcs08.  While this is not a poor fit overall, it is slightly larger than the 1\arcsec\ benchmark that is often used as the minimum level of acceptable rms error.  To try to improve the fit, we first attempt to add additional mass components to the system, in a similar fashion as the crown clump in the GTO model.  Unlike that situation though, there are no significant ``bad'' regions where the model predictions are highly offset from observation, and thus no obvious choice for where to position a new component.  Therefore, we generate several new trial configurations, placing a single additional potential at random points throughout the field -- but paying particular attention to the western side of the cluster where there are no large scale clumps currently.  As before, we again apply broad uniform priors on the component parameters, allowing for a considerable amount of freedom in the fit.  However, after running all configurations we find that the model rejects this new clump, assigning it an unrealistically low mass ($\sigma_0 < 100$ km s$^{-1}$) and leaving the average rms unchanged.

\subsubsection{External shear}
As an alternative, we run the modified \emph{gold} model again but with an external shear component taking the place of the new potential.  The results of this test are significantly different from our previous trials: the new model shows a marked improvement over the original, lowering the average rms to 0\farcs66.  While the new shear term does not significantly change the total enclosed mass of the cluster, it does shift mass between various large-scale components and also affects their shapes.  In particular, we notice that the two central cluster haloes (DM2 and DM3) have lower velocity dispersion parameters (by $\sim$12\%), while the northern BCG (BCG2) and a nearby cluster member that we model independently from the scaling relation (GAL2; ID 12084) increase by roughly the same amount ($\sim$15\%).  However, this is likely due to a degeneracy between the shear and the potentials, especially in the case of GAL2, which (while likely massive) does not appear to have an unusually large radius like a BCG, nor does it show exceptionally wide absorption features -- which would justify a high $\sigma_0$ value -- in its extracted spectrum.  At the same time, all large haloes become noticeably rounder with the shear model (a natural consequence of including an additional ``stretching'' term), with an average ellipticity $\epsilon = 0.35$ as compared to $\epsilon = 0.60$ in the shear-free model.  Even with this reduction, however, we note that the haloes still maintain the same orientation and continue to drive the boxy appearance of the central mass distribution.  

Although including shear improves the fit, the term itself presents a slight problem from a physical point of view: the nature of external shear (a constant, non-localized effect that uniformly distorts model constraints) is inherently non-physical, merely serving as a proxy for some unknown effect.  While a small amount of shear can account for localized perturbations such as the intrinsic variation in galaxy-scale masses, larger values ($\Gamma > 0.1$) point to a more significant issue, such as an unaccounted-for mass distribution or substructure -- which may be located a considerable distance away from the main cluster \citep[e.g.,][]{jau16}.  This is the case in our \emph{gold} model, where the magnitude of the shear term is $\Gamma = 0.128$.  We investigate the nature of this component in the next section (Section \ref{sec:substruc}) where we attempt to replace it with a more physically motivated feature.  For now, however, we maintain the shear term in our subsequent models due to the positive impact it has on the results.

\subsubsection{Incorporating further lens constraints}
\label{subsubsec:others}
\begin{figure*}
\centering{
\includegraphics[width=0.45\textwidth]{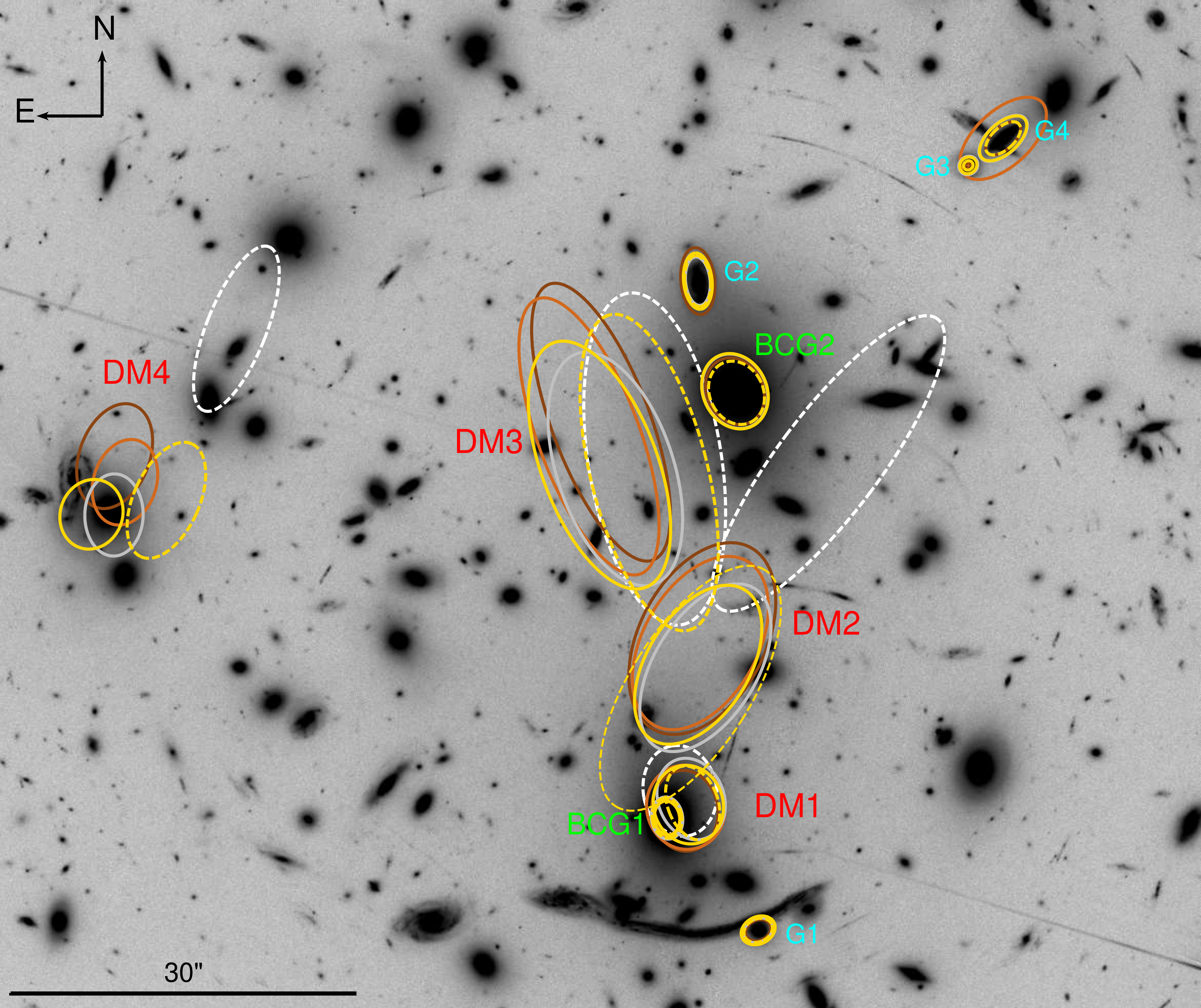}
\includegraphics[width=0.54\textwidth]{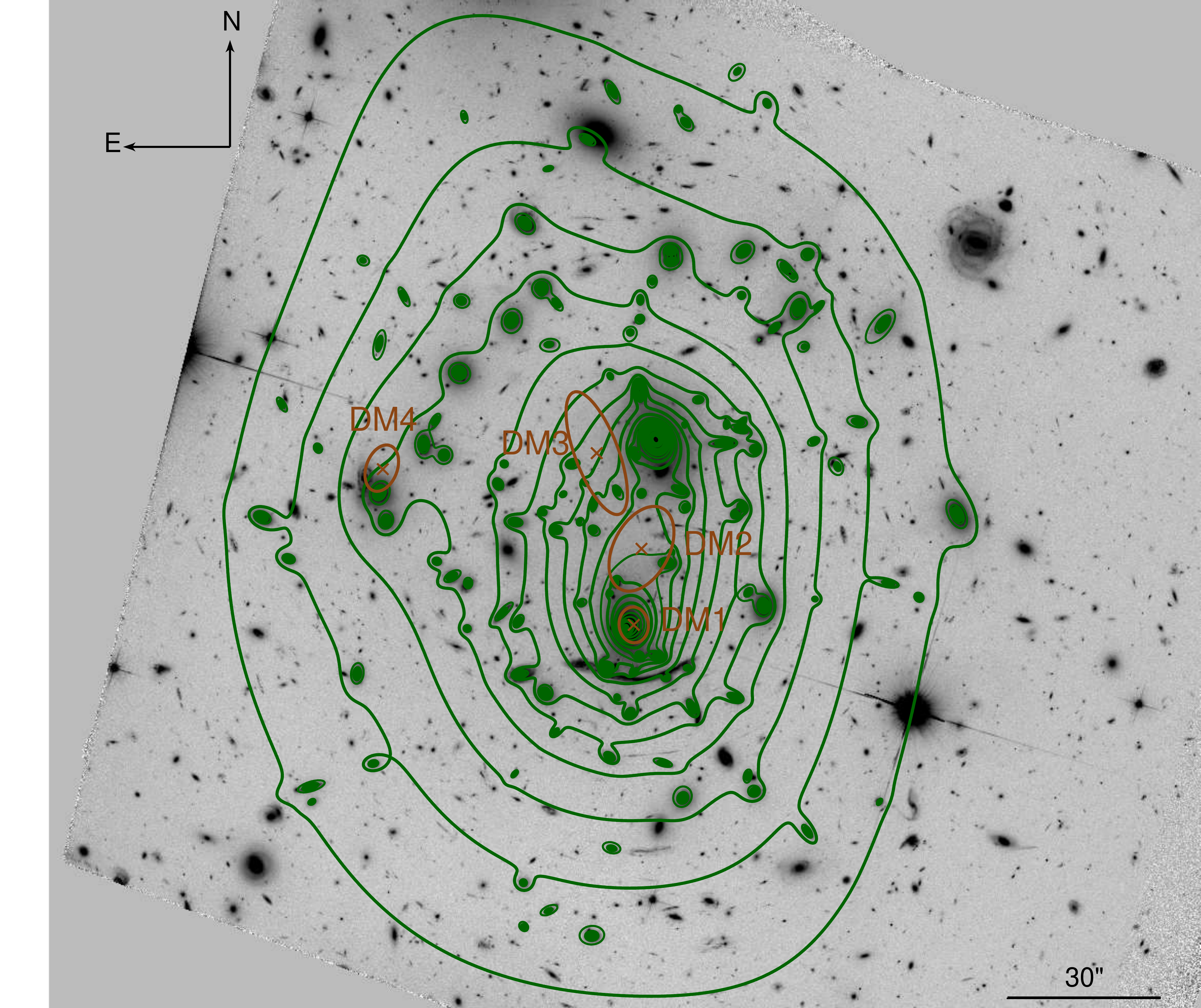}
}
\caption{Mass model properties. \textbf{Left:} Differences between the final optimized haloes of the \emph{gold}, \emph{silver}, \emph{bronze}, and \emph{copper} models (colour scheme matches Figure \ref{fig:colorImg}.)  For comparison, the final positions of the large-scale haloes in the GTO model are shown as dashed white ellipses, and those of the ``no-shear'' gold model are shown as dashed yellow ellipses.  The size of each ellipse is scaled to the halo's velocity dispersion, highlighting the relative mass of each component.  The large-scale components (DM1 to DM4) show little change from model to model -- thanks largely to the overwhelming number of \emph{gold}-class constraints -- though we do see a slight increase in ellipticity as lower-confidence multiple-image systems are added.  The crown clump (DM4) shows the most variation, though its change in position is still small (typically < 6\arcsec).  Smaller-scale masses show even less variation, and any noticeable change (such as the relative mass difference between G3 and G4) can be explained by model degeneracies.  Best-fitting values for all parameters are given in Table~\ref{tbl:ModParams}.  \textbf{Right:} Mass density contours of the best-fitting \emph{copper}-class model (green lines), along with the positions and orientations of the large-scale dark matter haloes (copper ellipses).  We again stress that the ellipses are only meant to show the relative strength and shape of each halo, and are not truly representative of their mass distributions. Contours appear at steps of $2.5\times10^8 M_\odot$ kpc$^{-2}$, starting at $5\times10^8 M_\odot$ kpc$^{-2}$.  Like the GTO model, the contours appear elliptical in the core, then gradually become boxy towards the outskirts.  As expected we see significant mass overdensities near the BCGs and the crown galaxies.}
\label{fig:MassProperties}
\end{figure*}

To explore the mass distribution further, we begin adding new constraints to the model, starting with the \emph{silver} images, followed by the \emph{bronze} and finally the \emph{copper} systems. Because the total number of \emph{gold} constraints is considerably larger than all other objects combined, the effects of the new systems are typically very small, though the total model rms error increases slightly, from 0\farcs70 in the \emph{silver} model to 0\farcs73 in the \emph{bronze} model and 0\farcs78 in the \emph{copper} model.  We do note, however, that incorporating new constraints near the central regions of the cluster seems to redistribute some mass from individual galaxies back to the large-scale haloes.  The final positions and orientations of the large-scale haloes in all models can be seen in Fig.~\ref{fig:MassProperties}.
From the figure, it is clear that -- despite having large freedom to vary -- the cluster haloes remain generally static as the model progresses from \emph{gold} to \emph{copper}.  While the crown clump potential (DM4) is the component most sensitive to newer constraints, its centroid moves by less than 6\arcsec\ over all models.  Given the typical centroid uncertainty of DM4 ($\sim$2.5\arcsec), this is consistent within $2\sigma$.  Likewise, the total mass of DM4 (within a 150 kpc aperture) varies by less than $\sim$10\%, also within the range of model uncertainty.

\begin{figure*}
\includegraphics[width=\textwidth]{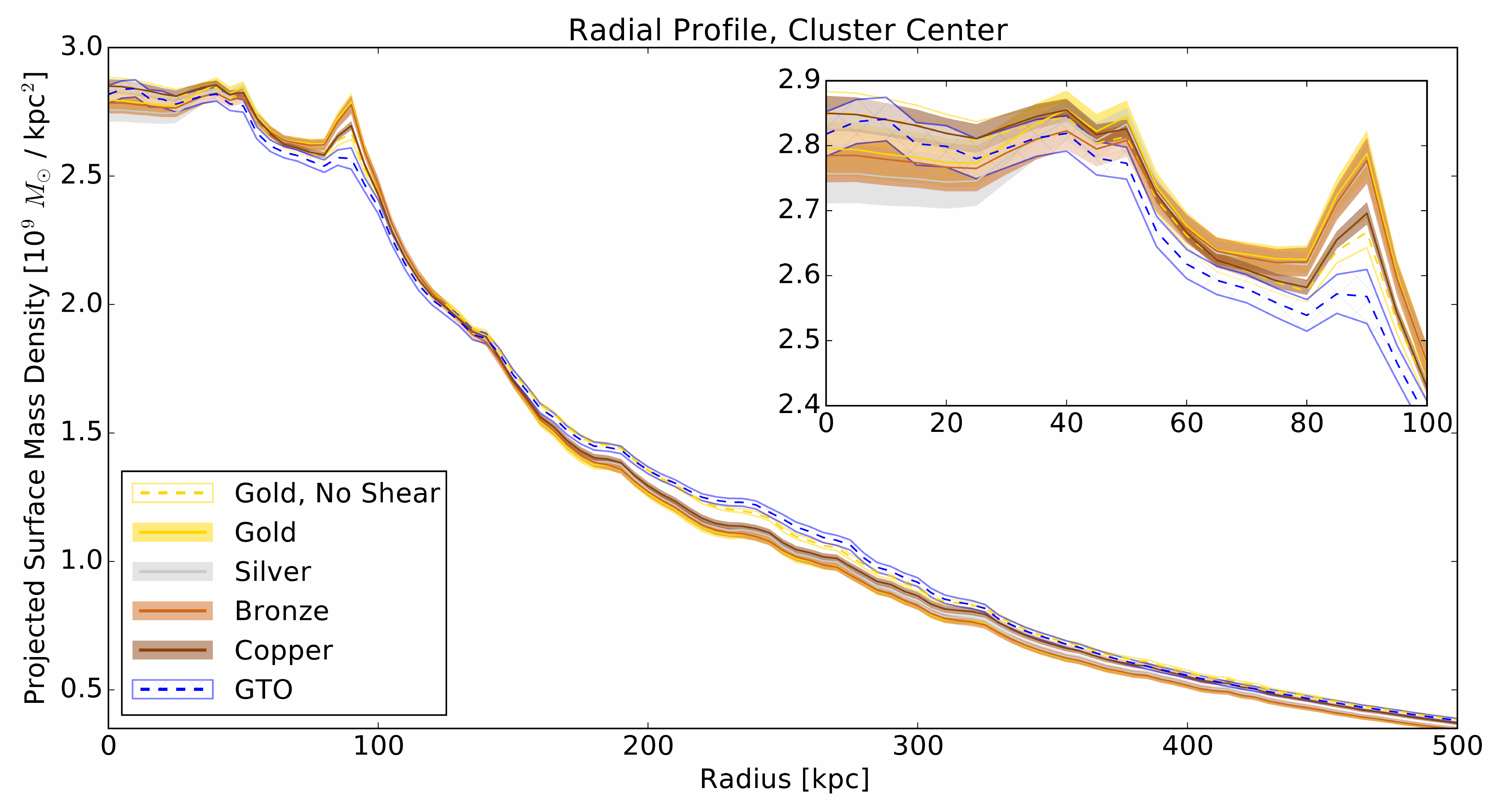}
\caption{Radial surface mass density profiles of all models.  In each case the best-fitting value is shown by a dark line, surrounded by a lighter shaded region representing model uncertainty.  The shape of the profile is similar for all models: the surface density is largely flat in the central cluster core then slowly diminishes at larger radii.  This is also consistent with other independent models of A370 \citepalias[e.g.][]{die18}, and we do expect a flat core, based on the high number of multiple images seen near the cluster centre (Section \ref{subsec:catMulti}).  The sharp peak at $\sim$90 kpc is due to the high mass of the BCGs, which are considerably overdense compared to other components.  The total magnitude of the density is also largely consistent between models in the inner cluster, given the measured errors, though the GTO and ``no-shear'' \emph{gold} models show higher densities at distances beyond 200 kpc.  This is because these models do not utilize a (massless) external shear term, and the discrepancy between the sheared and non-sheared models can point to the presence of additional mass components in the cluster outskirts.  In the inset plot we zoom in on the central 100 kpc of the cluster, better differentiating the profile shape of different models and highlighting their  good statistical agreement.}
\label{fig:RadProfs}
\end{figure*}

Similarly, we compute the radially-averaged surface mass density profile for each model, starting from the midpoint between the two BCGs ($\alpha$ = 39.970417, $\delta$ = -1.5768056) and continuing to distances well outside the cluster core (Fig. \ref{fig:RadProfs}.)  Like the individual mass components we again see a good agreement between models, both in terms of absolute value and overall profile shape.  In all cases the profile shows a flat central core, followed by a small bump at $\sim$90 kpc (corresponding to the location of the BCGs) and then a long, slowly decreasing tail.

Accounting for model uncertainties, the magnitude of the profiles are largely consistent with one another, though we note that the models that do not include external shear (the \emph{gold} no-shear model and the GTO model) have higher average densities in the cluster outskirts and beyond.  This is often a consequence of including a (massless) external shear term in the model, and can signal the presence of additional mass components at larger radii \citep[e.g.,][]{mah18}.  Measuring the logarithmic slope (d$\Sigma$) of the profile, we see different d$\Sigma$ values at different radii (though these values are again consistent between models), instead of a constant value indicating a pure power-law slope.  In the region best constrained by strong lensing (50 kpc $< R <$ 350 kpc) we find that the slope gradually steepens from d$\Sigma$ = -0.25 to d$\Sigma$ = -1.5.  This is generally consistent with an equivalent NFW halo model (and an exact match at $R \sim 200$ kpc) though our values are more extreme at the edges.  This can largely be attributed to the dPIE mass profile, which takes the form of a double broken power law that diverges from NFW.  Specifically, the dPIE surface profile has a flatter-than-NFW core region (d$\Sigma$ = 0), followed by an isothermal middle (d$\Sigma$ = -1), and a steeper-than-NFW outer region (d$\Sigma$ = -3).  Regardless of these slight changes, however, we note that our slope values are fully consistent with predictions made by $\Lambda$CDM structure formation models \citep[see, e.g.,][]{tas04}.

Because the \emph{copper} model is broadly similar to the other models and includes the largest number of lensing constraints, we subsequently adopt it as our best-fitting ``fiducial'' model, and it is the one we use to test the effects of substructure in the next section.  The major parameters of each model can be seen in Table \ref{tbl:ModParams}, and we highlight the 2D mass contours of the \emph{copper} model in Fig.~\ref{fig:MassProperties}.

Additionally, we use the \emph{copper} model to test the properties of the remaining lens systems presented in \citetalias{kaw18}.  This is done by feeding the already-optimized model and the candidate multiple-image positions to \texttt{LENSTOOL}, allowing the software to make model-based predictions of their counterimages.  By comparing the differences between the model predictions and the observed positions of the systems, we measure their rms compatibility with the model.  Since all candidates lack a known redshift, for the purposes of this test we assume the best-fitting redshift values measured in the \citetalias{kaw18} model.  After running the test, we find that the candidate multiple-image systems have an average rms error of 1\farcs5.  While this is inherently large, the result is unsurprising since the model is not specifically optimized for these objects and there is a large uncertainty on their redshift estimates.  Therefore, we conclude that the \citetalias{kaw18} systems are at least marginally consistent with the model, and some (if not all) could truly provide additional model constraints.  We note that follow-up data, such as secure spectroscopic redshifts, would help this investigation considerably. 

As a final check for missed lensing constraints, we also use \texttt{LENSTOOL} to predict the existence of counterimages in the \emph{copper} model for all remaining spectroscopically-confirmed galaxies lying behind the cluster ($z > 0.4$), finding that all such candidates have only single-image configurations.

\subsubsection{Complex multiple-image groups}
\label{subsubsec:complexGroups}
In addition to teaching us about the cluster mass distribution, our best-fitting model also provides information about the multiple-image galaxies themselves.  As mentioned in Section \ref{subsec:catMulti}, predictions made by a developed lens model can reveal the presence of additional faint counterimages initially missed by observations.  This is the case for Images 15.6 and 15.7, which lie too close to a cluster member (Object 12084) to be directly seen in imaging, and are too faint to be automatically detected by \texttt{MUSELET}.  Only by following-up a prediction from the initial \emph{gold}-class model are we able to observe faint traces of emission in narrow-band imaging, which we identify as Lyman $\alpha$ after manually extracting spectra from the MUSE cube, thus confirming the nature of the counterimage pair.  

However, model predictions can also help to ``match'' constraint systems together, signaling that objects initially thought to be independent constraints are actually part of the same system.  A prime example of this is System 37, a pair of images that is -- coincidentally -- also located near Object 12084.  Initially identified as a simple galaxy-galaxy lens pair in the north of the cluster, our updated mass model predicts three additional counterimages in the south, which happen to coincide with the positions of System 19.  Likewise, the three images in System 19 also predict additional counterimages which match to the positions of System 37.  While the redshifts of the two systems are slightly different ($z_{\rm Sys19} = 5.6493;\ z_{\rm Sys37} = 5.6489$) they are within measurement uncertainty, and the spectra of both show similar shapes in their Lyman $\alpha$ profiles (Fig. \ref{fig:SpecFigs}).  Unfortunately, contamination from intracluster light (from the BCG and Object 12084) make an appearance match between Systems 19 and 37 impossible in the HFF imaging.  However, given the strong spectral and model evidence, it is very likely that the two systems are one and the same.  To test this, we run a new \emph{gold} model where we explicitly designate the images of System 37 to be part of System 19 (as Images 19.4 and 19.5, respectively), and find that the end result is identical (within uncertainty) to the model where the two systems are kept separate. This further supports our theory.  Although we could in principle change the labeling of System 37 based on this test, we choose to keep the original numbering scheme (with separate Systems 19 and 37) in order to keep our results in line with others based on the public modeling challenge. 

Similarly, we use model predictions to investigate Systems 7 and 10, which may also be linked together.  This possibility is presented in \citet{die18}, who find that by decreasing the mass-to-light ratio of the two BCGs (which subsequently reduces the dark matter content of the galaxies), their model is able to fit all counterimages of both systems using a single source galaxy.  This is a particularly intriguing possibility, not only because a combined System 7+10 would create yet another system with a large number of counterimages ($n > 5$) and one with multiple radial arcs, but also because the proposed solution  -- a BCG that is significantly depleted of dark matter -- would be at odds with N-body simulations, and may challenge our understanding of BCG formation history \citep[e.g.,][]{new13a,new13b,lap15}.

Although we keep the two systems separate in our \emph{copper} model, we find that they fall very close to one another in the source plane, with predicted counterimages of System 7 overlapping the positions of System 10 and vice-versa.  Thus, given their identical redshifts and similar {\it HST} colours and appearance, the joint-image theory seems plausible.  Unlike \citetalias{die18}, our model does not need to alter the mass content of the BCGs for this match to work, as we still fit massive haloes around each galaxy.  However, our results do modify the configuration of System 10: rather than a merging pair of images near the northern BCG (Images 10.1 and 10.2), the model instead predicts a single, highly stretched component at the same location.  Because the actual image seen in the HFF data is considerably distorted, it is difficult to identify any distinct morphological markers which could verify whether we are seeing a single galaxy image or a merging pair.  We note, though, that a merging-pair solution would require the critical curve of our model (which would, in that case, bisect the two images) to shift from its current position by a large amount ($> 2\arcsec$), which is significantly greater than the rms uncertainty of the fit.

\section{Substructure: a replacement for external shear?}
\label{sec:substruc}

\begin{figure*}
  \includegraphics[width=0.495\textwidth]{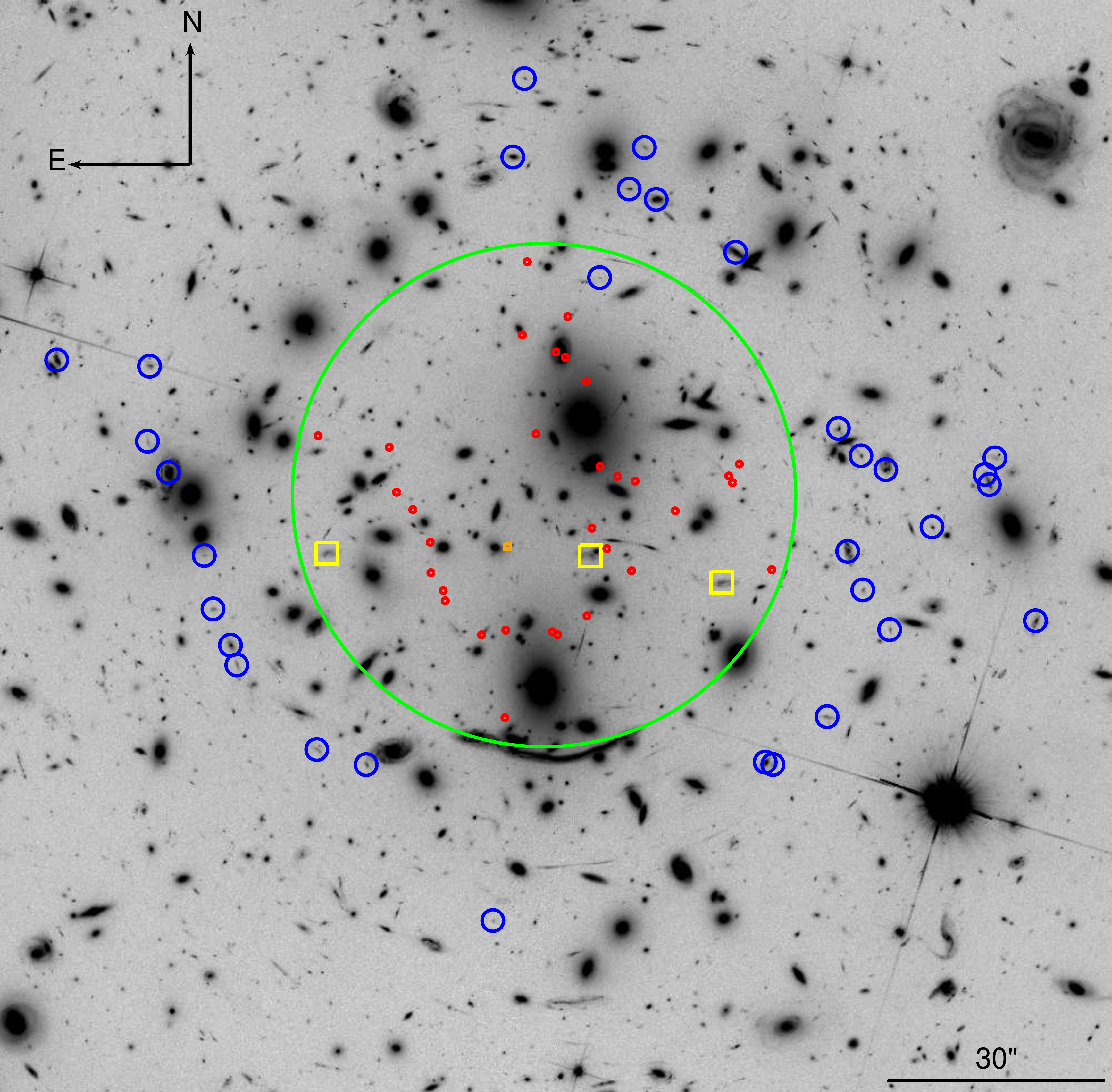}
  \includegraphics[width=0.495\textwidth]{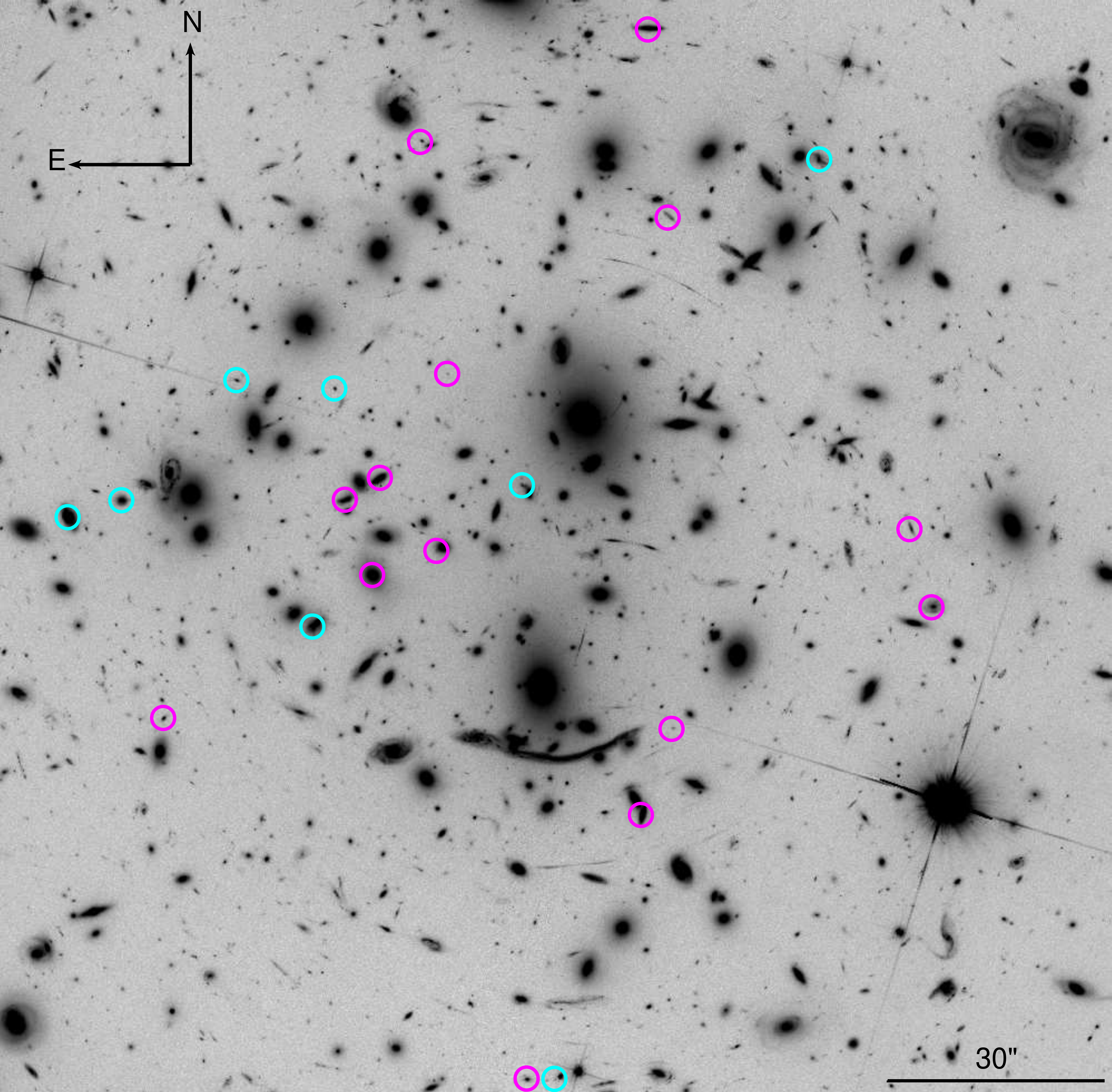}
\caption{Line-of-sight substructure candidates.  \textbf{Left:} Objects in the overdense spectral bin ($z \sim 1.05$) behind A370.  Image-plane positions of all galaxies are shown as blue circles, while their subsequent source-plane locations (assuming the best-fitting \emph{copper} model) are shown in red.  The three counterimages of System 6 (which also falls in this redshift range) and its source-plane location are instead shown as yellow and orange squares.  We investigate the effects of line-of-sight substructure by including these systems as additional components in the mass model (Section \ref{sec:substruc}.)  The green circle highlights the extent of the group in the source plane, and serves as the centroid of a group-scale mass component.  \textbf{Right:} Substructure candidates in the foreground of A370.  We identify two distinct groups in the redshift catalog, at $z \sim 0.256$ (cyan circles) and $z \sim 0.326$ (magenta circles).  The positions of the galaxies are consistent with the orientation of the external shear term in the model, and we note that many fall near the location of the crown mass clump, raising the possibility that they could be responsible for one or both components.  However, much like their background counterparts, these systems have low mass and we find that they do not significantly affect the total mass model.}
\label{fig:subGroups}
\end{figure*}

As previously mentioned, a large external shear term can signal the presence of substructure or other unidentified mass in the vicinity of the main cluster. To investigate this possibility in A370, we look for conceivable substructures in the data, with the goal of replacing the shear with a more physically-motivated component.  Of all possible candidates, the most obvious choice is the moderate overdensity of galaxies at $z \sim 1$, containing 33 galaxies spanning the narrow redshift range $1.03 < z < 1.09$, including the intriguing ring-like galaxy (ID 10287) first identified in \citet{sou99}.  At first glance, this appears to agree well with the background group identified in \citetalias{die18} (30 objects between $z = 1.0$ and $z = 1.1$) using GLASS spectroscopy.  However, after cross-checking with MUSE data we find that several members of the \citetalias{die18} group have been misidentified or have incorrect redshifts.  Specifically, six of these galaxies are more clearly identified in MUSE as low-redshift ($z<0.6$) systems (MUSE IDs:  5263, 7115, 9336, 9914, 10544, and 14179), while four other sets of ``close-pair'' galaxies are simply separate components of highly clumpy spirals.  Furthermore, we also notice that three members of the group are in fact the three images of Multiple Image System 6 ($z = 1.0633$), which should only count as a single object.  Eliminating the false detections reduces the total number of GLASS systems to 18.  Of these, 15 lie in the MUSE footprint and we are able to confirm that their redshifts are indeed in the correct range.  The remaining three galaxies lie slightly outside of the MUSE field, so we are unable to independently verify their redshifts.  However, looking at these systems in the HFF data, we find that two objects (GLASS IDs 1982 and 3004) appear as faint blue clumps with similar colours and morphologies to many confirmed galaxies in the group, while the third (GLASS ID 1230) is red and elliptical, with an appearance closer to the galaxies rejected by MUSE.  Therefore, we include the two blue objects in the final group but reject the red galaxy, which we note is flagged in GLASS as having a low spectral quality ($Q=1$).  Adding in the remaining MUSE-identified systems that do not appear in the GLASS catalog (18 galaxies), we include 35 galaxies in total for the final substructure sample.  The observed positions of these objects are shown in Fig.~\ref{fig:subGroups}.

As a first test we simply include all group members as additional galaxy-scale potentials, while leaving the external shear term in place.  Mirroring the procedure for the cluster members, we employ a second mass-traces-light scaling relation (Equation \ref{eqn:scaling}) for the background group, in order to keep the total number of model parameters manageable.  We normalize this relation to the median redshift of the group ($z = 1.0494$), determining a normalization constant ($m_{F814W}= 21.45$) based on the behavior of $L^*$ with redshift presented in \citet{lin06}.  Since this model includes masses at two distinct redshifts, we use a modified version of \texttt{LENSTOOL} to optimize the distribution, designed to account for multiple lensing deflections along the line of sight.  Specifically, the ``multi-plane'' setting of \texttt{LENSTOOL} first transforms the observed coordinates of the background group members (which are themselves deflected by the cluster) to their intrinsic source positions before including the mass in the model fit, providing a more accurate description of the mass distribution.  This in turn creates a more realistic light path for the observed system: flux from the distant lensing constraints first reaches the plane of the background group where it is lensed and deflected by the group mass.  The deflected source light and undeflected group light then travel together to the cluster plane, where everything is subsequently deflected again by the cluster mass.  

After optimizing the new configuration, we find that the additional potentials do slightly improve the fit (rms = 0\farcs75; Table~\ref{tbl:ModParams}) but the model assigns only a small amount of mass to the new potentials and maintains a high-magnitude external shear term.  Therefore, to test for possible degeneracies between shear and substructure mass, we run a second version of the model which eliminates the shear term entirely.  Without the shear, the model quickly reverts back to a configuration closer to the no-shear \emph{gold}-class model, with a large rms value and a poorer overall fit.  This suggests that the background galaxies alone are not massive enough to account for the observed model distortions; this result is perhaps unsurprising, given that the identified group galaxies are typically much smaller and fainter than the cluster members.

Alternatively, we model the group with a large-scale mass component.  Taking advantage of the \texttt{LENSTOOL} multi-plane ray-tracing capability, we replace the individual galaxies in the $z \sim 1$ scaling relation with a single, massive potential representing the group halo.  This not only allows us to test for an alternative to shear but also assess the likelihood that the group members are truly gravitationally bound.  Since \texttt{LENSTOOL} currently forces background components to have a fixed lens-plane position, we place the halo at the group centroid, determined by averaging over the individual coordinates of all group members (Fig.~\ref{fig:subGroups}).  Similarly, we fix its redshift to the group median, $z = 1.054$.  Given the uniform spatial distribution of the group members we assume a circular rather than an elliptical shape for this component, though we maintain a radial profile that behaves like a dPIE potential.  We note that these parameter choices are largely driven by the MUSE data, and that it is possible additional group members may reside outside of the MUSE footprint, leading to a change in centroid and redshift.  However, given the flexibility of the multi-plane model -- by calculating the group centroid in the image plane, the component still moves relative to the cluster in the source plane -- the effects of these possible systematics should have only a small effect on the results.

After running two new models in this configuration -- with and without the external shear term -- we again find that the background group has little impact on the overall model, failing to significantly reduce the shear term when it is included or provide a viable alternative in its absence.  Furthermore, the low total mass of the component ($\sigma_0 \sim 200$ km s$^{-1}$) makes it unlikely to be large enough to bind the individual group members together, implying that the ``group'' is simply a chance alignment of several unrelated galaxies.  We note, though, that the extended nature of the profile makes it more susceptible to degeneracies with other components.  This is particularly true of the central large-scale potentials (DM2 and DM3) that make up the cluster mass distribution, which are both largely coincident on the group position.  Even accounting for possible degeneracies, however, the fact remains that no combination of background galaxies or halo terms is able to significantly alter the external shear component, strongly suggesting that these masses are not the main driver of the additional model perturbations.  We are therefore forced to look at other possibilities.

The two small overdensities in the foreground of A370 (Section \ref{subsec:catOthers}) stand as another logical choice, as both groups show some evidence of clustering (Fig.~\ref{fig:subGroups}), and many of the galaxies are located to the east or west of the main cluster centre -- regions where additional mass would be expected given the orientation of the external shear term ($\Gamma_{\theta} \sim -18\degr$).  However, most objects in these groups appear small and faint in HFF imaging, implying that they have low masses and thus a limited range of influence on the model.  While low-mass systems make for poor lenses in any circumstance, lying close to an already massive lensing cluster enhances the effect: even assuming these objects have masses equal to a typical cluster member, their deflection angles relative to the cluster plane are only expected to be 0\farcs3.  Since the foreground galaxies are on average over 3\arcsec\ away from ``nearby'' lensing constraints, it is unlikely that they will significantly perturb the constraint coordinates, especially not in a coherent fashion that mimics external shear.  Furthermore, the sparse number of objects in each group (8 at $z = 0.256$ and 14 at $z = 0.326$) suggests that any halo binding the objects together is itself a low-mass system that is far less significant than the existing physical components.  Nevertheless, we run an additional model that includes all foreground galaxies in both groups.  Given the small line-of-sight separation between these objects and the main cluster, we simply add them directly to the existing $L^*$-based scaling relation, instead of generating a new, separate relation.  However, to account for the fact that foreground objects have a less efficient lensing effect, we renormalize their fluxes by the ratio of luminosity distances ($R = D_{L,\rm group}/D_{L,\rm cluster})$ making them slightly fainter (and thus, less massive) during the fitting process.  As expected, we find that the model is unchanged relative to the \emph{copper}-class model, and we reject the foreground deflectors as a suitable replacement for the shear.  In doing so, we effectively eliminate all sources of line-of-sight substructure as strong model perturbers, though we note that they do have some influence on the final fit.  This agrees with the results presented in \citet{chi18}, who find that including line-of-sight galaxies slightly improves the model of MACS0416, but not enough to significantly modify the final parameters of the main cluster mass components.

While our current analysis shows that line-of-sight substructures do not play a large role in the A370 mass model, it does not address physically related structures further away from the cluster core.  As shown in \citet{jau16}, these components can dramatically impact the cluster mass environment, and in some cases act in place of external shear \citep[e.g., ][]{mah18}.  Though outside the focus of this paper, follow-up on such structures would be particularly interesting.  Indeed, there is already some evidence that additional mass structures in the outskirts of A370 may exist: lower-resolution imaging data taken with CFH12K \citep{hoe07}\footnote{\url{http://home.strw.leidenuniv.nl/~hoekstra/Projects/CCCP_CFH12k_images.html}} on the Canada-France-Hawaii Telescope (CFHT) show several concentrations of galaxies with similar colours to cluster members (Fig.\ \ref{fig:CFHT}), and early CFHT/PUMA and ESO/PUMA2 spectroscopy \citep{for86} tentatively measure redshifts for a few of these objects, placing them at the same distance as the cluster \citep{mel88}.  The east- or west-of-centre locations of these concentrations are also promising in terms of external shear replacement, since despite being at greater distances from the cluster, their potentially much larger masses could have a stronger effect.  The current low-resolution imaging and sparse spectroscopy do not allow us to investigate this possibility, though future data will improve these prospects significantly.  The upcoming, wider field-of-view BUFFALO imaging of A370 will provide deep, multi-band imaging of these groups, while an upcoming wide-area MUSE mosaic designed to overlap the BUFFALO imaging will bring high-resolution spectroscopy, providing secure redshifts for all group members.  With this additional data in hand we will be in a much better position to investigate substructure in the A370 field.

\begin{figure}
\includegraphics[width=0.5\textwidth]{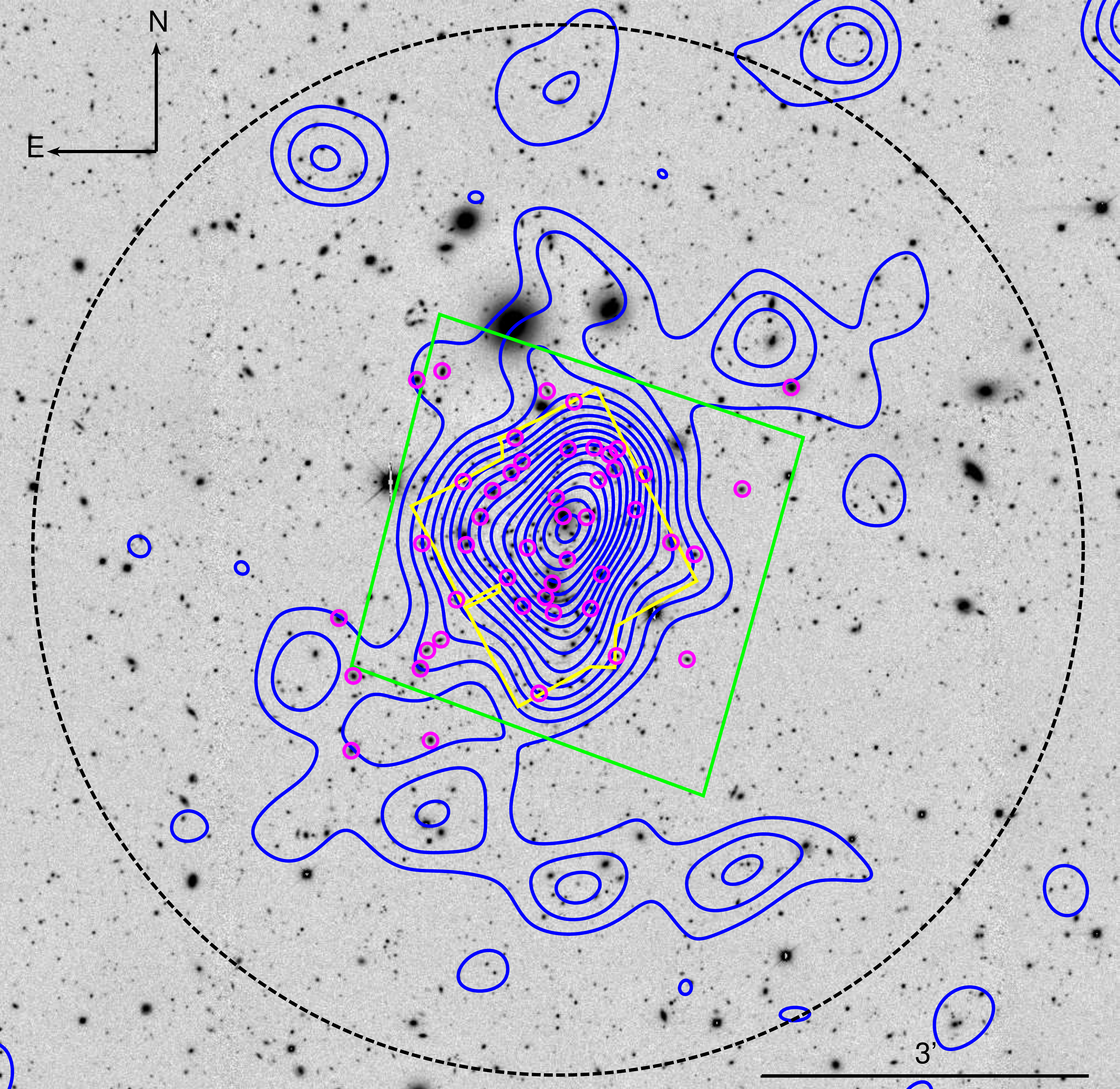}
\caption{CFH12K imaging of A370 and its surroundings.  Using multi-band (B+R+I) photometry we identify members of the cluster red-sequence, smoothing the light from these galaxies to observe any extended structure they create (blue contours).  While the majority of the light falls in the core of the cluster (seen at the image centre), we do find other concentrations at extended distances -- especially in the northwest and southeast.  Low-resolution PUMA spectroscopy taken throughout the field reveals that at least some of the galaxies in these extended regions have confirmed cluster redshifts (magenta circles), suggesting they could be additional substructure candidates.  The black dashed circle has a radius of 1.5 Mpc and serves as a rough estimate of the A370 virial radius, showing that these structures fall well within the theoretical cluster boundary.  For reference, the extent of the MUSE and HFF footprints are shown in yellow and green, respectively.}
\label{fig:CFHT}
\end{figure}

\section{Conclusions}
\label{sec:end}

In this work we have continued our investigation into the strong-lensing cluster Abell 370, expanding on our initial GTO study \citep{lag17} thanks to a larger mosaic of high-resolution MUSE spectroscopy and deeper \emph{HST} imaging.  As part of these efforts we have constructed a comprehensive redshift catalog covering the full line-of-sight region in the vicinity of the cluster core.  This catalog, which is significantly more complete than the GTO edition, has considerable legacy value in its own right, and we highlight the fact that many of the redshifts published in this catalog -- already provided internally to other lens modeling groups as part of the Hubble Frontier Fields modeling challenge -- have proven useful in other modeling efforts \citep[e.g.,][]{die18,str18}.  Our main results are as follows:

\begin{itemize}

\item The new redshift catalog contains 584 entries, including 506 independent objects when accounting for multiple-imaging effects.  With 11 stars, 57 foreground galaxies, 244 cluster members, 155 background galaxies, and 39 multiply-imaged systems, this catalog provides a 3-dimensional picture of the central cluster and its immediate surroundings.  \\

\item The MUSE spectroscopic data cover nearly the entire multiple-image region of A370, allowing the identification of several new multiply-imaged galaxies, more than doubling the number of spectroscopically-confirmed systems.  These include a number of faint or continuum-free Lyman $\alpha$ emitters that are ``invisible'' in broad-band imaging.  In addition, we confirm and refine the redshifts of non-spectroscopic galaxies identified in earlier works.  \\

\item Thanks to the significant increase in multiple-image constraints, we are able to probe the A370 mass distribution out to larger radii, and with much greater accuracy.  Given the abundance of high-confidence ``\emph{gold}-class'' spectroscopic constraints, we find little difference between our initial (\emph{gold}) model and subsequent trials (\emph{silver}, \emph{bronze}, and \emph{copper}) using additional, lower-confidence systems -- though these small variations demonstrate the level of systematic uncertainty.  Our final \emph{copper}-class model contains 45 background galaxies creating 136 individual multiple-image constraints.  The best-fitting rms is 0\farcs78, a significant improvement over the GTO model (rms = 0\farcs94).\\

\item Broadly, this model confirms the results of our GTO work: namely the boxy shape of the central mass distribution and the need for an additional mass clump (the ``crown'' clump) in the northeast of the cluster.  However, to achieve an acceptable fit level, we must include a significant external shear component, which is inherently non-physical.  The presence of shear suggests additional mass not incorporated into the current model and could indicate the presence of local substructure.\\

\item Using the redshift catalog, we identify three possible substructure candidates along the immediate line of sight: two small concentrations of galaxies in the foreground ($z = 0.256$ and $z = 0.326$) and one larger group in the background ($z = 1.049$).  While the relative locations of these objects are consistent with the orientation of the best-fitting shear term, subsequent analysis shows that they are not massive enough to completely account for the shear.  Nevertheless, we find that including the galaxies in background group as individual haloes in the mass model slightly improves the overall fit, reducing the total rms to 0\farcs75.\\

\item Although we rule out line-of-sight substructure as the cause of shear, we are able to identify promising substructure candidates at the cluster redshift but at much greater distances from the core.  An upcoming extension of our current MUSE mosaic designed to provide spectroscopic follow-up of the BUFFALO imaging campaign will target these regions, providing valuable insight into both the nature of these mass concentrations and the overall matter distribution of A370 itself.  

\end{itemize}

These results showcase the advantages of our updated MUSE data set: with deeper spectroscopic mosaic we discover several new multiply-imaged constraints and evidence of structure along the line of sight, both of which improve the lens model and reveal a more complex mass distribution in the form of external shear.  However, they also highlight limitations to the current approach: despite a larger coverage area, we still focus our efforts primarily on the multiple-image region alone, preventing us from definitively identifying the source of the external shear term and leaving an important question unresolved.  It is clear, therefore, that there is a benefit in acquiring additional MUSE data in the outskirts of the cluster.  Spectroscopic coverage in these regions will unambiguously identify additional substructures and give a rough estimate of their masses, eliminating the uncertainty in the current model.  At the same time, an extended MUSE footprint can aid complementary studies to our own, such as a weak-lensing analysis of the mass distribution.  In particular, spectroscopic data will help to calibrate the selection function of background galaxies, removing foreground interlopers that dilute the weak-lensing signal.  Additionally, these data provide a more precise measure of the 3D separation between cluster and background objects, better normalizing the derived weak-lensing mass.  

While joint strong- and weak-lensing techniques have already been used to investigate cluster properties \citep[e.g.,][]{jau16,str18}, these studies often lack significant, wide-ranging spectroscopic coverage, especially in the weak-lensing regime.  Given the different data requirements for these techniques, it is likely that a ``wedding cake'' style survey -- deeper coverage in the cluster core to identify strong lensing constraints, plus shallower coverage in the surrounding areas to identify substructure and calibrate weak lensing constraints -- represents the most efficient method for analyzing the cluster mass with such data.  Indeed, this is the strategy we are adopting in our efforts to extend the current A370 MUSE mosaic, which we will report on in future work.  

Finally, we also point out that this type of coverage would be equally useful for the other FF clusters, or indeed any massive cluster with significant lensing features.  A joint analysis of several clusters would be less susceptible to the effects of cosmic variance, providing more robust constraints on the substructure fraction and mass assembly history of the universe's large-scale structure. With wide-area spectral data now more easily available thanks to MUSE and other instruments, obtaining such a sample is well within the realm of possibility, and would be well worth the effort.

\section*{Acknowledgments}
We thank the referee for a thorough read-through of the manuscript and for providing helpful comments and suggestions, which improved the clarity and content of the work.  DJL, JR, BC, GM, DC, JM and VP acknowledge support from the ERC Starting Grant 336736-CALENDS.  FEB acknowledges support from the CONICYT grants Basal-CATA PFB-06/2007 and AFB-170002, FONDECYT Regular 1141218, Programa de Cooperaci{\'{o}n Cient{\'{\i}}fica ECOS-CONICYT C16U02, and the Ministry of Economy, Development, and Tourism's Millennium Science Initiative through grant IC120009, awarded to The Millennium Institute of Astrophysics.  Some of the data presented in this paper were based on observations made with the NASA/ESA Hubble Space Telescope and obtained from the Mikulski Archive for Space Telescopes (MAST) at the Space Telescope Science Institute (STScI). STScI is operated by the Association of Universities for Research in Astronomy, Inc., under NASA contract NAS 5-26555.  Also based on observations made with ESO Telescopes at the La Silla Paranal Observatory under programme IDs 094.A-0115 and 096.A-0710.  This work utilizes gravitational lensing models produced by PIs Brada\v{c}, Natarajan \& Kneib (CATS), Merten \& Zitrin, Sharon, Williams, Keeton, Bernstein and Diego, and the GLAFIC group. This lens modeling was partially funded by the HST Frontier Fields program conducted by STScI.

  



\appendix
\section{Lens Models}

The best-fitting parameters of all mass models generated by this work are presented in Table \ref{tbl:ModParams}.  In addition to the final parameters, we also provide several goodness-of-fit criteria in the leftmost column.  These include the total model rms (in arcseconds), the $\chi^2$ per degree of freedom ($\nu$), the maximum model likelihood ($\mathcal{L}$), and model evidence ($\mathcal{E}$).  Furthermore, we calculate the Bayesian Information Criterion (BIC), taking into account the total number of model constraints ($n$) and the total number of fit parameters ($k$).

\begin{table*}
  \centering
  \caption{Lens Models and Best-Fitting Parameters} 
  \label{tbl:ModParams}
  \begin{tabular}{lcrrrrrrr}
    \hline
    Model Name & Component & $\Delta\alpha^{\rm ~a}$& $\Delta\delta^{\rm ~a}$ & $\varepsilon^{\rm ~b}$ & $\theta$ & $r_{\rm core}$ & $r_{\rm cut}$ & $\sigma_0$\\
    (Fit Statistics)$^{\rm ~d}$ &  & (\arcsec) & (\arcsec) & & ($\deg$) & (kpc) & (kpc) & (km s$^{-1}$)\\
    \hline
    Gold, No Shear & DM1 & $  2.21^{+  0.12}_{ -0.10}$ & $  1.33^{+  0.05}_{ -0.06}$ & $ 0.40^{+ 0.03}_{-0.03}$ &   $-69.6^{+  1.5}_{ -1.3}$ & $ 14.7^{+  1.0}_{ -1.5}$ & $[800.0]^{\rm ~c}$ & $394^{+15}_{-9}$ \\
& DM2 & $  2.01^{+  0.10}_{ -0.23}$ & $ 11.35^{+  0.31}_{ -0.38}$ & $ 0.69^{+ 0.02}_{-0.01}$ & $-122.3^{+  0.4}_{ -0.6}$ & $137.4^{+  0.2}_{ -1.3}$ & $[800.0]$ & $1039^{+6}_{-14}$ \\
rms = 1\farcs08 & DM3 & $ -1.59^{+  0.26}_{ -0.56}$ & $ 30.15^{+  1.34}_{ -0.82}$ & $ 0.77^{+ 0.01}_{-0.02}$ & $104.1^{+  0.5}_{ -0.4}$ & $164.1^{+  1.7}_{ -2.3}$ & $[800.0]$ & $1030^{+19}_{-8}$ \\
$\chi^2/\nu$ = 5.04 & DM4 & $-43.69^{+  1.47}_{ -0.27}$ & $ 27.74^{+  0.57}_{ -0.78}$ & $ 0.56^{+ 0.03}_{-0.03}$ & $ 65.7^{+  1.5}_{ -1.2}$ & $ 73.6^{+  2.8}_{ -3.8}$ & $[800.0]$ & $544^{+11}_{-10}$ \\
$\log~(\mathcal{L}) = -276.79$ & BCG1 & $[ -0.01]$ & $[  0.02]$ & $[0.30]$ & $[-81.9]$ & $[  0.1]$ & $ 57.6^{+  4.1}_{ -5.1}$ & $224^{+9}_{-6}$ \\
$\log~(\mathcal{E}) = -351.89$ & BCG2 & $[  5.90]$ & $[ 37.24]$ & $[0.20]$ & $[-63.9]$ & $[  0.1]$ & $ 77.6^{+  6.0}_{ -6.8}$ & $388^{+6}_{-9}$ \\
BIC = 738.54 & GAL1 & $[  7.92]$ & $[ -9.76]$ & $[0.26]$ & $[ 25.7]$ & $[  0.1]$ & $  1.7^{+  1.0}_{ -1.0}$ & $169^{+2}_{-1}$ \\
n = 130; k = 38 & GAL2 & $[  2.66]$ & $[ 46.84]$ & $[0.60]$ & $[-84.3]$ & $[  0.2]$ & $ 16.0^{+  1.1}_{ -1.4}$ & $216^{+5}_{-3}$ \\
& GAL3 & $[ 26.25]$ & $[ 56.94]$ & $[0.10]$ & $[ 37.0]$ & $[  0.1]$ & $ 5.3^{+  0.7}_{ -0.6}$  & $62^{+23}_{-28}$ \\
& GAL4 & $[ 29.29]$ & $[ 59.30]$ & $[0.57]$ & $[ 42.5]$ & $[  0.2]$ & $  3.9^{+  1.0}_{ -1.0}$ & $185^{+10}_{-12}$ \\
& L$^{*}$ galaxy &  & & & & $[0.15]$ & $ 10.8^{+  0.6}_{ -1.2}$ & $152^{+2}_{-1}$\\
\hline
Gold, Shear & DM1 & $  1.84^{+  0.30}_{ -0.21}$ & $  1.20^{+  0.18}_{ -0.17}$ & $ 0.13^{+ 0.04}_{-0.03}$ & $-73.9^{+  2.8}_{ -3.6}$ & $ 19.1^{+  2.0}_{ -2.1}$ & $[800.0]$ & $490^{+20}_{-16}$ \\
& DM2 & $  2.70^{+  0.24}_{ -0.31}$ & $ 13.39^{+  0.80}_{ -0.62}$ & $ 0.50^{+ 0.02}_{-0.03}$ & $-122.0^{+  1.0}_{ -1.4}$ & $ 91.9^{+  4.6}_{ -3.0}$ & $[800.0]$ & $827^{+14}_{-19}$ \\
rms = 0\farcs66 & DM3 & $ -5.92^{+  0.58}_{ -0.54}$ & $ 30.81^{+  0.89}_{ -0.82}$ & $ 0.67^{+ 0.02}_{-0.01}$ & $110.7^{+  1.1}_{ -0.9}$ & $182.2^{+  4.1}_{ -4.7}$ & $[800.0]$ & $992^{+17}_{-29}$ \\
$\chi^2/\nu = 1.89$ & DM4 & $-50.23^{+  0.79}_{ -0.92}$ & $ 26.51^{+  1.15}_{ -0.58}$ & $ 0.13^{+ 0.05}_{-0.07}$ & $ 69.6^{+  8.1}_{ -7.9}$ & $ 56.3^{+  8.5}_{ -9.4}$ & $[800.0]$ & $437^{+26}_{-16}$ \\
$\log~(\mathcal{L} = -129.82)$ & BCG1 & $[ -0.01]$ & $[  0.02]$ & $[0.30]$ & $[-81.9]$ & $[  0.1]$ & $ 71.0^{+ 14.4}_{-13.4}$ & $191^{+26}_{-31}$ \\
$\log~(\mathcal{E} = -171.88)$ & BCG2 & $[  5.90]$ & $[ 37.24]$ & $[0.20]$ & $[-63.9]$ & $[  0.1]$ & $125.7^{+  8.0}_{ -7.3}$ & $461^{+8}_{-8}$ \\
BIC = 454.34 & GAL1 & $[  7.92]$ & $[ -9.76]$ & $[0.26]$ & $[ 25.7]$ & $[  0.1]$ & $  1.4^{+  1.0}_{ -1.0}$ & $201^{+12}_{-21}$ \\
n = 130; k = 40 & GAL2 & $[  2.66]$ & $[ 46.84]$ & $[0.60]$ & $[-84.3]$ & $[  0.2]$ & $ 15.5^{+  3.1}_{ -2.0}$ & $238^{+16}_{-13}$ \\
& GAL3 & $[ 26.25]$ & $[ 56.94]$ & $[0.10]$ & $[ 37.0]$ & $[  0.1]$ & $ 4.3^{+  1.7}_{ -2.3}$ & $112^{+36}_{-44}$ \\
& GAL4 & $[ 29.29]$ & $[ 59.30]$ & $[0.57]$ & $[ 42.5]$ & $[  0.2]$ & $  2.3^{+  1.0}_{ -1.0}$ & $250^{+44}_{-30}$ \\
& L$^{*}$ galaxy &  & & & & $[0.15]$ & $ 16.0^{+  1.0}_{ -1.1}$ & $162^{+4}_{-5}$\\
& Shear &  &  & $0.128^{+  0.005}_{ -0.005}$ & $-19.7^{+  1.5}_{ -0.8}$ &  &  &  \\
\hline
Silver, Shear & DM1 & $  2.08^{+  0.12}_{ -0.26}$ & $  1.69^{+  0.12}_{ -0.15}$ & $ 0.24^{+ 0.03}_{-0.04}$ & $-68.8^{+  3.3}_{ -1.6}$ & $ 21.7^{+  2.4}_{ -1.7}$ & $[800.0]$ & $484^{+16}_{-14}$ \\
& DM2 & $  3.33^{+  0.64}_{ -0.40}$ & $ 13.14^{+  0.93}_{ -0.76}$ & $ 0.54^{+ 0.03}_{-0.04}$ & $-121.7^{+  1.2}_{ -1.4}$ & $105.3^{+  3.4}_{ -4.5}$ & $[800.0]$ & $842^{+26}_{-44}$ \\
rms = 0\farcs70 & DM3 & $ -4.54^{+  0.49}_{ -0.51}$ & $ 30.47^{+  1.20}_{ -1.12}$ & $ 0.63^{+ 0.03}_{-0.03}$ & $108.4^{+  1.2}_{ -0.8}$ & $160.3^{+  3.1}_{ -2.8}$ & $[800.0]$ & $985^{+32}_{-41}$ \\
$\chi^2/\nu = 2.17$ & DM4 & $-48.26^{+  0.67}_{ -1.03}$ & $ 26.51^{+  0.83}_{ -0.80}$ & $ 0.33^{+ 0.05}_{-0.06}$ & $ 88.8^{+  5.2}_{ -3.4}$ & $ 60.3^{+  4.7}_{ -7.6}$ & $[800.0]$ & $445^{+20}_{-20}$ \\
$\log~(\mathcal{L}) = -161.20$ & BCG1 & $[ -0.01]$ & $[  0.02]$ & $[0.30]$ & $[-81.9]$ & $[  0.1]$ & $ 69.4^{+  6.4}_{ -7.5}$ & $235^{+20}_{-15}$ \\
$\log~(\mathcal{E}) = -256.96$ & BCG2 & $[  5.90]$ & $[ 37.24]$ & $[0.20]$ & $[-63.9]$ & $[  0.1]$ & $102.4^{+  6.2}_{ -6.2}$ & $456^{+10}_{-7}$ \\
BIC = 541.68 & GAL1 & $[  7.92]$ & $[ -9.76]$ & $[0.26]$ & $[ 25.7]$ & $[  0.1]$ & $  1.8^{+  1.0}_{ -1.0}$ & $173^{+5}_{-6}$ \\
n = 146; k = 44 & GAL2 & $[  2.66]$ & $[ 46.84]$ & $[0.60]$ & $[-84.3]$ & $[  0.2]$ & $ 14.3^{+  2.5}_{ -1.4}$ & $205^{+5}_{-5}$ \\
& GAL3 & $[ 26.25]$ & $[ 56.94]$ & $[0.10]$ & $[ 37.0]$ & $[  0.1]$ & $  3.4^{+  0.8}_{ -0.5}$  & $123^{+27}_{-15}$ \\
& GAL4 & $[ 29.29]$ & $[ 59.30]$ & $[0.57]$ & $[ 42.5]$ & $[  0.2]$ & $  1.6^{+  1.0}_{ -1.0}$ & $256^{+27}_{-17}$ \\
& L$^{*}$ galaxy &  & & & & $[0.15]$ & $ 16.2^{+  1.3}_{ -0.7}$ & $161^{+7}_{-1}$\\
& Shear &  &  & $0.107^{+  0.004}_{ -0.004}$ & $-19.9^{+  0.9}_{ -1.4}$ &  &  &  \\
\hline
Bronze, Shear & DM1 & $  1.62^{+  0.22}_{ -0.23}$ & $  0.96^{+  0.22}_{ -0.13}$ & $ 0.08^{+ 0.02}_{-0.04}$ & $-76.3^{+  2.1}_{ -1.3}$ & $ 25.4^{+  1.8}_{ -1.0}$ & $[800.0]$ & $538^{+11}_{-9}$ \\
& DM2 & $  2.89^{+  0.33}_{ -0.21}$ & $ 15.29^{+  0.68}_{ -0.55}$ & $ 0.47^{+ 0.01}_{-0.02}$ & $-119.7^{+  0.7}_{ -0.9}$ & $105.7^{+  3.8}_{ -3.0}$ & $[800.0]$ & $914^{+12}_{-17}$ \\
rms = 0\farcs73 & DM3 & $ -6.83^{+  0.32}_{ -0.47}$ & $ 33.30^{+  0.92}_{ -0.83}$ & $ 0.79^{+ 0.03}_{-0.03}$ & $110.6^{+  0.7}_{ -0.8}$ & $195.5^{+ 10.2}_{-20.5}$ & $[800.0]$ & $896^{+25}_{-26}$ \\
$\chi^2/\nu = 2.23$ & DM4 & $-47.29^{+  0.73}_{ -0.99}$ & $ 29.31^{+  0.79}_{ -0.59}$ & $ 0.32^{+ 0.05}_{-0.04}$ & $ 72.6^{+  4.0}_{ -1.2}$ & $ 64.3^{+  3.0}_{ -4.4}$ & $[800.0]$ & $478^{+14}_{-13}$ \\
$\log~(\mathcal{L}) = -180.41$ & BCG1 & $[ -0.01]$ & $[  0.02]$ & $[0.30]$ & $[-81.9]$ & $[  0.1]$ & $ 37.8^{+  4.5}_{ -7.8}$ & $214^{+13}_{-10}$ \\
$\log~(\mathcal{E}) = -282.62$ & BCG2 & $[  5.90]$ & $[ 37.24]$ & $[0.20]$ & $[-63.9]$ & $[  0.1]$ & $148.0^{+  9.5}_{-12.7}$ & $457^{+10}_{-5}$ \\
BIC = 588.64 & GAL1 & $[  7.92]$ & $[ -9.76]$ & $[0.26]$ & $[ 25.7]$ & $[  0.1]$ & $  1.9^{+  1.0}_{ -1.0}$ & $197^{+8}_{-10}$ \\
n = 158; k = 45 & GAL2 & $[  2.66]$ & $[ 46.84]$ & $[0.60]$ & $[-84.3]$ & $[  0.2]$ & $  5.9^{+  1.0}_{ -1.0}$ & $231^{+26}_{-6}$ \\
& GAL3 & $[ 26.25]$ & $[ 56.94]$ & $[0.10]$ & $[ 37.0]$ & $[  0.1]$ & $  5.9^{+  1.6}_{ -0.5}$ & $6^{+20}_{-5}$ \\
& GAL4 & $[ 29.29]$ & $[ 59.30]$ & $[0.57]$ & $[ 42.5]$ & $[  0.2]$ & $  0.5^{+  1.0}_{ -1.0}$ & $459^{+71}_{-13}$ \\
&L$^{*}$ galaxy &  & & & & $[0.15]$ & $ 19.0^{+  4.4}_{ -1.0}$ & $140^{+2}_{-2}$\\
& Shear &  &  & $0.114^{+  0.004}_{ -0.003}$ & $-18.6^{+  1.3}_{ -1.0}$ &  &  &  \\
\hline
\end{tabular}
\medskip\\
\begin{flushleft}
$^{\rm a}$ $\Delta\alpha$ and $\Delta\delta$ are measured relative to the reference coordinate point: ($\alpha$ = 39.97134, $\delta$ = -1.5822597)\\[1pt]
  $^{\rm b}$ Ellipticity ($\varepsilon$) is defined to be $(a^2-b^2) / (a^2+b^2)$, where $a$ and $b$ are the semi-major and semi-minor axes of the ellipse\\[1pt]
   $^{\rm c}$ Quantities in brackets are fixed parameters\\[1pt]
   $^{\rm d}$ Statistics notes: $\mathcal{L}$ represents the model likelihood and $\mathcal{E}$ the model evidence.  BIC is the Bayesian Information Criterion.  The total number of model   \\~\ ~\ constraints is given by $n$, while $k$ represents the total number of model parameters.
\end{flushleft}

\end{table*}

\begin{table*}\ContinuedFloat
  \centering
  \caption{Lens Models and Best-Fitting Parameters (continued)}
  \label{tbl:ModParams}
  \begin{tabular}{lcrrrrrrr}
    \hline
    Model Name & Component & $\Delta\alpha^{\rm ~a}$& $\Delta\delta^{\rm ~a}$ & $\varepsilon^{\rm ~b}$ & $\theta$ & $r_{\rm core}$ & $r_{\rm cut}$ & $\sigma_0$\\
    (Fit Statistics) &  & (\arcsec) & (\arcsec) & & ($\deg$) & (kpc) & (kpc) & (km s$^{-1}$)\\
    \hline
Copper, Shear & DM1 & $  1.52^{+  0.21}_{ -0.15}$ & $  0.66^{+  0.15}_{ -0.13}$ & $ 0.20^{+ 0.03}_{-0.03}$ & $-83.6^{+  4.1}_{ -3.0}$ & $ 17.0^{+  1.1}_{ -1.6}$ & $[800.0]$ & $480^{+10}_{-9}$ \\
& DM2 & $  3.06^{+  0.21}_{ -0.21}$ & $ 15.68^{+  0.41}_{ -0.35}$ & $ 0.46^{+ 0.02}_{-0.02}$ & $-117.7^{+  0.9}_{ -1.7}$ & $107.5^{+  2.9}_{ -2.6}$ & $[800.0]$ & $1007^{+20}_{-29}$ \\
rms = 0\farcs78 & DM3 & $ -5.78^{+  0.74}_{ -0.92}$ & $ 34.55^{+  1.38}_{ -1.56}$ & $ 0.80^{+ 0.02}_{-0.02}$ & $110.7^{+  0.5}_{ -0.9}$ & $198.4^{+  8.1}_{ -6.6}$ & $[800.0]$ & $872^{+30}_{-39}$ \\
$\chi^2/\nu = 2.45$ & DM4 & $-48.20^{+  0.97}_{ -1.00}$ & $ 31.59^{+  0.72}_{ -0.70}$ & $ 0.41^{+ 0.03}_{-0.03}$ & $ 69.4^{+  2.2}_{ -1.3}$ & $ 88.0^{+  2.6}_{ -3.7}$ & $[800.0]$ & $551^{+4}_{-12}$ \\
$\log~(\mathcal{L}) = -227.91$ & BCG1 & $[ -0.01]$ & $[  0.02]$ & $[0.30]$ & $[-81.9]$ & $[  0.1]$ & $ 29.6^{+  4.3}_{ -5.0}$ & $186^{+5}_{-4}$ \\
$\log~(\mathcal{E}) = -338.36$ & BCG2 & $[  5.90]$ & $[ 37.24]$ & $[0.20]$ & $[-63.9]$ & $[  0.1]$ & $ 91.7^{+  5.2}_{ -4.8}$ & $410^{+5}_{-3}$ \\
BIC = 695.20 & GAL1 & $[  7.92]$ & $[ -9.76]$ & $[0.26]$ & $[ 25.7]$ & $[  0.1]$ & $  1.8^{+  1.0}_{ -1.0}$ & $149^{+7}_{-10}$ \\
n = 182; k = 46 & GAL2 & $[  2.66]$ & $[ 46.84]$ & $[0.60]$ & $[-84.3]$ & $[  0.2]$ & $  9.6^{+  1.4}_{ -1.0}$ & $282^{+12}_{-11}$ \\
& GAL3 & $[ 26.25]$ & $[ 56.94]$ & $[0.10]$ & $[ 37.0]$ & $[  0.1]$ & $  6.4^{+  0.8}_{ -0.9}$ & $27^{+8}_{-7}$ \\
& GAL4 & $[ 29.29]$ & $[ 59.30]$ & $[0.57]$ & $[ 42.5]$ & $[  0.2]$ & $  4.1^{+  1.0}_{ -1.0}$ & $184^{+14}_{-13}$ \\
& L$^{*}$ galaxy &  & & & & $[0.15]$ & $ 15.0^{+  1.4}_{ -0.8}$ & $149^{+1}_{-1}$\\
&Shear &  &  & $0.096^{+  0.004}_{ -0.003}$ & $-18.3^{+  1.1}_{ -1.2}$ &  &  &  \\
\hline
Copper, Shear+Group & DM1 & $  1.58^{+  0.12}_{ -0.11}$ & $  0.93^{+  0.15}_{ -0.13}$ & $ 0.17^{+ 0.01}_{-0.02}$ & $-88.9^{+  5.1}_{ -3.9}$ & $ 15.8^{+  1.1}_{ -1.1}$ & $[800.0]$ & $487^{+9}_{-8}$ \\
& DM2 & $  2.59^{+  0.14}_{ -0.15}$ & $ 15.68^{+  0.22}_{ -0.18}$ & $ 0.42^{+ 0.01}_{-0.02}$ & $-116.9^{+  0.7}_{ -0.6}$ & $108.9^{+  3.7}_{ -1.2}$ & $[800.0]$ & $1009^{+8}_{-12}$ \\
rms = 0\farcs75 & DM3 & $ -6.63^{+  0.52}_{ -0.56}$ & $ 35.76^{+  1.37}_{ -0.94}$ & $ 0.84^{+ 0.01}_{-0.01}$ & $110.9^{+  0.4}_{ -0.6}$ & $215.9^{+  7.3}_{ -6.8}$ & $[800.0]$ & $857^{+13}_{-21}$ \\
$\chi^2/\nu = 2.30$ & DM4 & $-48.20^{+  0.67}_{ -0.50}$ & $ 31.87^{+  1.10}_{ -0.40}$ & $ 0.32^{+ 0.02}_{-0.03}$ & $ 71.2^{+  1.4}_{ -1.7}$ & $ 85.4^{+  2.2}_{ -2.9}$ & $[800.0]$ & $543^{+7}_{-10}$ \\
$\log~(\mathcal{L}) = -215.84$ & BCG1 & $[ -0.01]$ & $[  0.02]$ & $[0.30]$ & $[-81.9]$ & $[  0.1]$ & $ 22.6^{+  7.7}_{ -7.9}$ & $174^{+4}_{-7}$ \\
$\log~(\mathcal{E}) = -260.28$ & BCG2 & $[  5.90]$ & $[ 37.24]$ & $[0.20]$ & $[-63.9]$ & $[  0.1]$ & $126.2^{+ 10.4}_{ -5.9}$ & $407^{+2}_{-4}$ \\
BIC = 681.47 & GAL1 & $[  7.92]$ & $[ -9.76]$ & $[0.26]$ & $[ 25.7]$ & $[  0.1]$ & $  1.7^{+  1.0}_{ -1.0}$ & $157^{+15}_{-8}$ \\
n =182; k = 48 & GAL2 & $[  2.66]$ & $[ 46.84]$ & $[0.60]$ & $[-84.3]$ & $[  0.2]$ & $ 12.6^{+  1.0}_{ -1.1}$ & $288^{+9}_{-5}$ \\
& GAL3 & $[ 26.25]$ & $[ 56.94]$ & $[0.10]$ & $[ 37.0]$ & $[  0.1]$ & $  4.9^{+  2.3}_{ -2.7}$ & $28^{+13}_{-8}$ \\
& GAL4 & $[ 29.29]$ & $[ 59.30]$ & $[0.57]$ & $[ 42.5]$ & $[  0.2]$ & $  4.5^{+  1.0}_{ -1.0}$ & $165^{+12}_{-13}$ \\
& L$^{*}$ galaxy (cluster)&  & & & & $[0.15]$ & $ 16.0^{+ 0.8}_{ -0.8}$ & $153^{+1}_{-1}$\\
& L$^{*}$ galaxy (group)&  & & & & $[0.15]$ & $ 31.1^{+ 21.5}_{ -1.2}$ & $80^{+35}_{-24}$\\
& Shear &  &  & $0.104^{+  0.003}_{ -0.004}$ & $-19.0^{+  1.2}_{ -0.6}$ &  &  &  \\
\hline

\end{tabular}
\end{table*}

\section{Multiple-Image Cutout Figures}
\setcounter{figure}{1}

This section presents cutout images of all multiple-image systems used in this work, to better show their morphology and (in the case of systems with spectroscopic redshifts) their relative line strengths.  Systems with confirmed spectroscopic redshifts are shown in Fig.~\ref{fig:SpecFigs}, while those without can be seen in Fig.~\ref{fig:NoSpecFigs}.

\begin{figure*} \ContinuedFloat
\centering{
\includegraphics[width=0.45\textwidth]{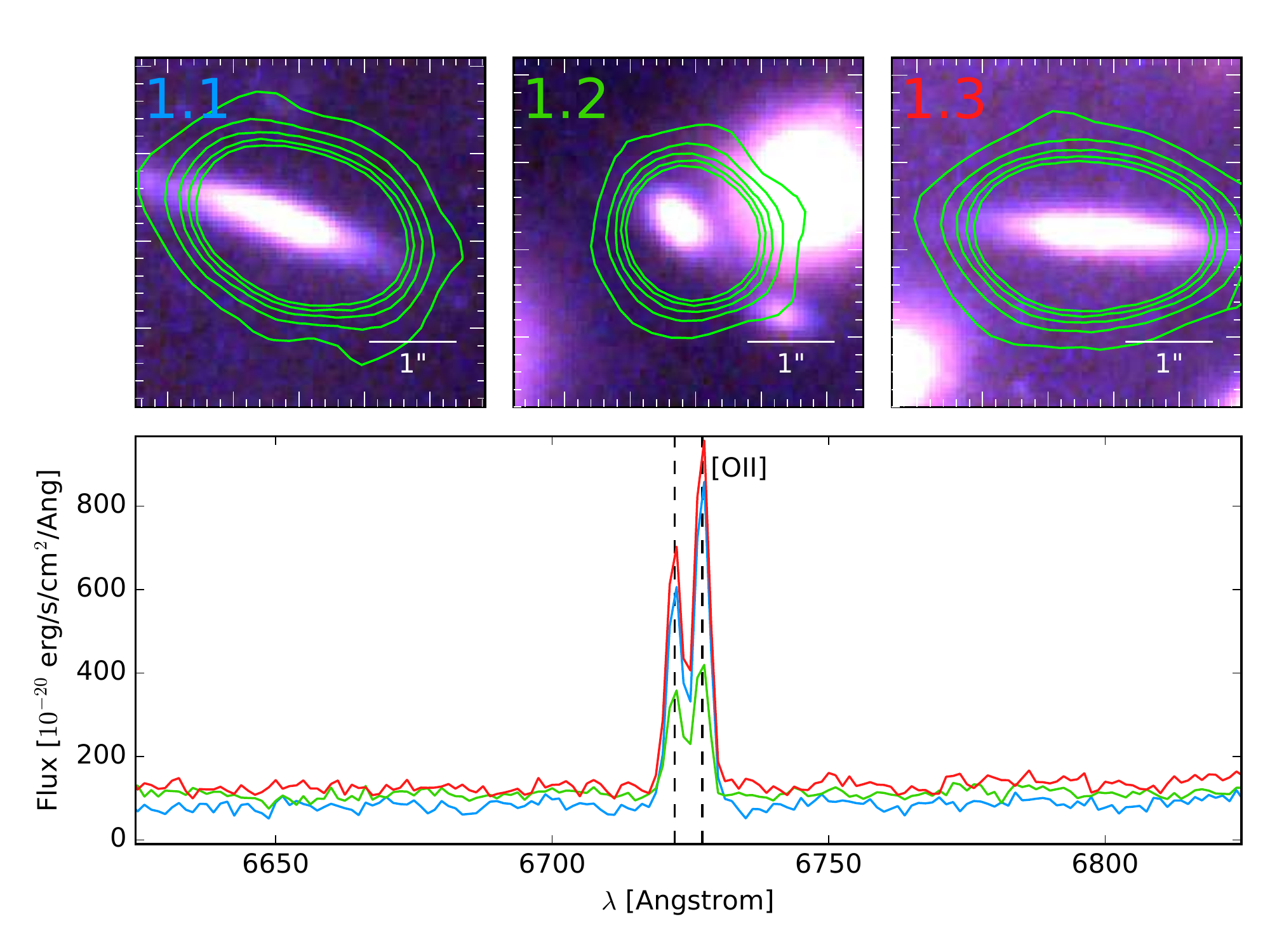}
\includegraphics[width=0.45\textwidth]{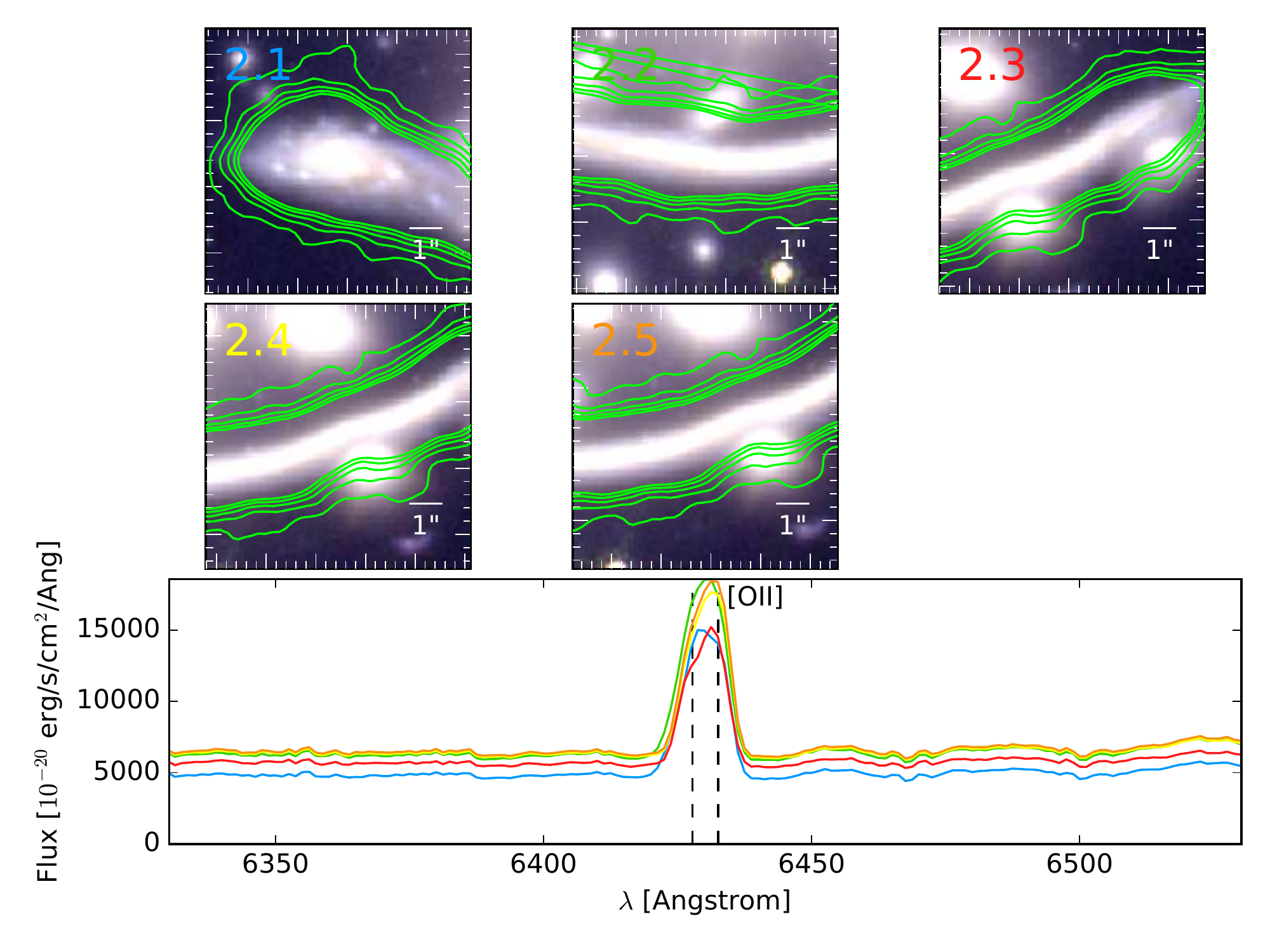}
}
\centering{
\includegraphics[width=0.45\textwidth]{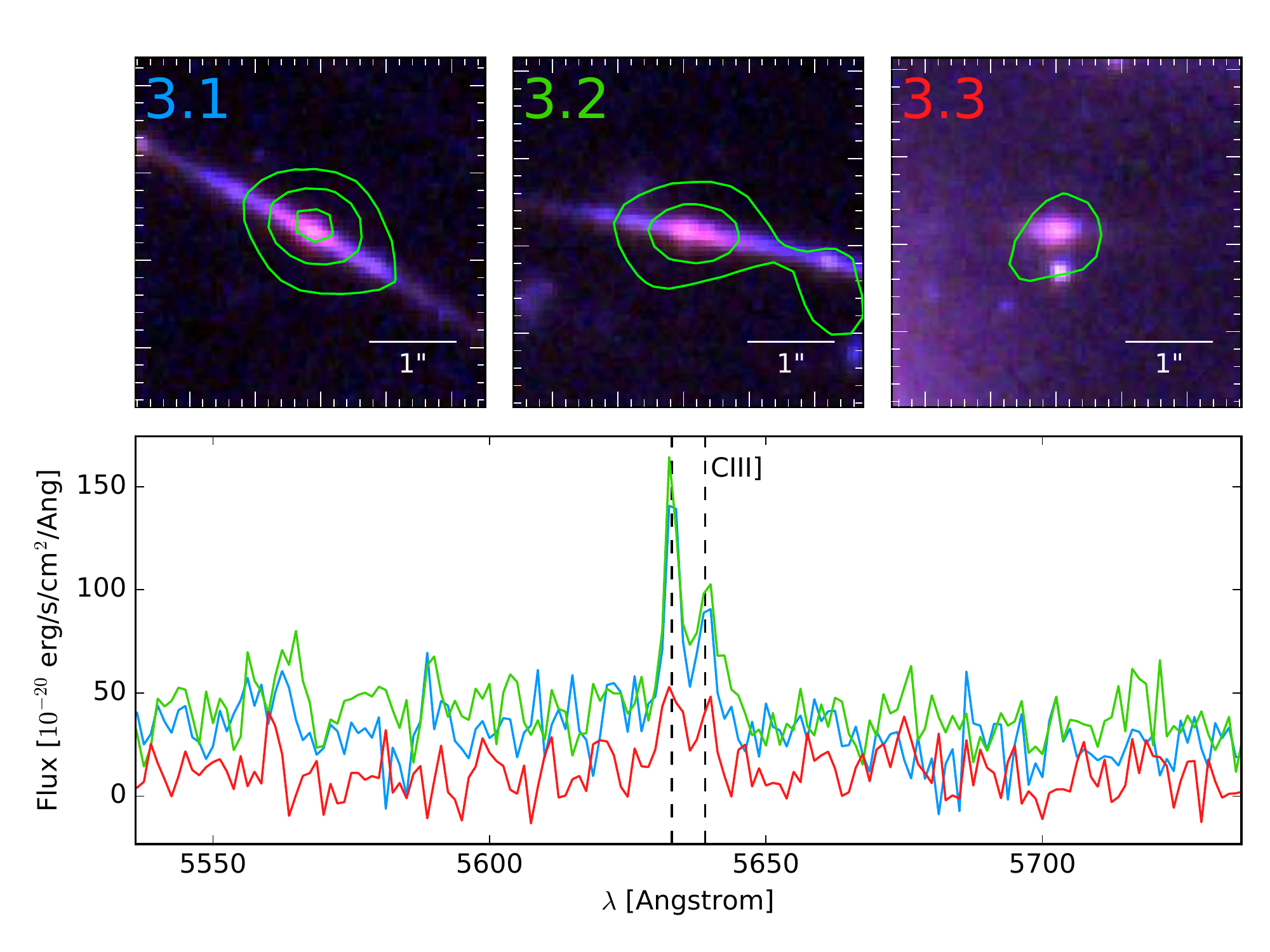}
\includegraphics[width=0.45\textwidth]{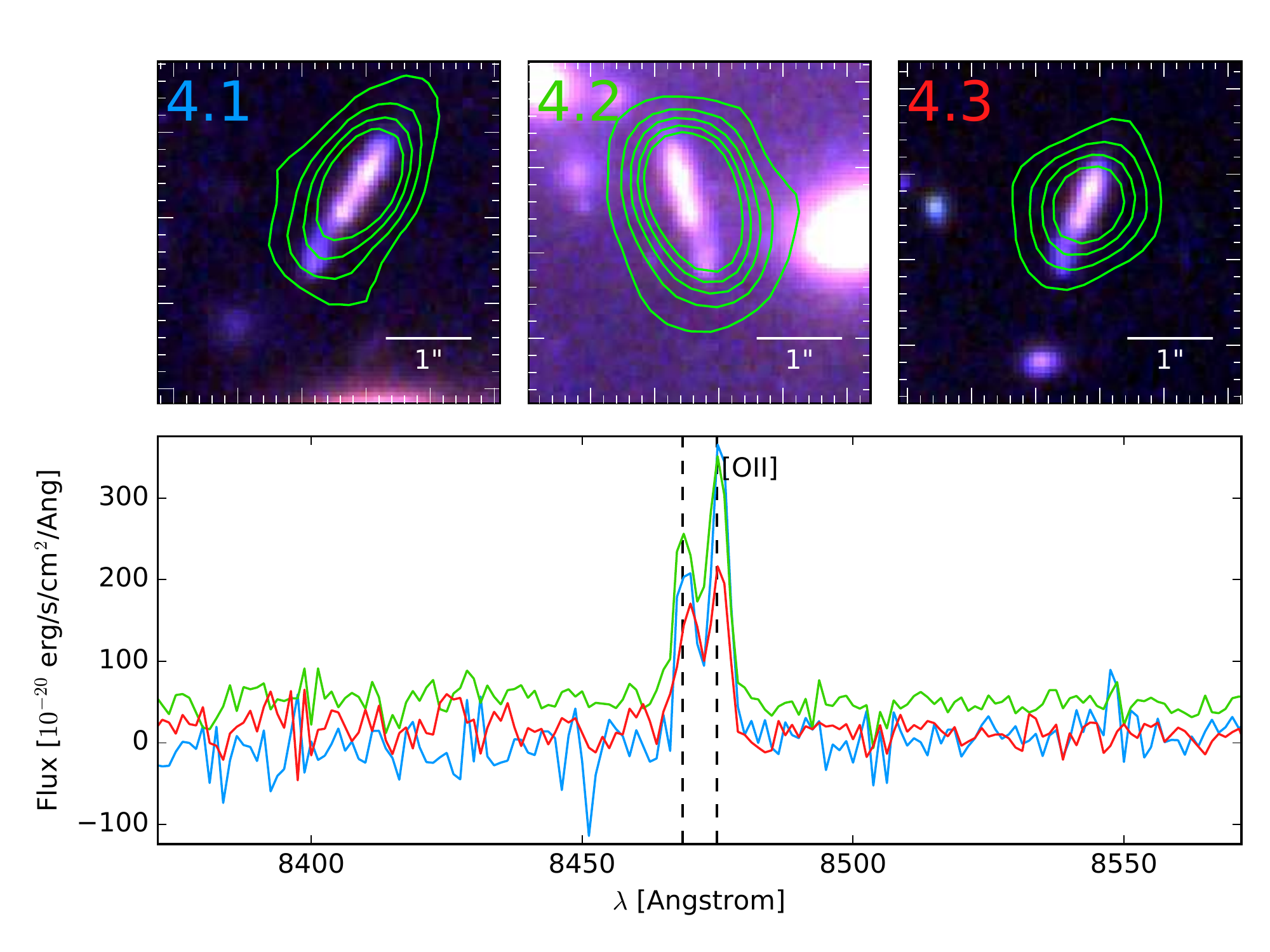}
}
\centering{
\includegraphics[width=0.45\textwidth]{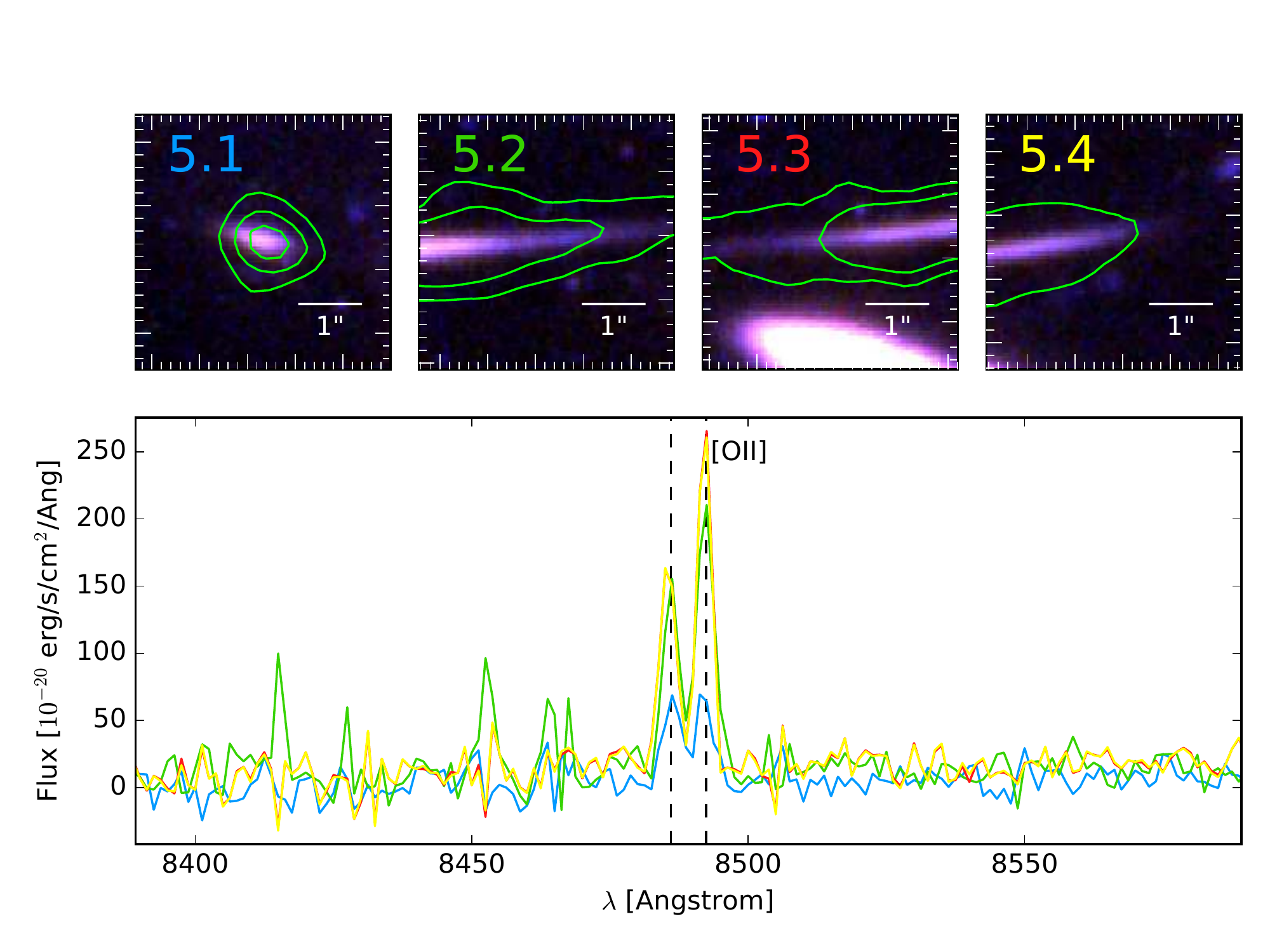}
\includegraphics[width=0.45\textwidth]{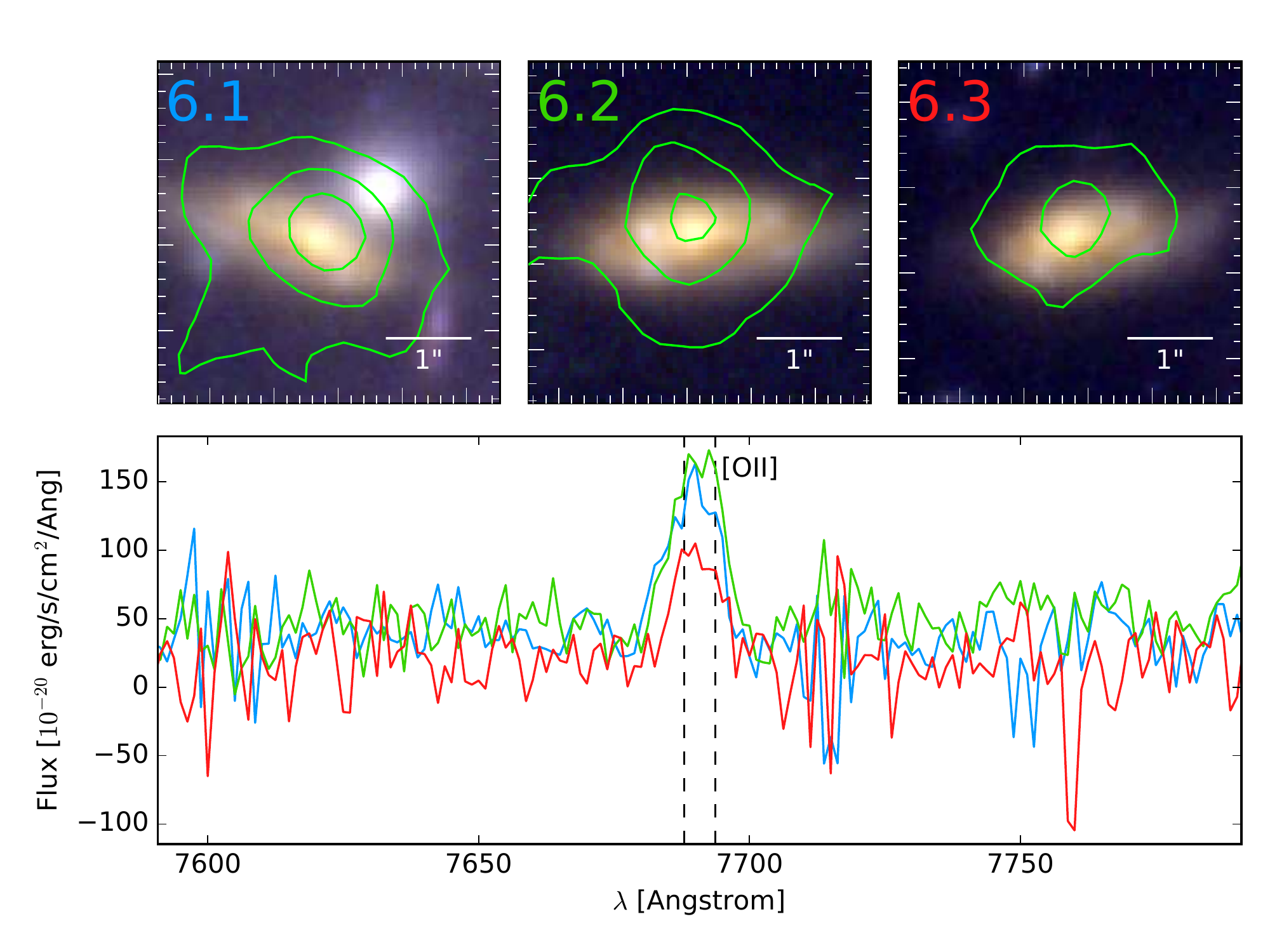}
}
\caption{Cutout images and associated spectra of all spectroscopically-identified multiple-image systems.  RGB images are taken from the HFF data, using the $F814W$, $F125W$, and $F160W$ filters.  Green contours are calculated on narrow-band imaging and represent the 1- through 5$\sigma$ regions of flux according to the Median Absolute Deviation (MAD) statistic of the cutout region.  System 38 (a Lyman Break galaxy) is instead identified from continuum flux, which is represented by a red contour.  The emission feature used to generate the contour is shown in the spectra below the images.  The redshift of System 41 (only weakly identified, and only in Image 41.1) is too tentative to be used as a redshift constraint, though we include it here for completeness and note that the model-predicted redshift of the system (Table \ref{tbl:Multi-Images}) is very close to the measured value ($z = 1.2777$).  Additionally, the observed redshift of System 42 (a galaxy-galaxy lensed spiral arm) appears offset from its systemic value (used in the model), due to a large rotation curve.}
\label{fig:SpecFigs}
\end{figure*}

\begin{figure*} \ContinuedFloat
\centering{
\includegraphics[width=0.45\textwidth]{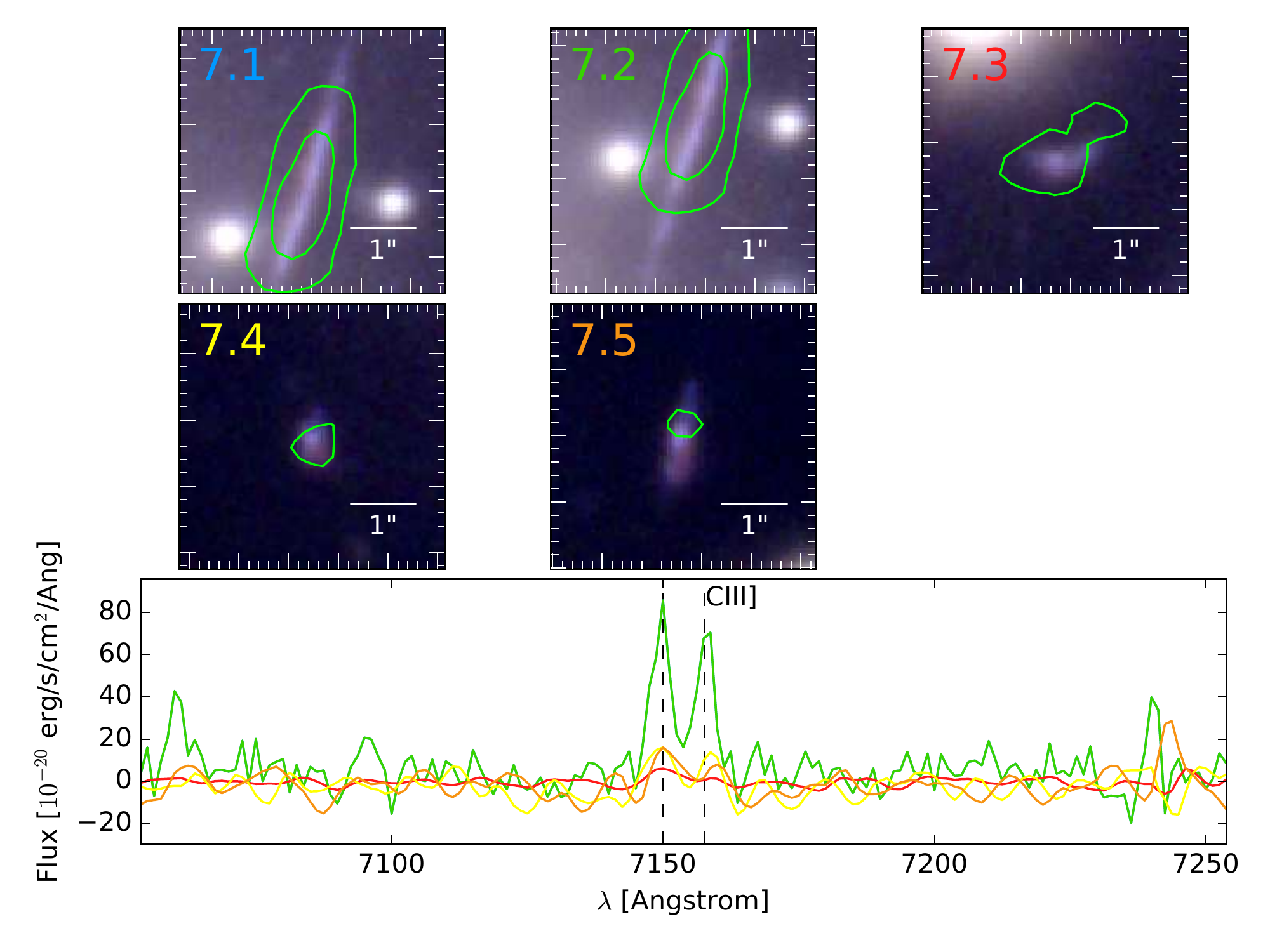}
\includegraphics[width=0.45\textwidth]{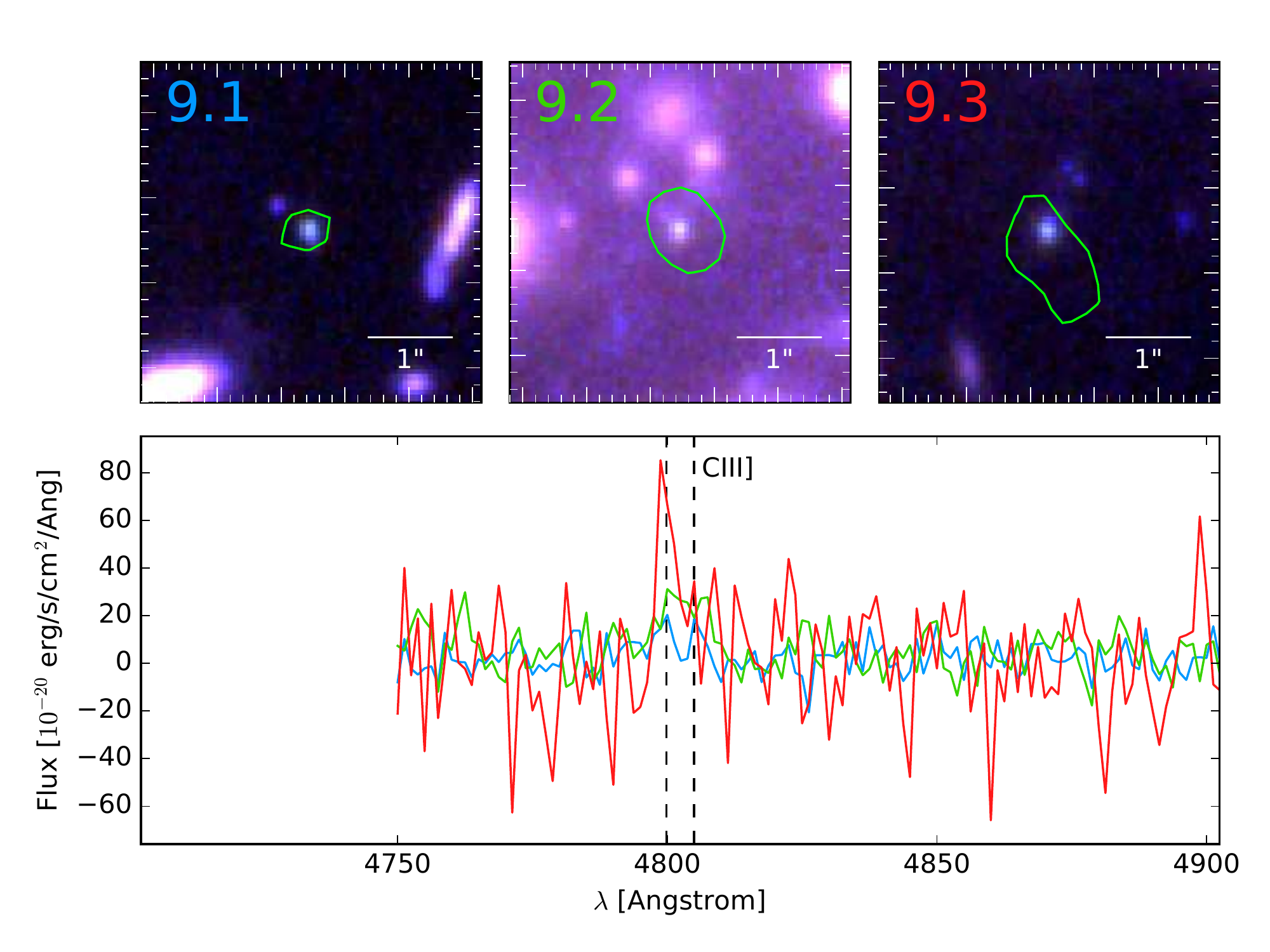}
}
\centering{
\includegraphics[width=0.45\textwidth]{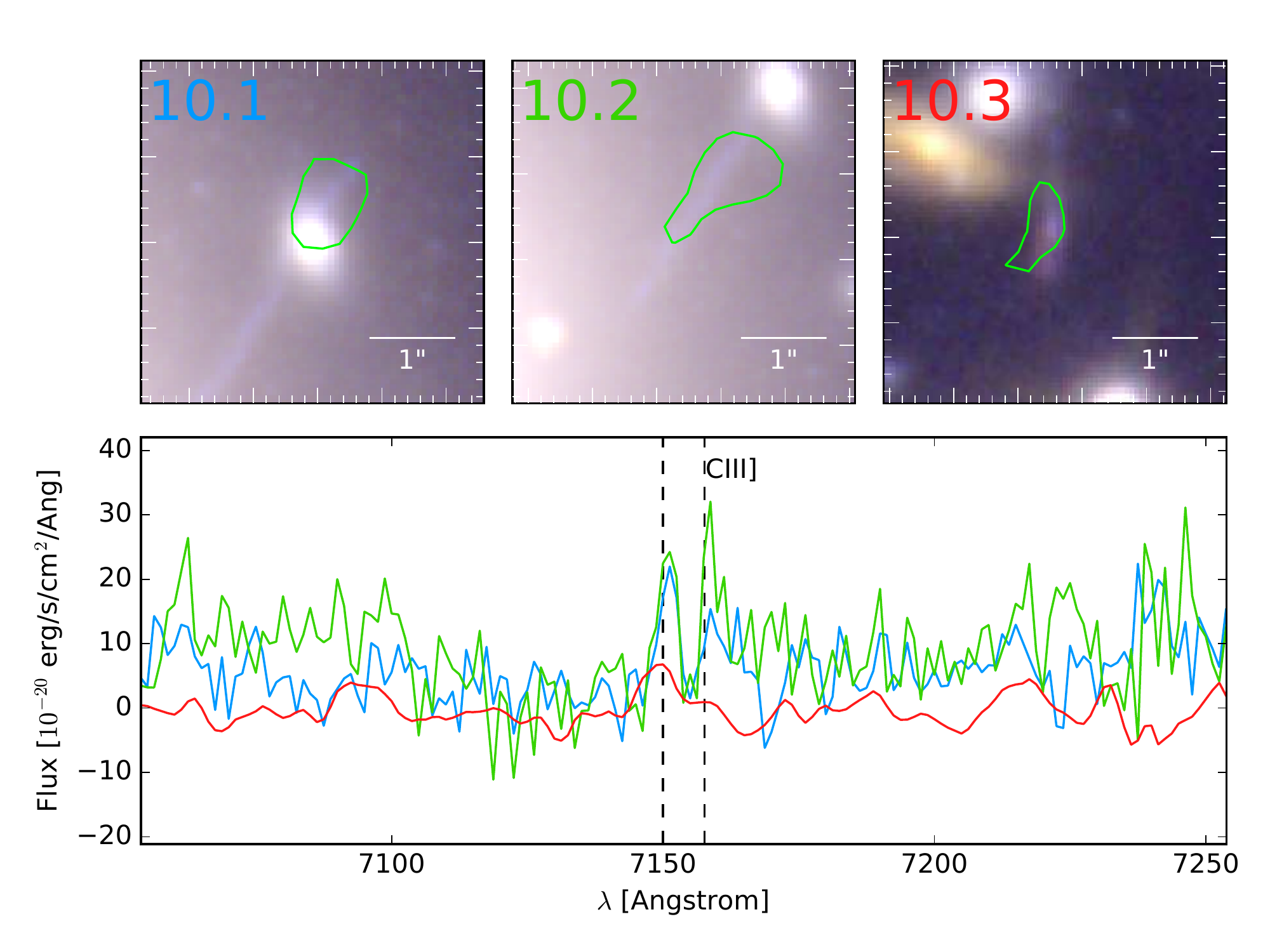}
\includegraphics[width=0.45\textwidth]{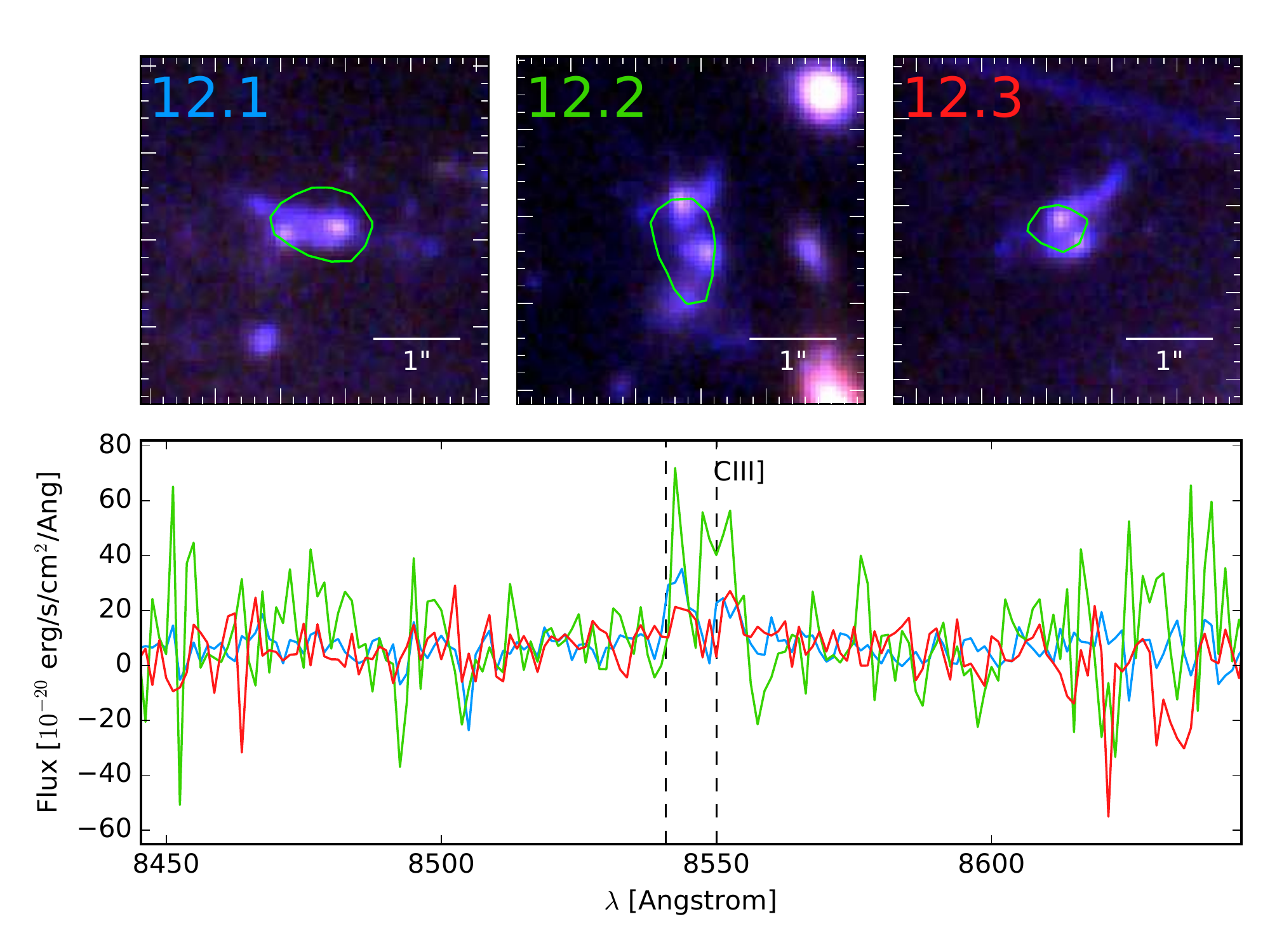}
}
\centering{
\includegraphics[width=0.45\textwidth]{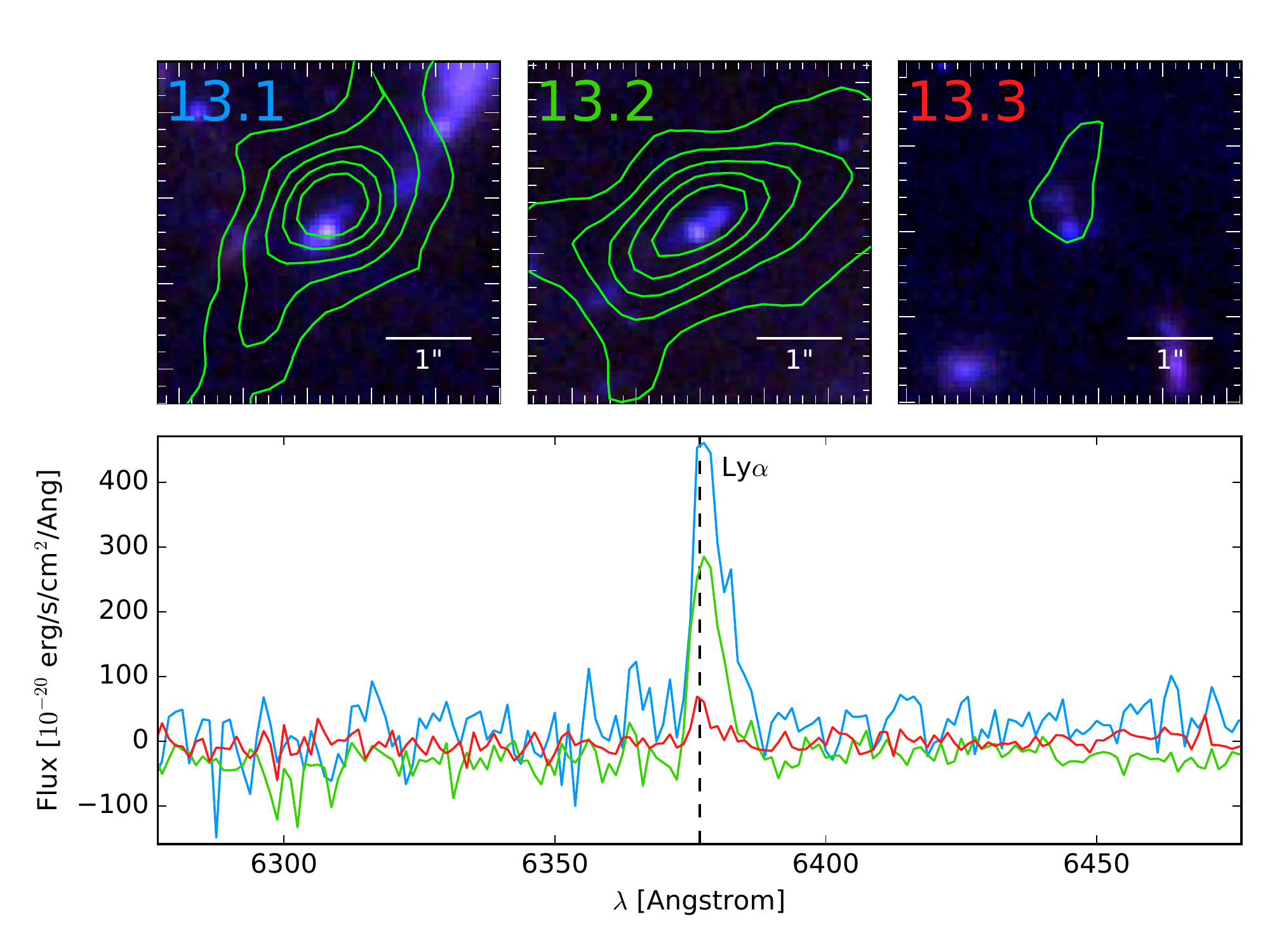}
\includegraphics[width=0.45\textwidth]{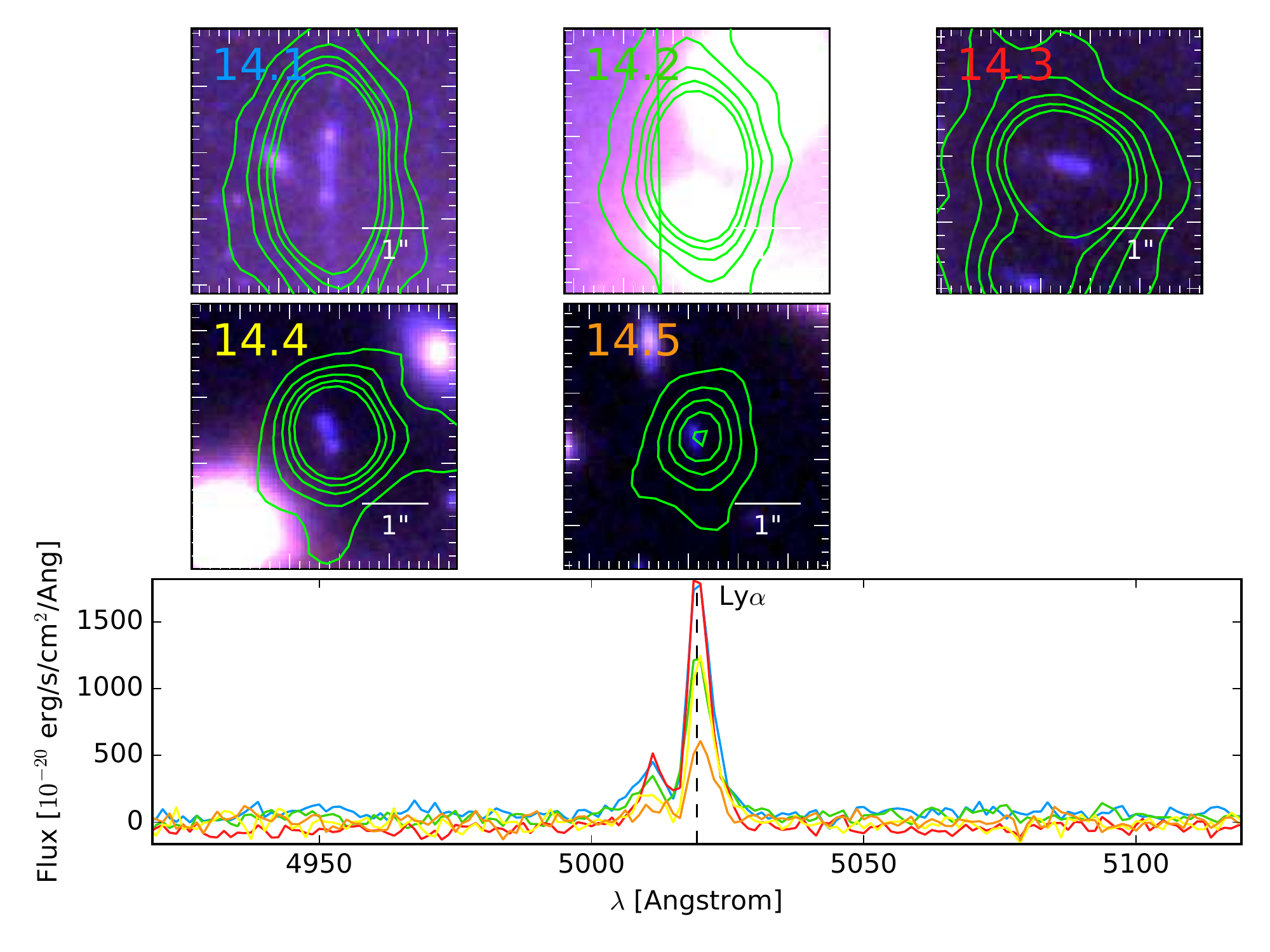}
}
\centering{
\includegraphics[width=0.45\textwidth]{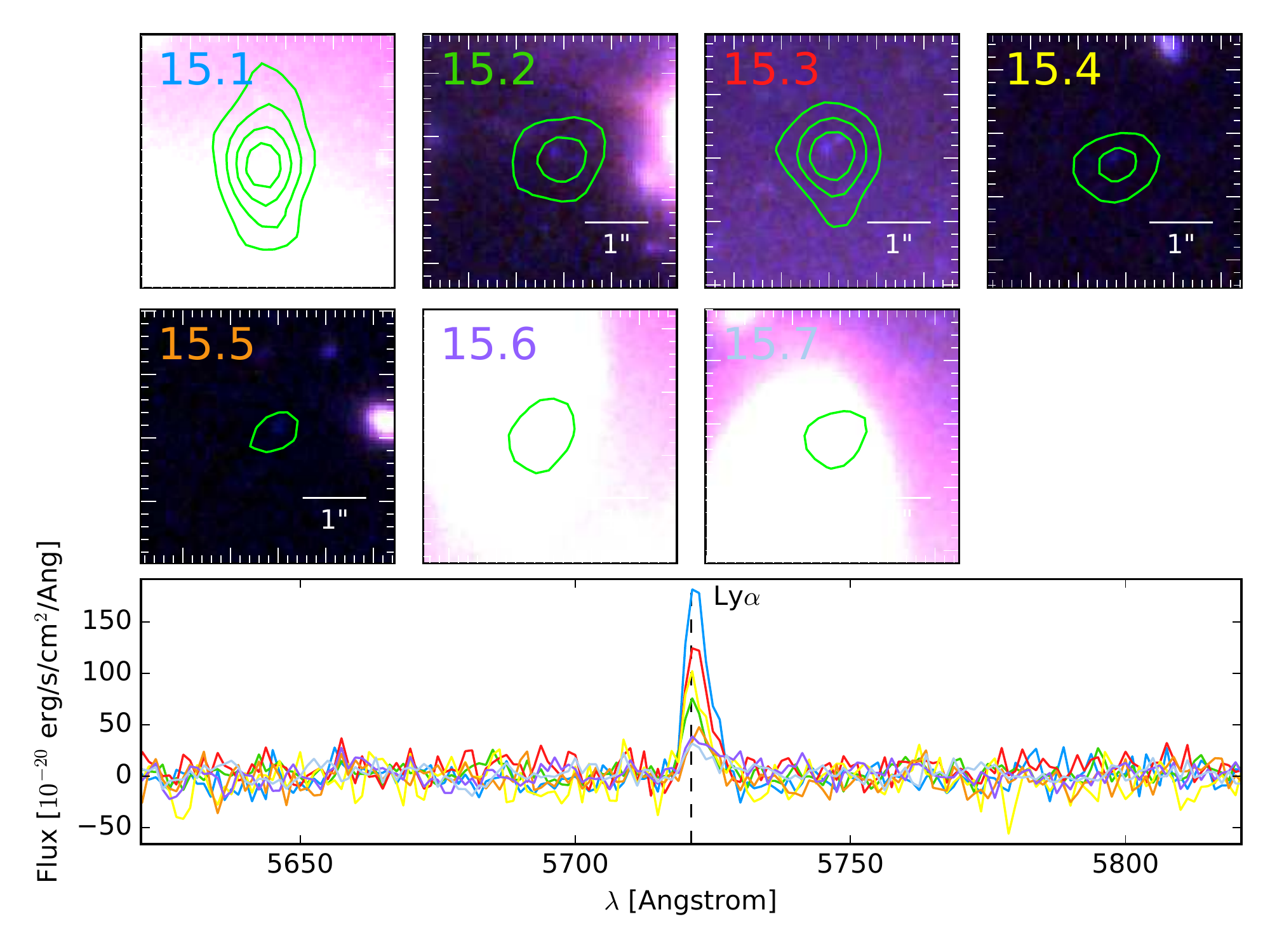}
\includegraphics[width=0.45\textwidth]{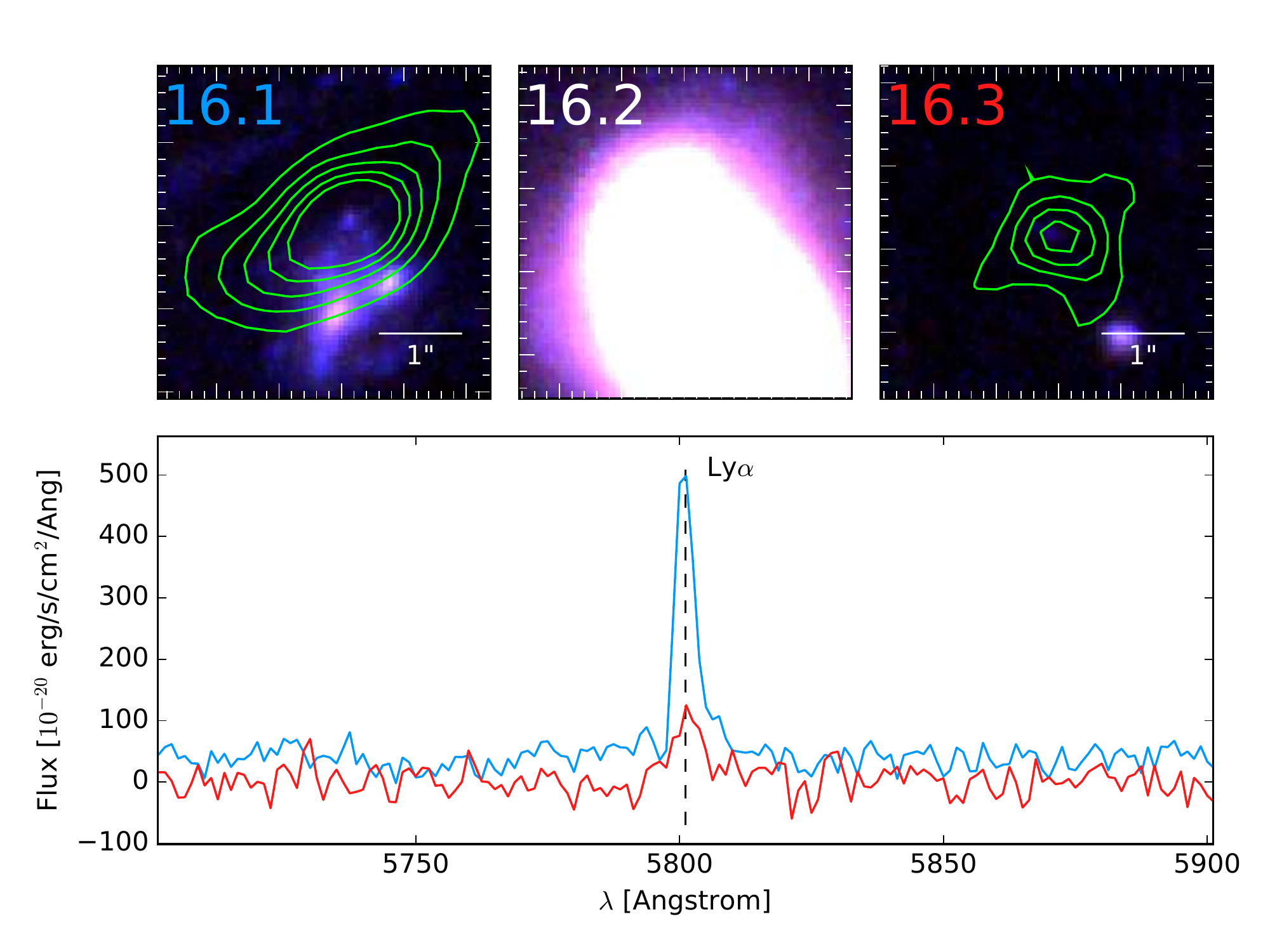}
}
\caption{Multiple-Images with spectroscopic redshifts (continued)}
\end{figure*}

\begin{figure*} \ContinuedFloat
\centering{
\includegraphics[width=0.45\textwidth]{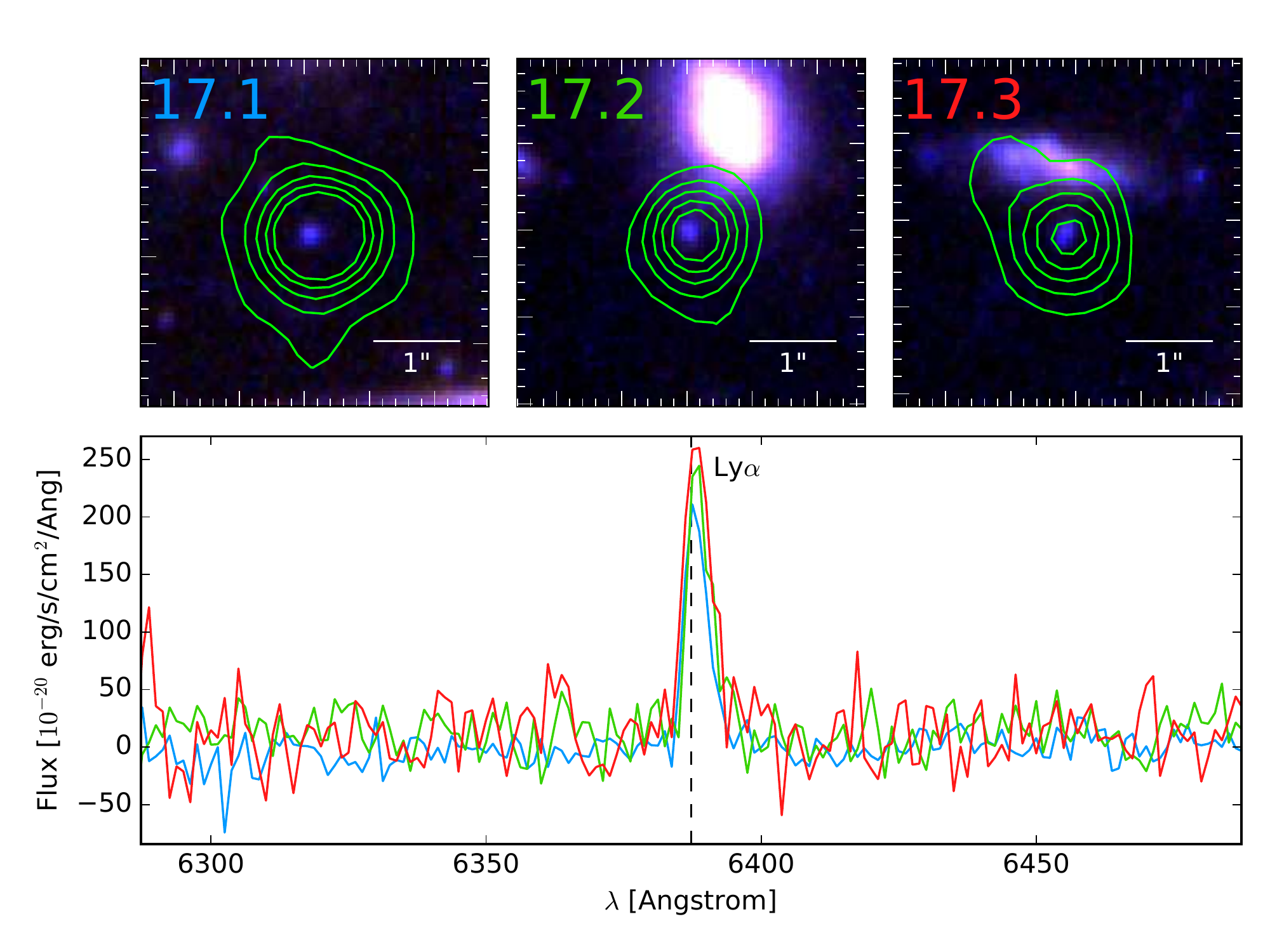}
\includegraphics[width=0.45\textwidth]{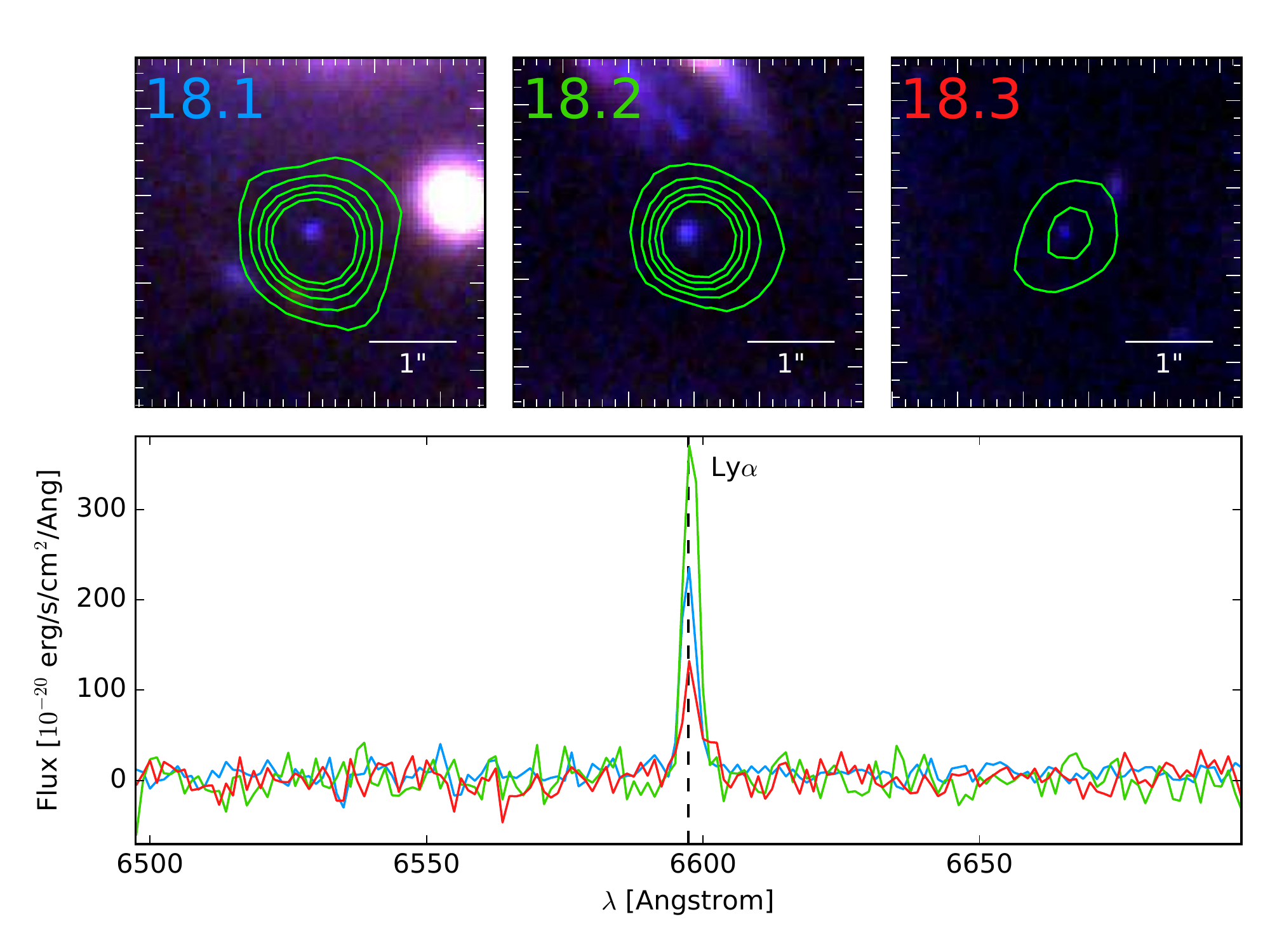}
}
\centering{
\includegraphics[width=0.45\textwidth]{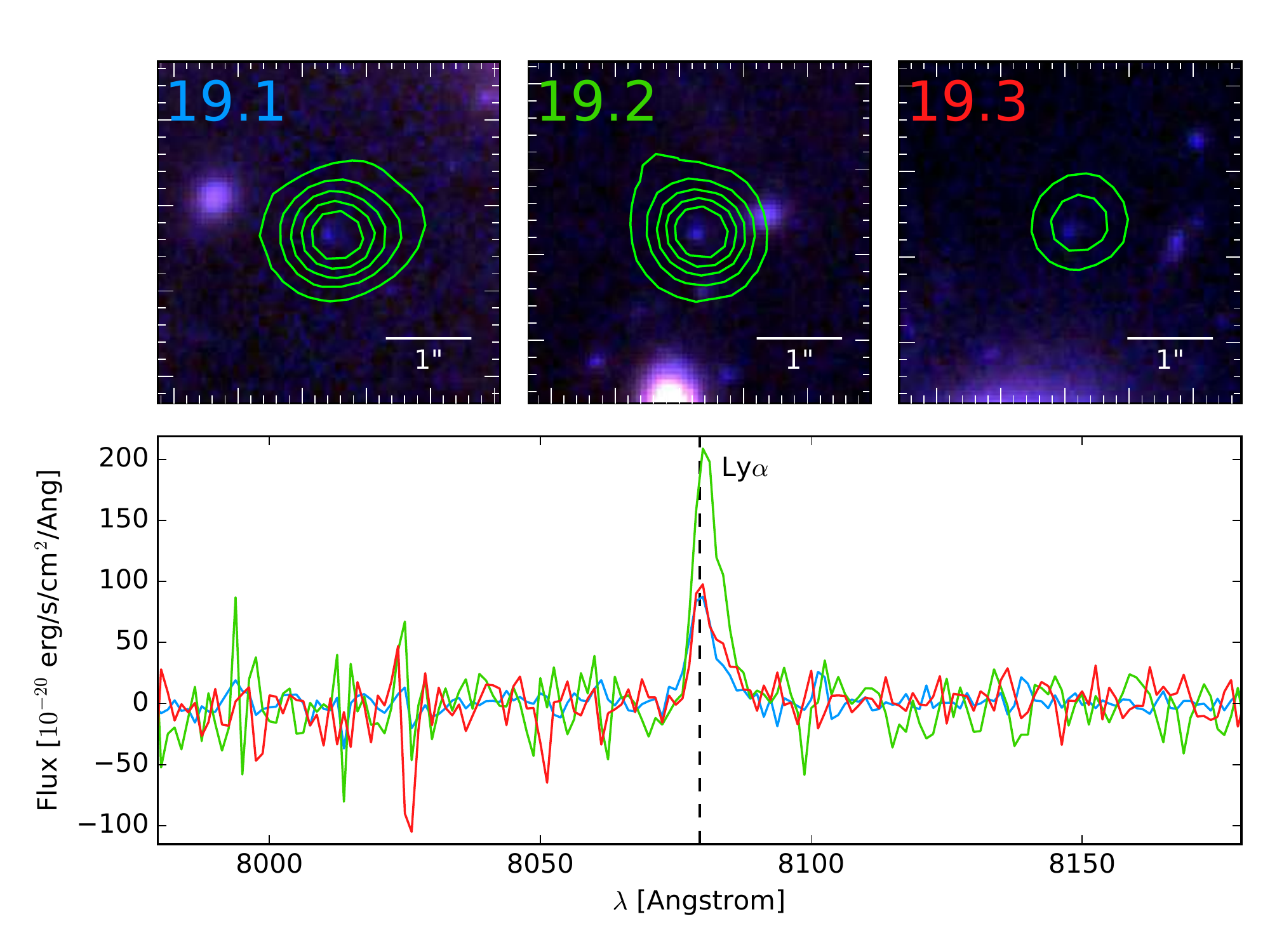}
\includegraphics[width=0.45\textwidth]{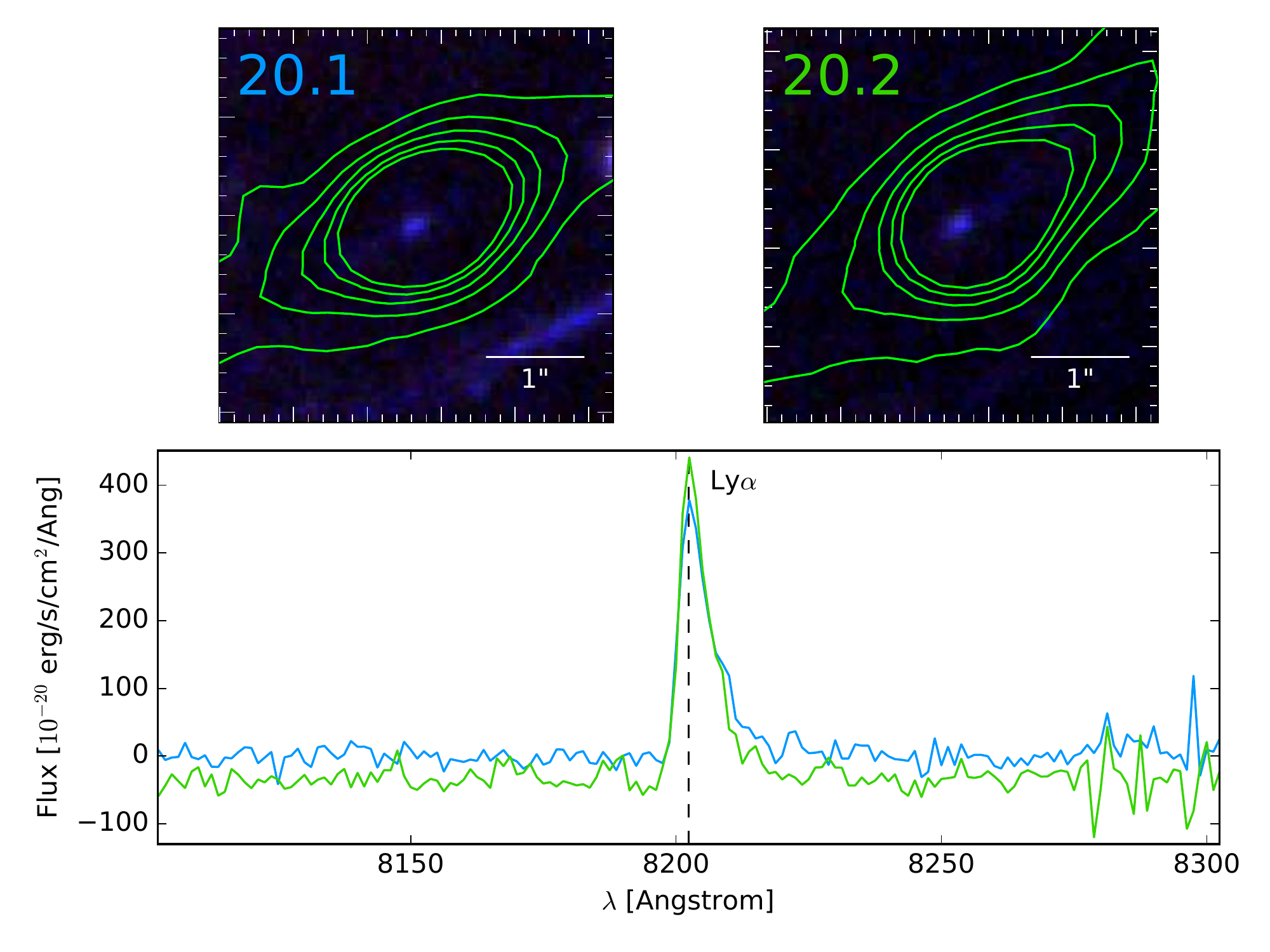}
}
\centering{
\includegraphics[width=0.45\textwidth]{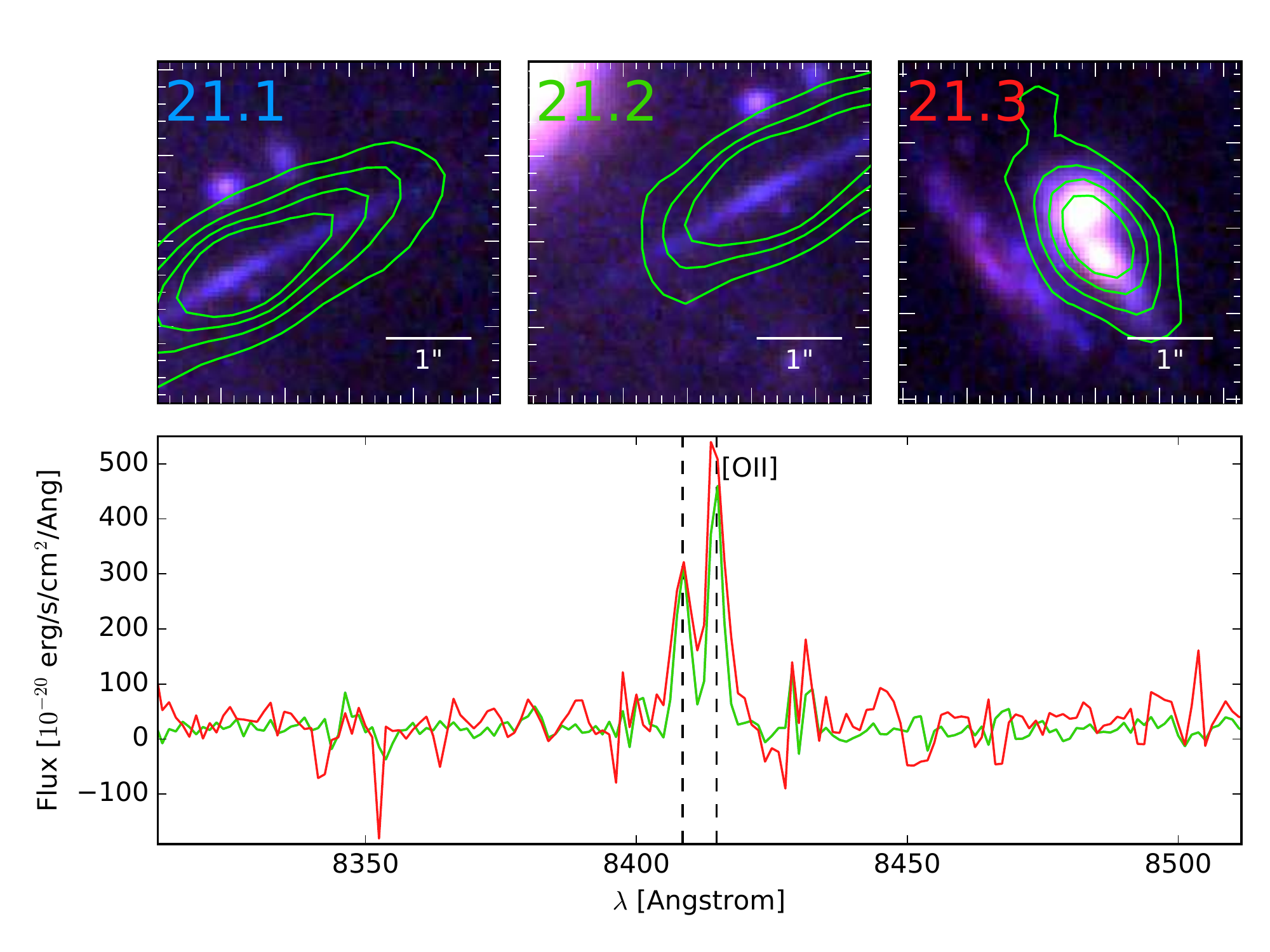}
\includegraphics[width=0.45\textwidth]{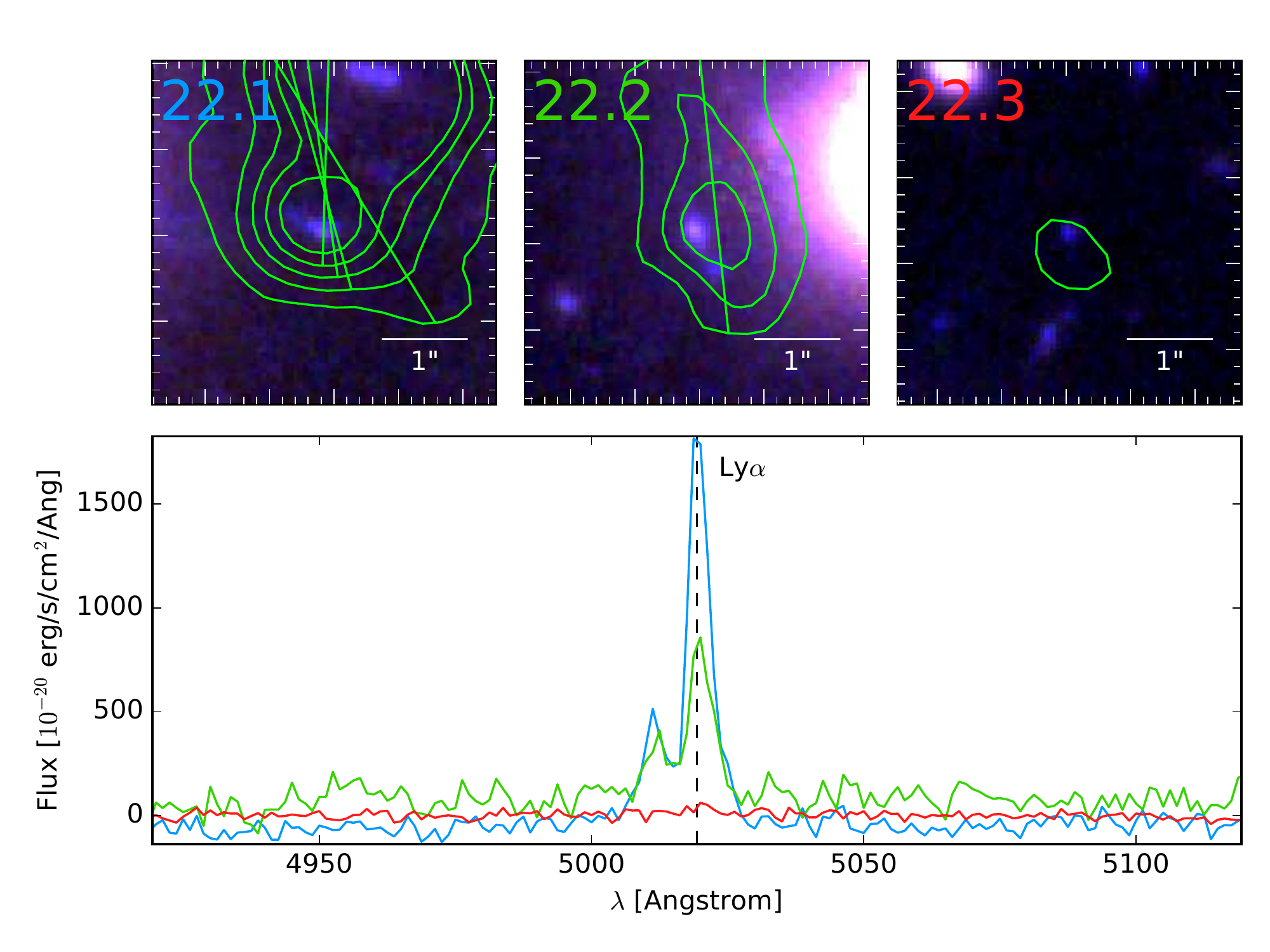}
}
\centering{
\includegraphics[width=0.45\textwidth]{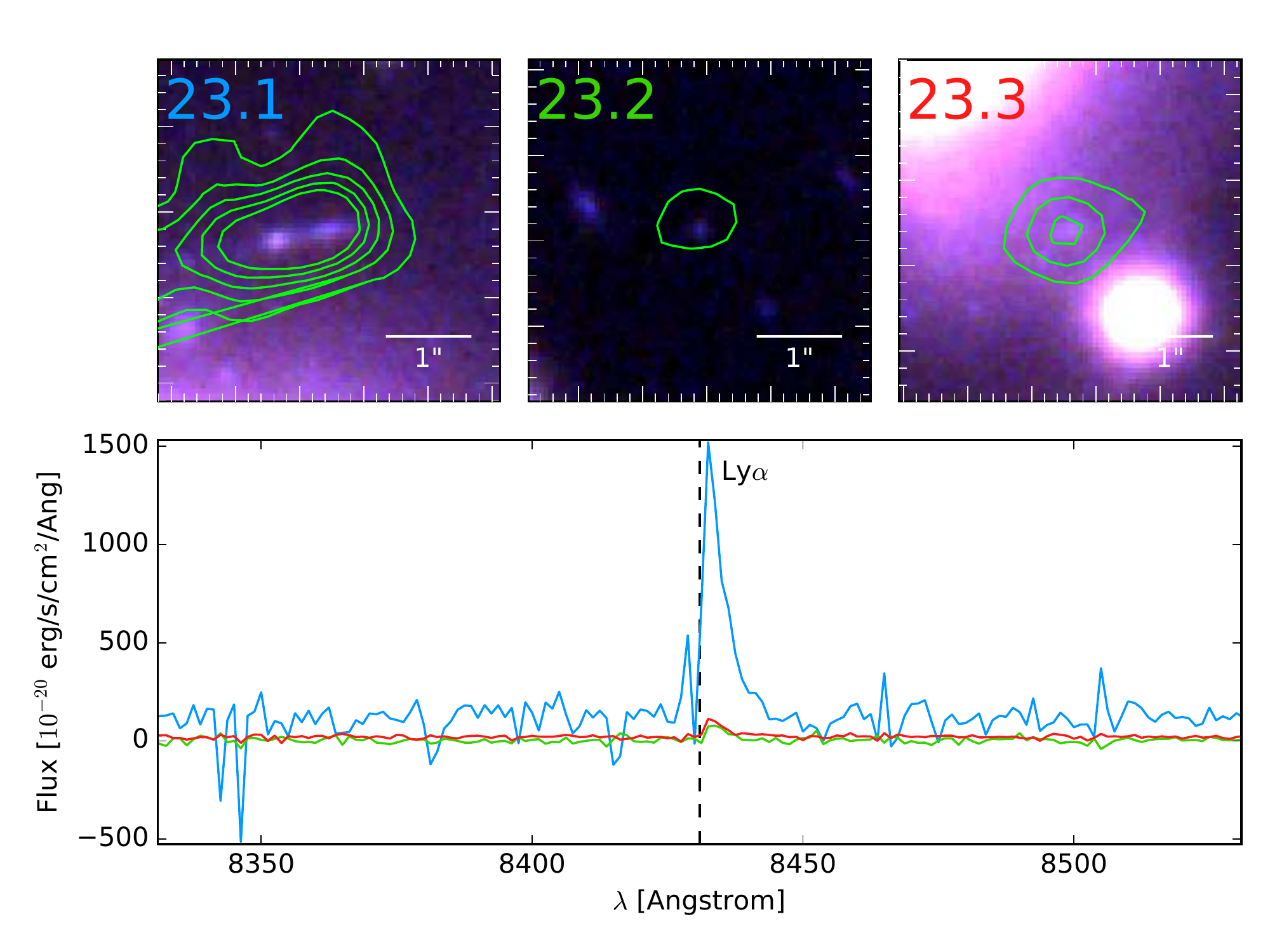}
\includegraphics[width=0.45\textwidth]{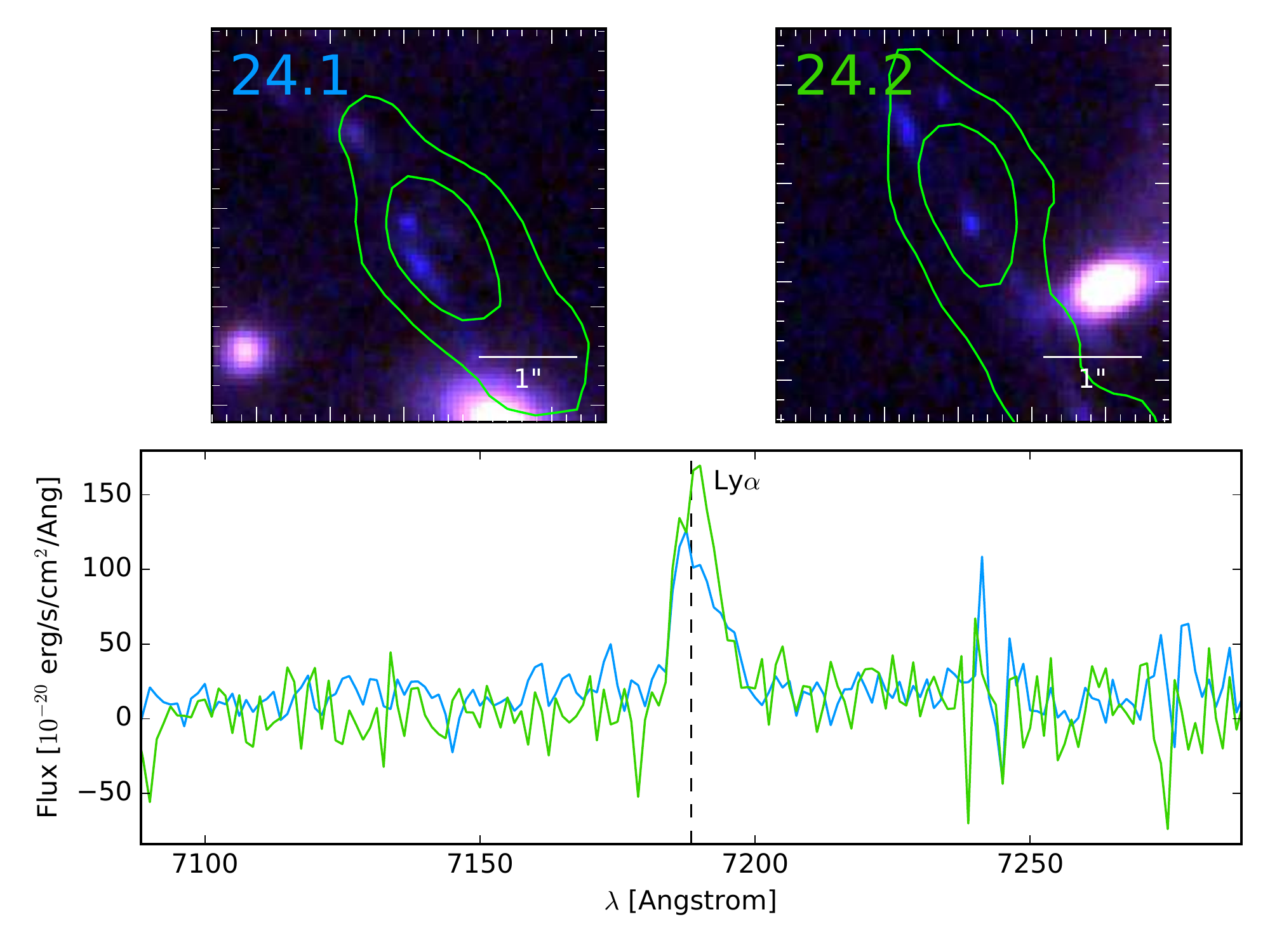}
}
\caption{Multiple-Images with spectroscopic redshifts (continued)}
\end{figure*}

\begin{figure*} \ContinuedFloat
\centering{
\includegraphics[width=0.45\textwidth]{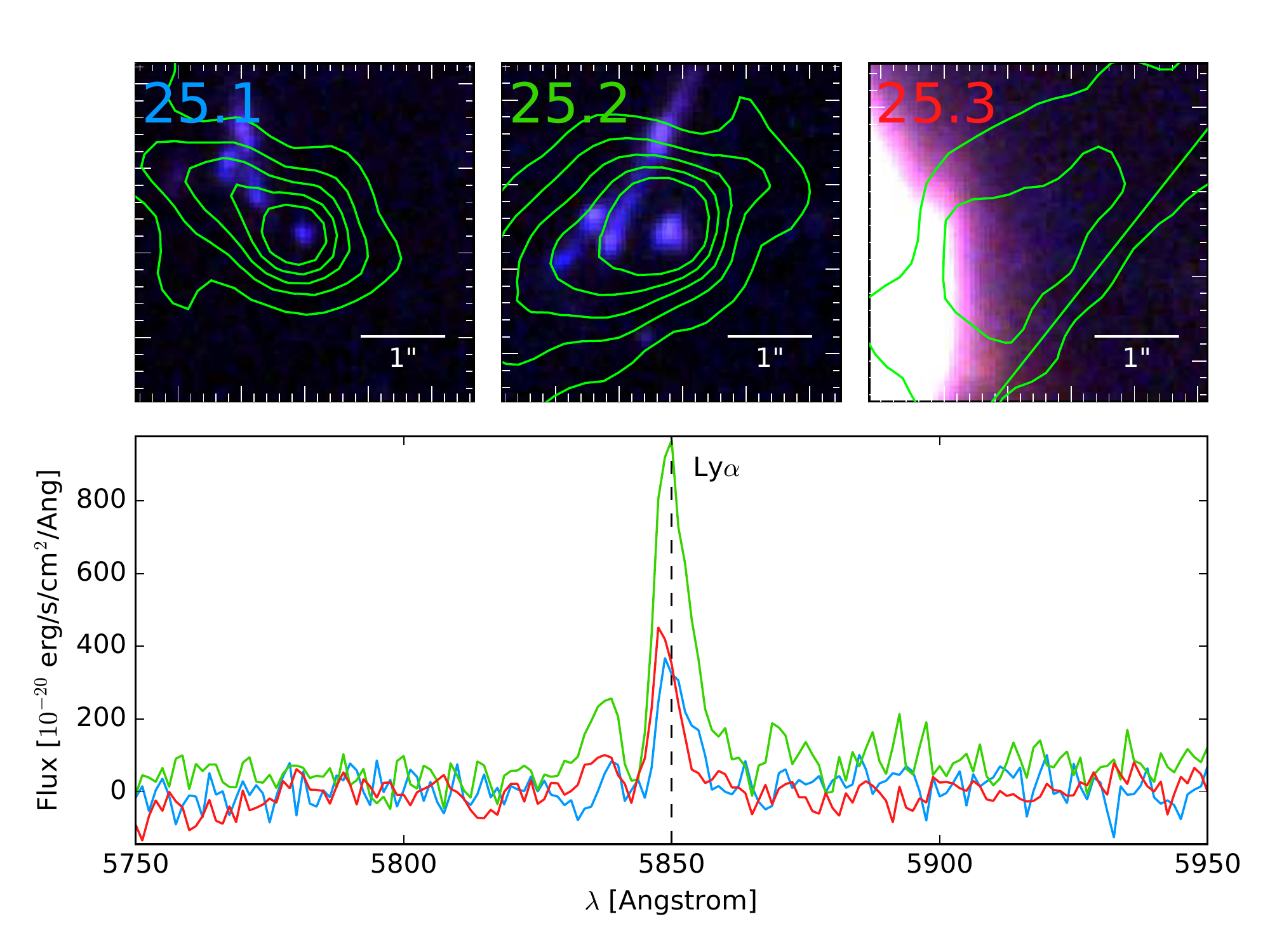}
\includegraphics[width=0.45\textwidth]{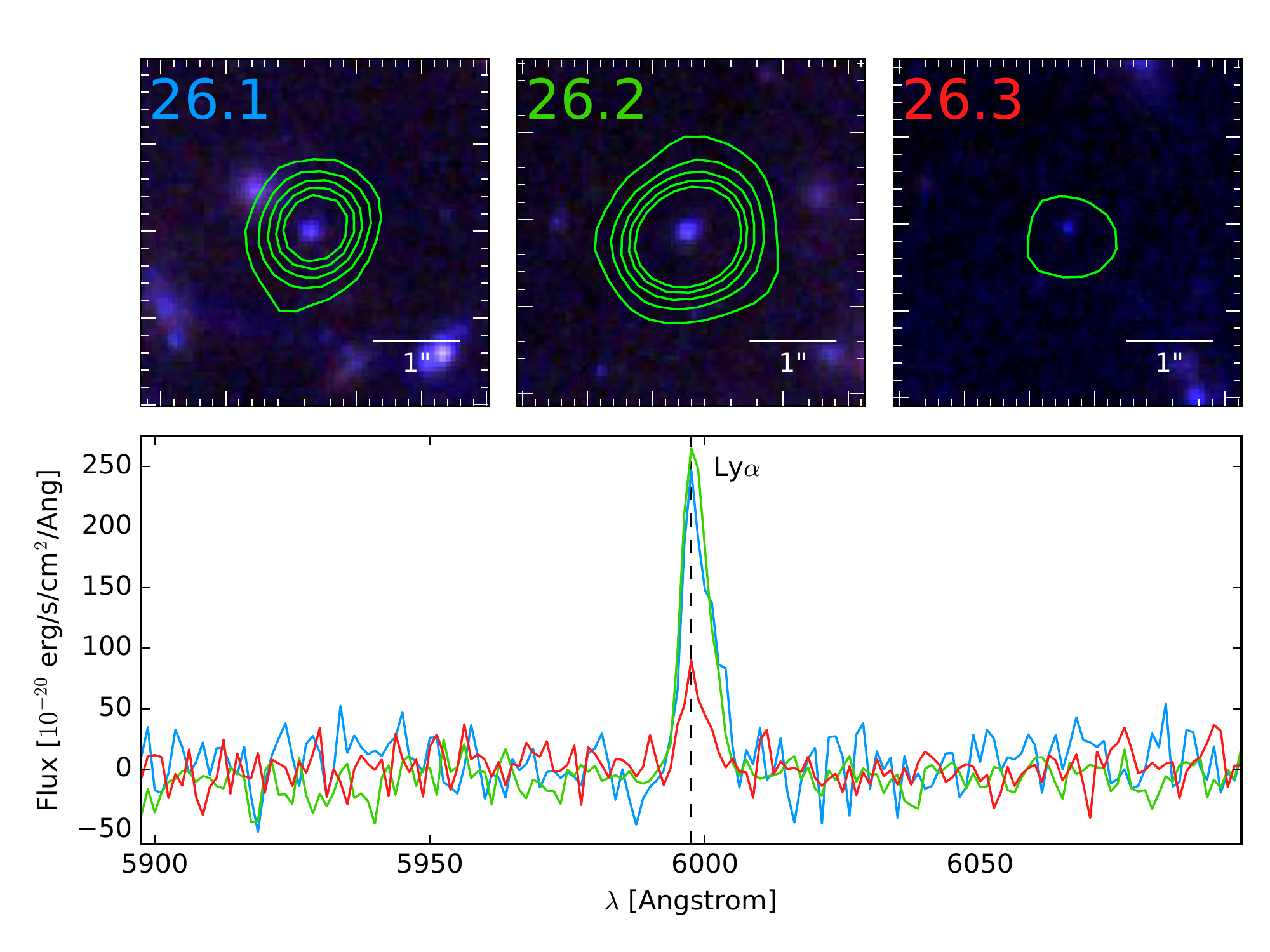}
}
\centering{
\includegraphics[width=0.45\textwidth]{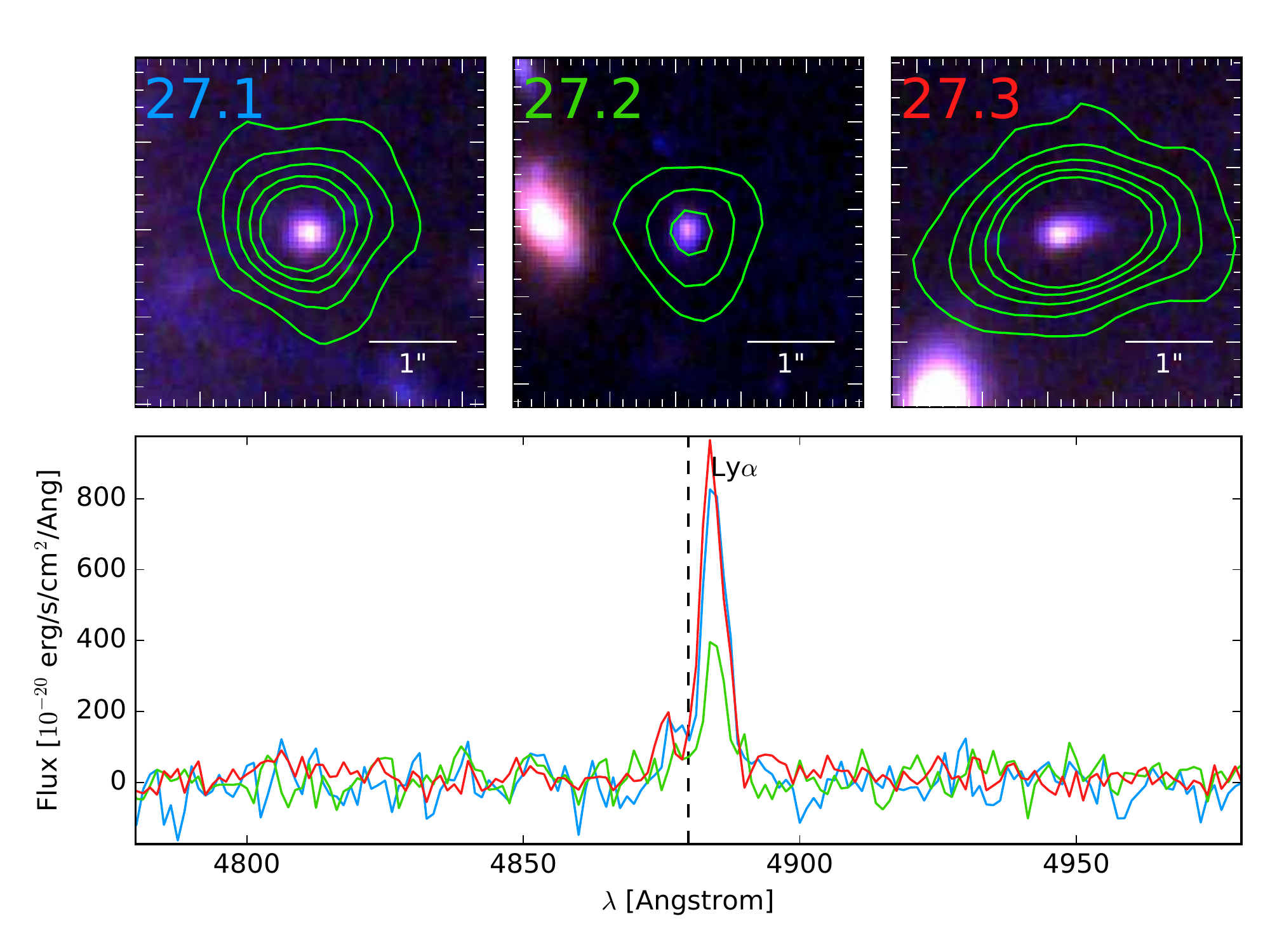}
\includegraphics[width=0.45\textwidth]{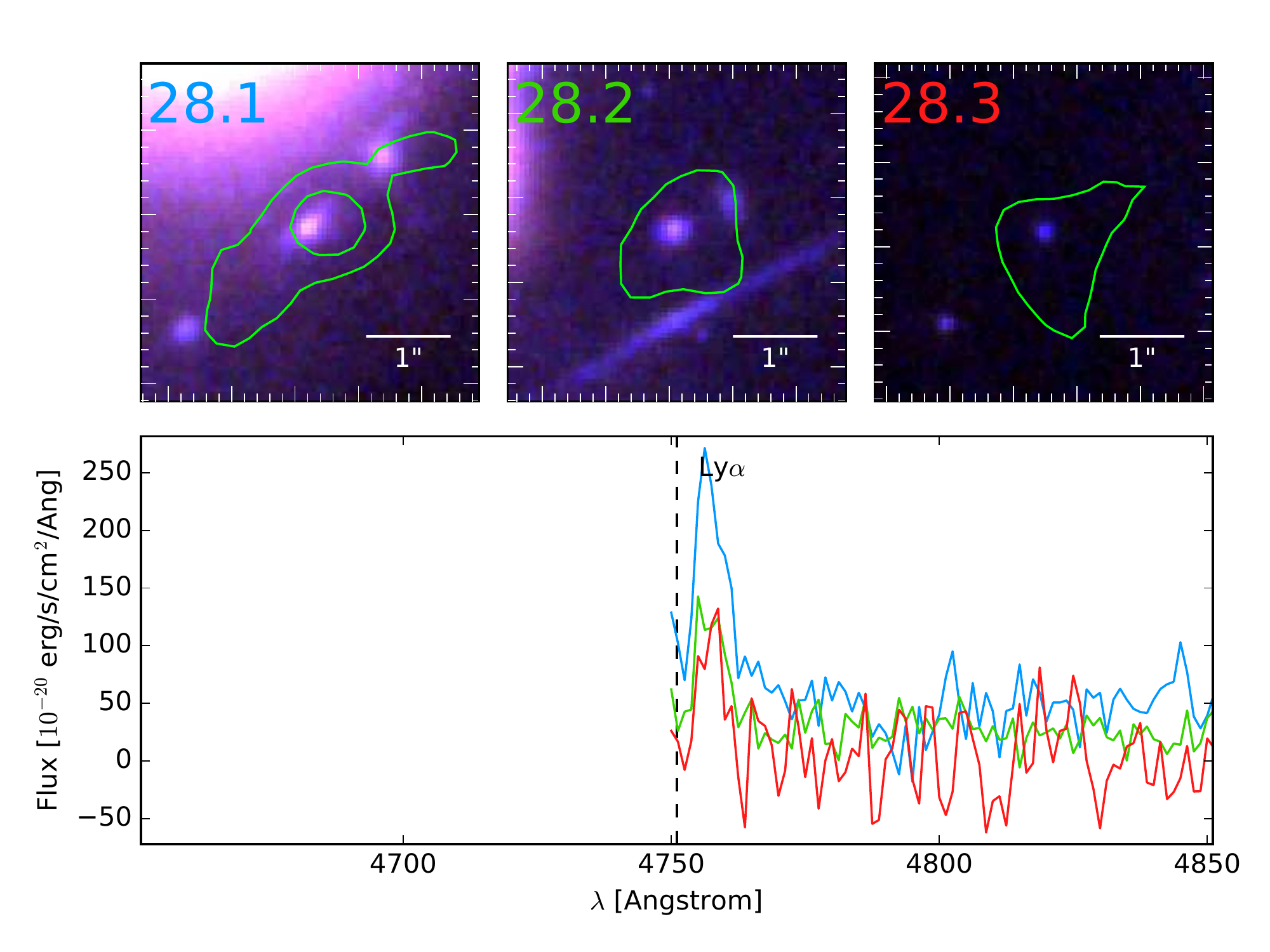}
}
\centering{
\includegraphics[width=0.45\textwidth]{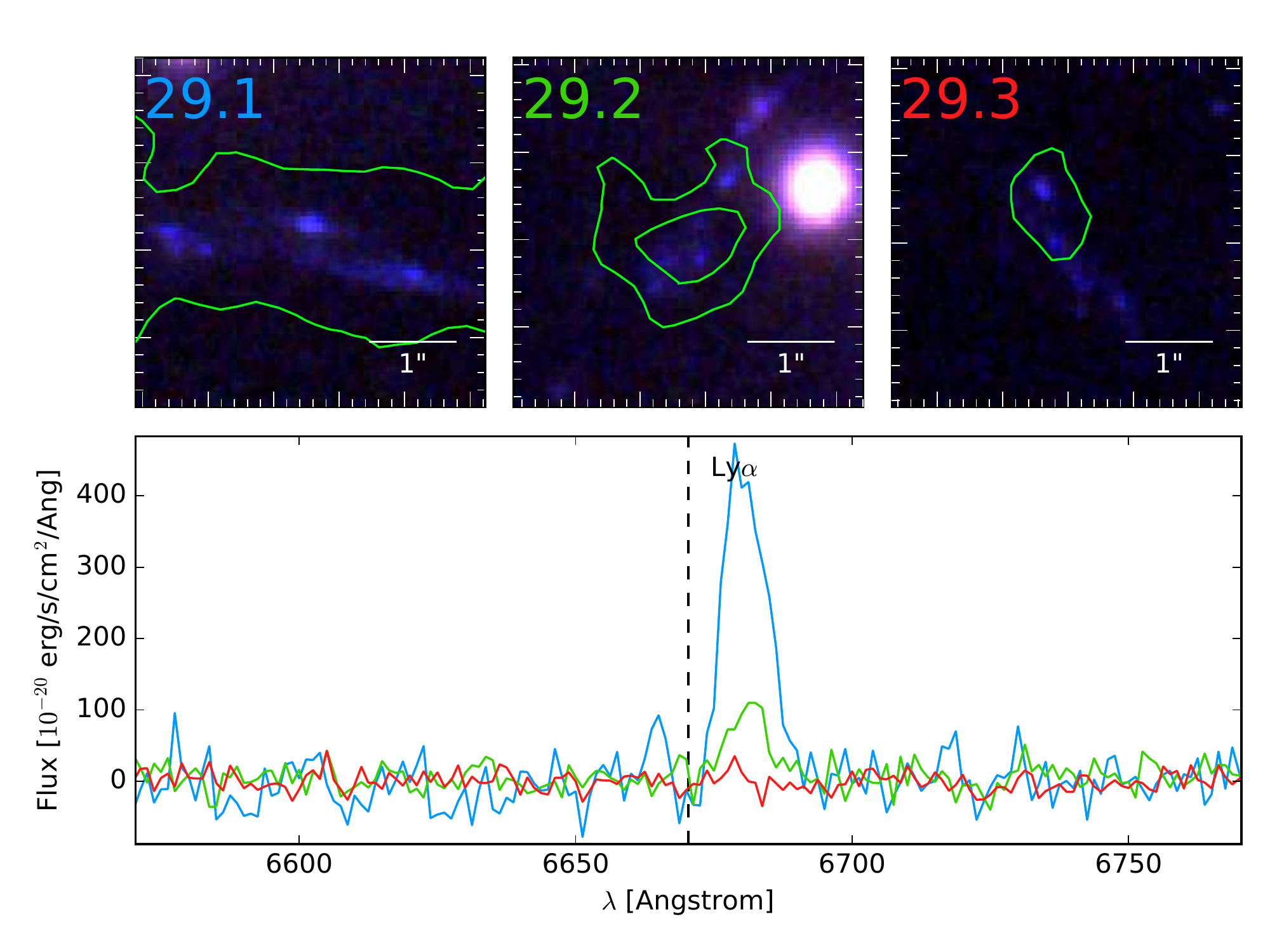}
\includegraphics[width=0.45\textwidth]{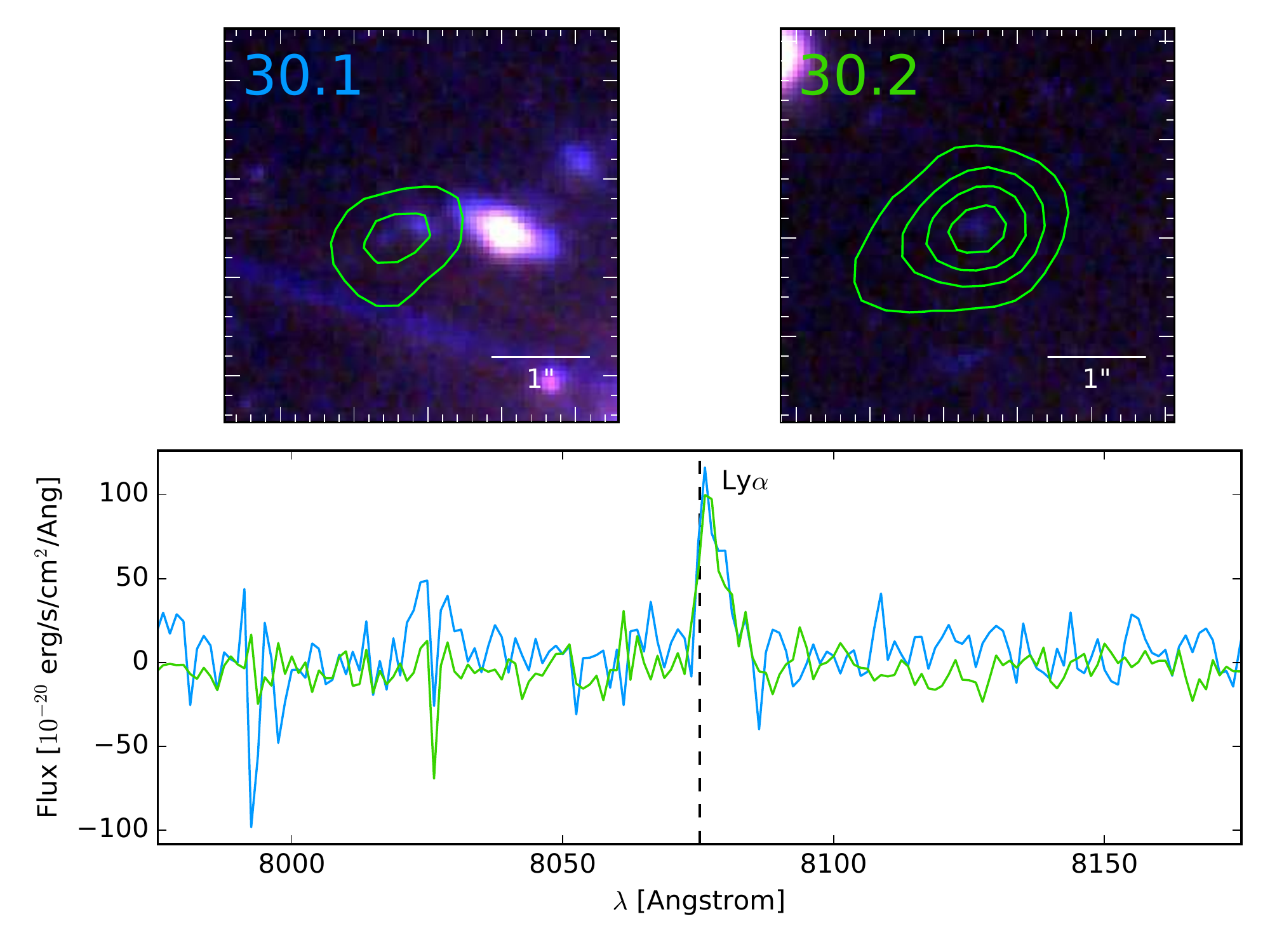}
}
\centering{
\includegraphics[width=0.45\textwidth]{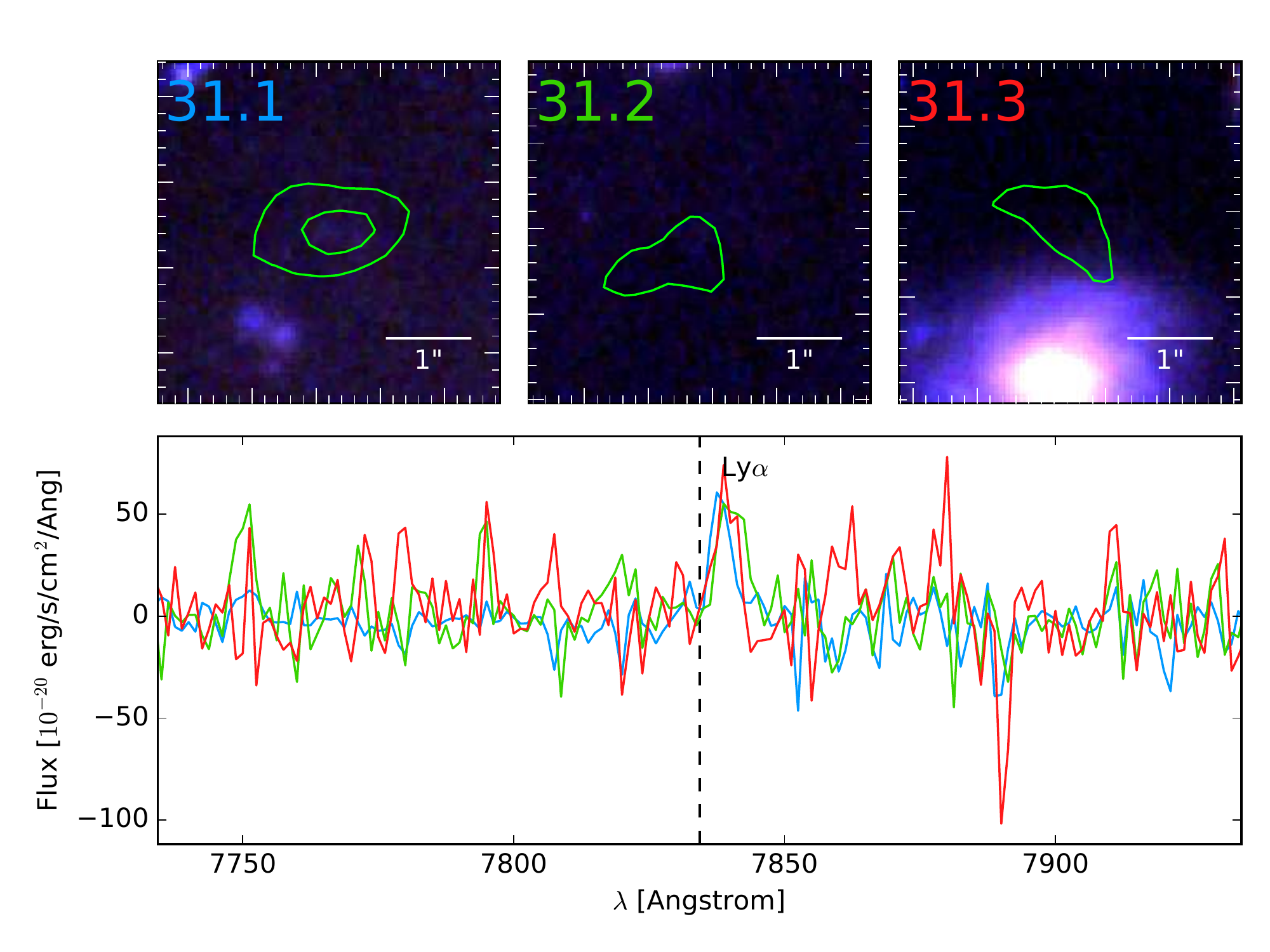}
\includegraphics[width=0.45\textwidth]{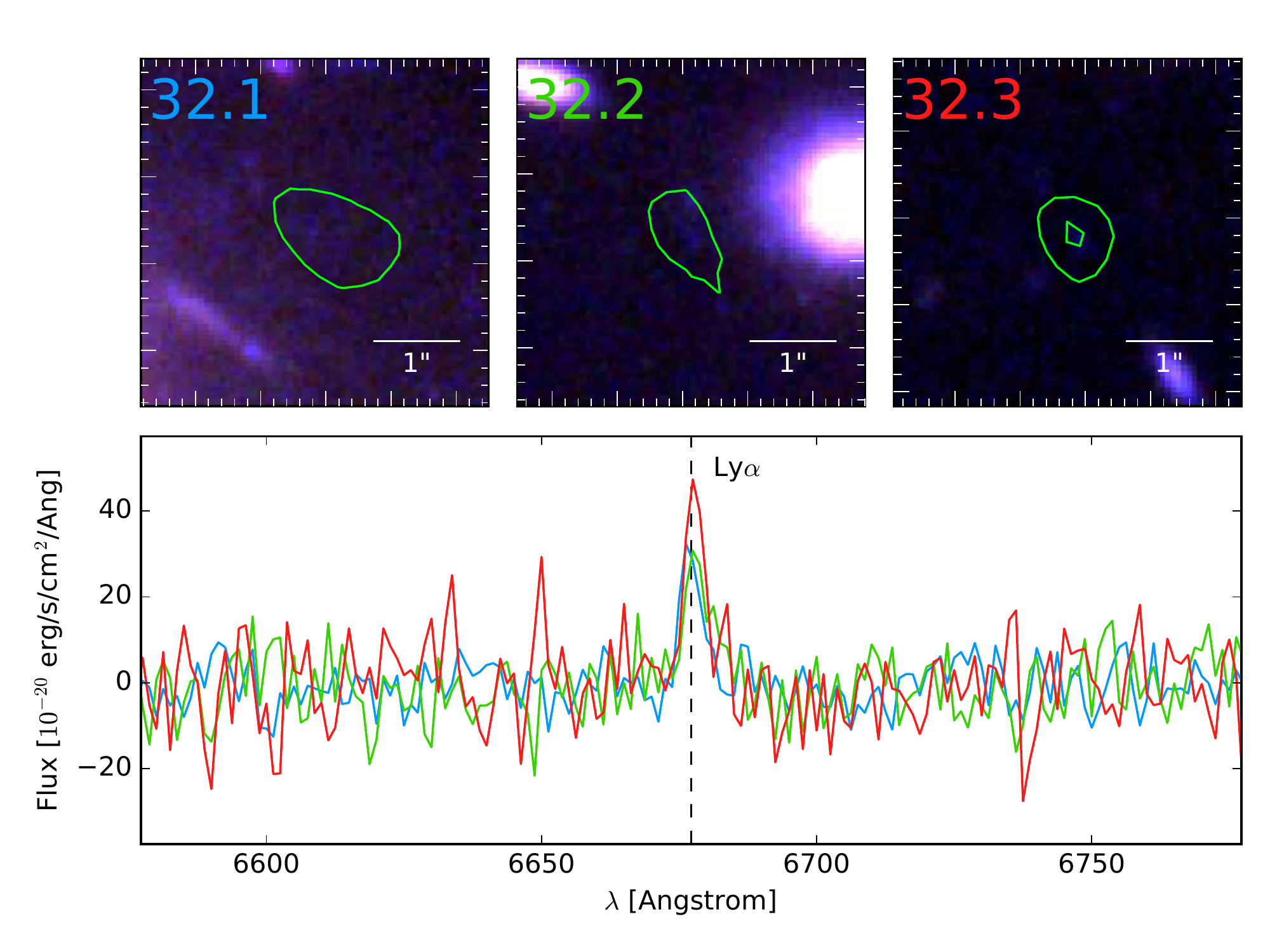}
}
\caption{Multiple-Images with spectroscopic redshifts (continued)}
\end{figure*}

\begin{figure*} \ContinuedFloat
\centering{
\includegraphics[width=0.45\textwidth]{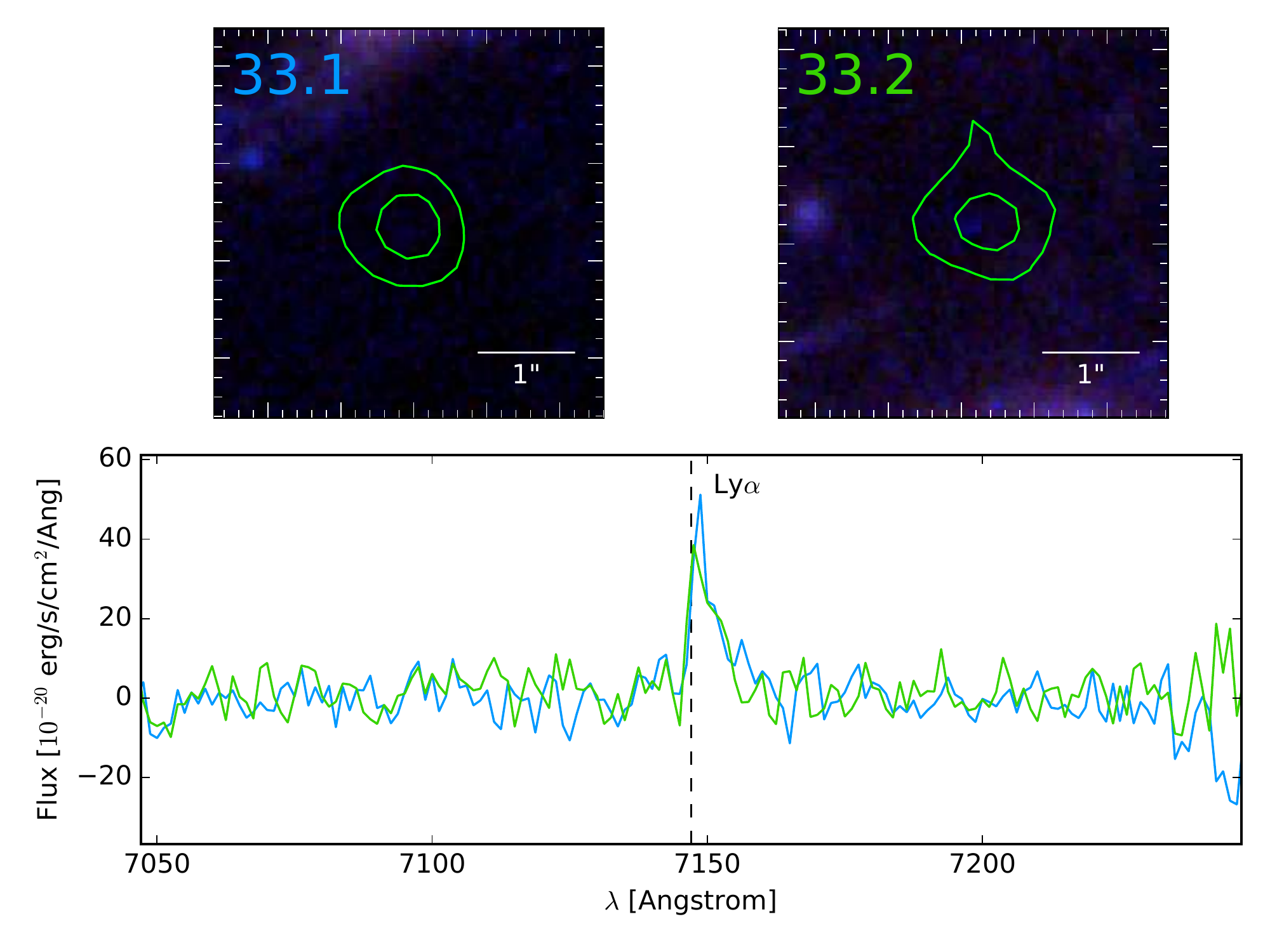}
\includegraphics[width=0.45\textwidth]{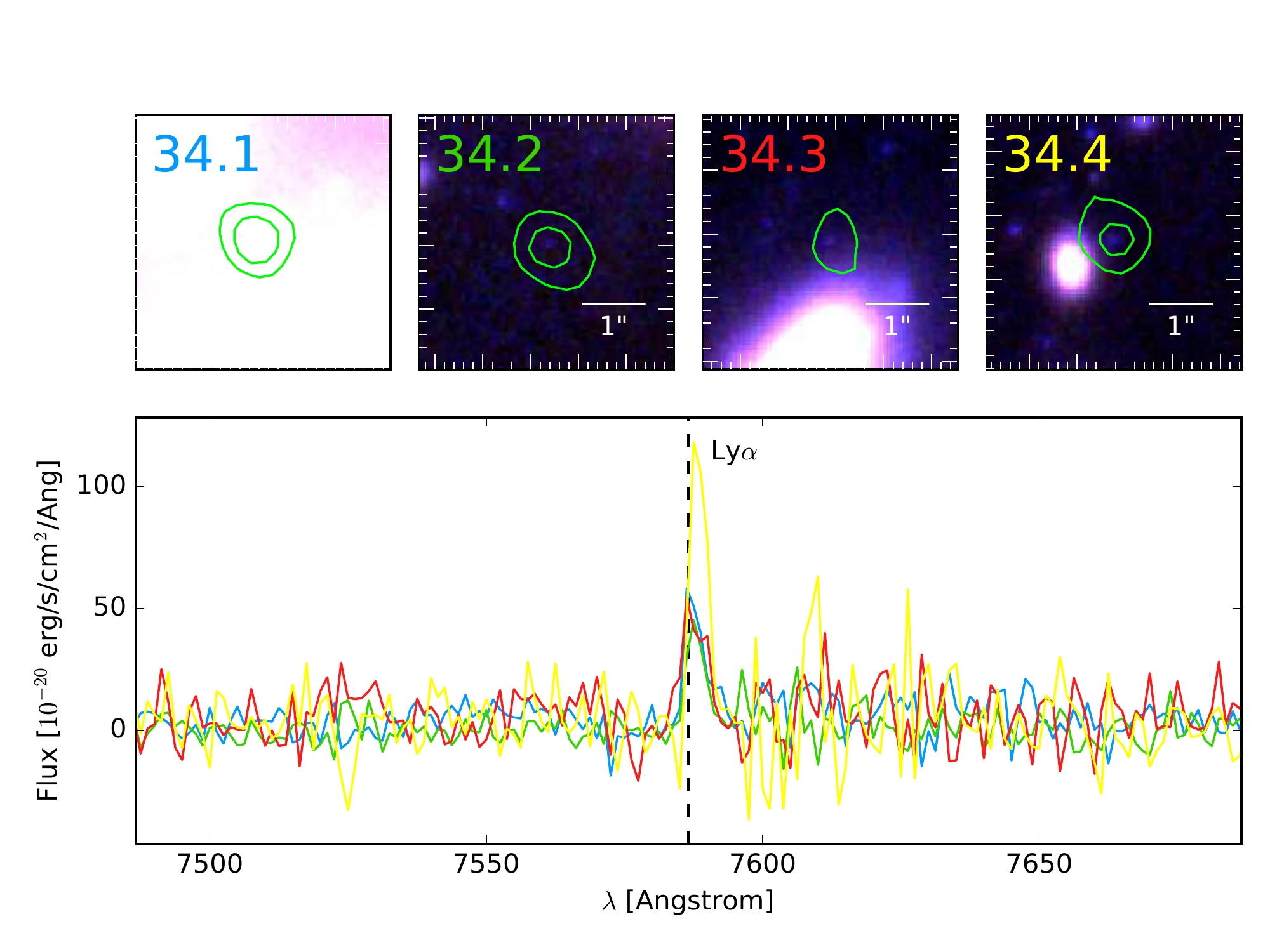}
}
\centering{
\includegraphics[width=0.45\textwidth]{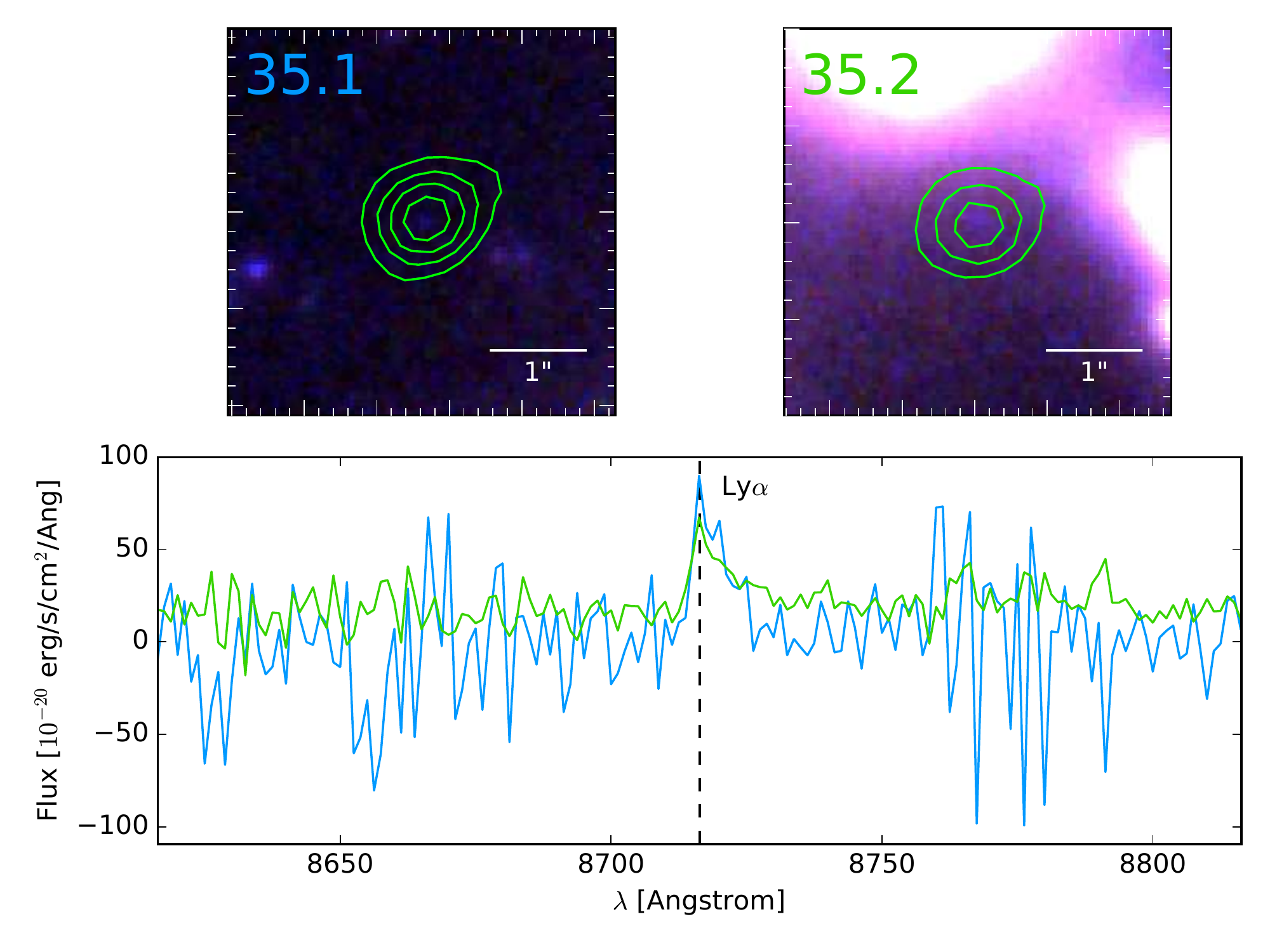}
\includegraphics[width=0.45\textwidth]{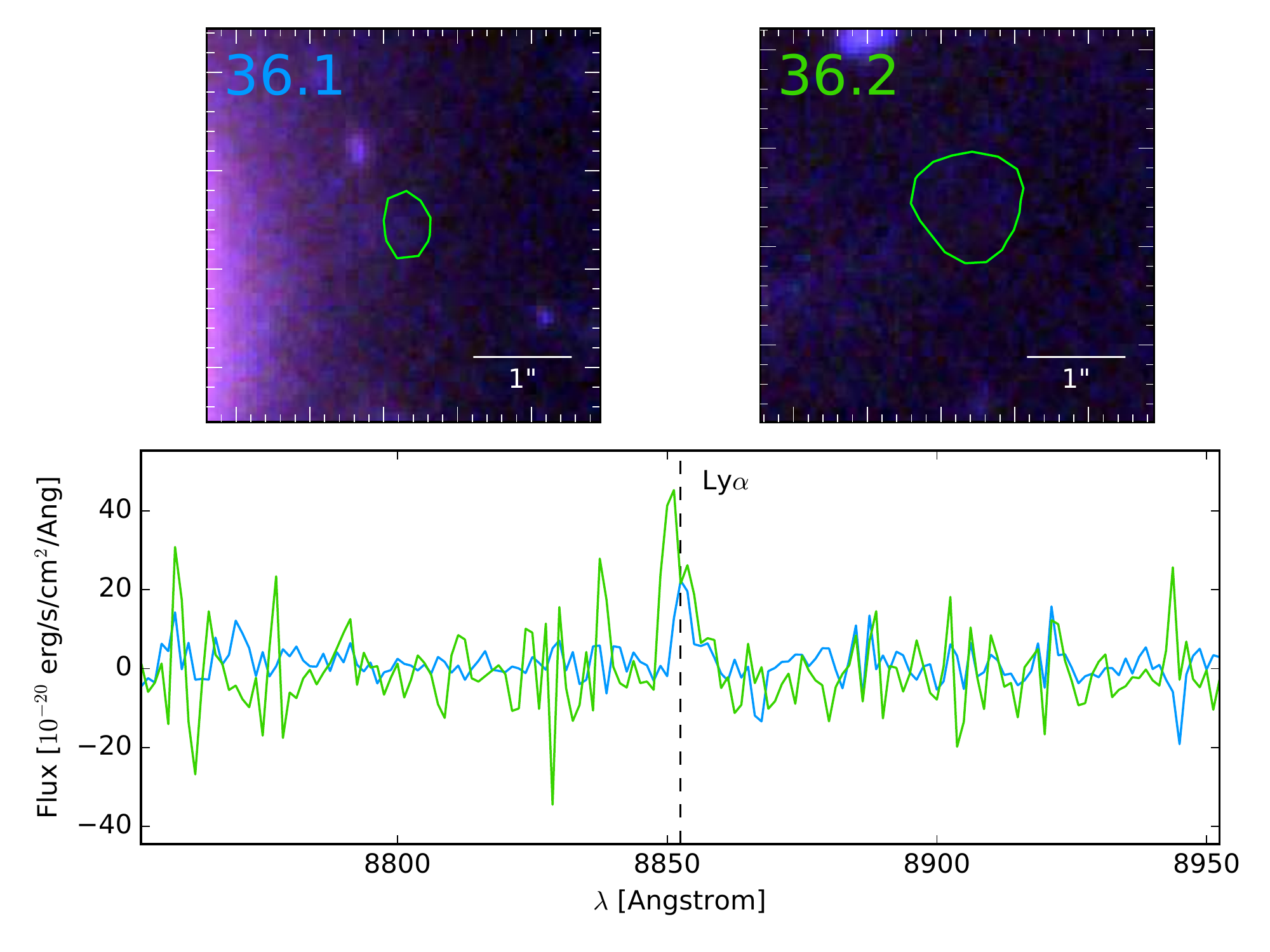}
}
\centering{
\includegraphics[width=0.45\textwidth]{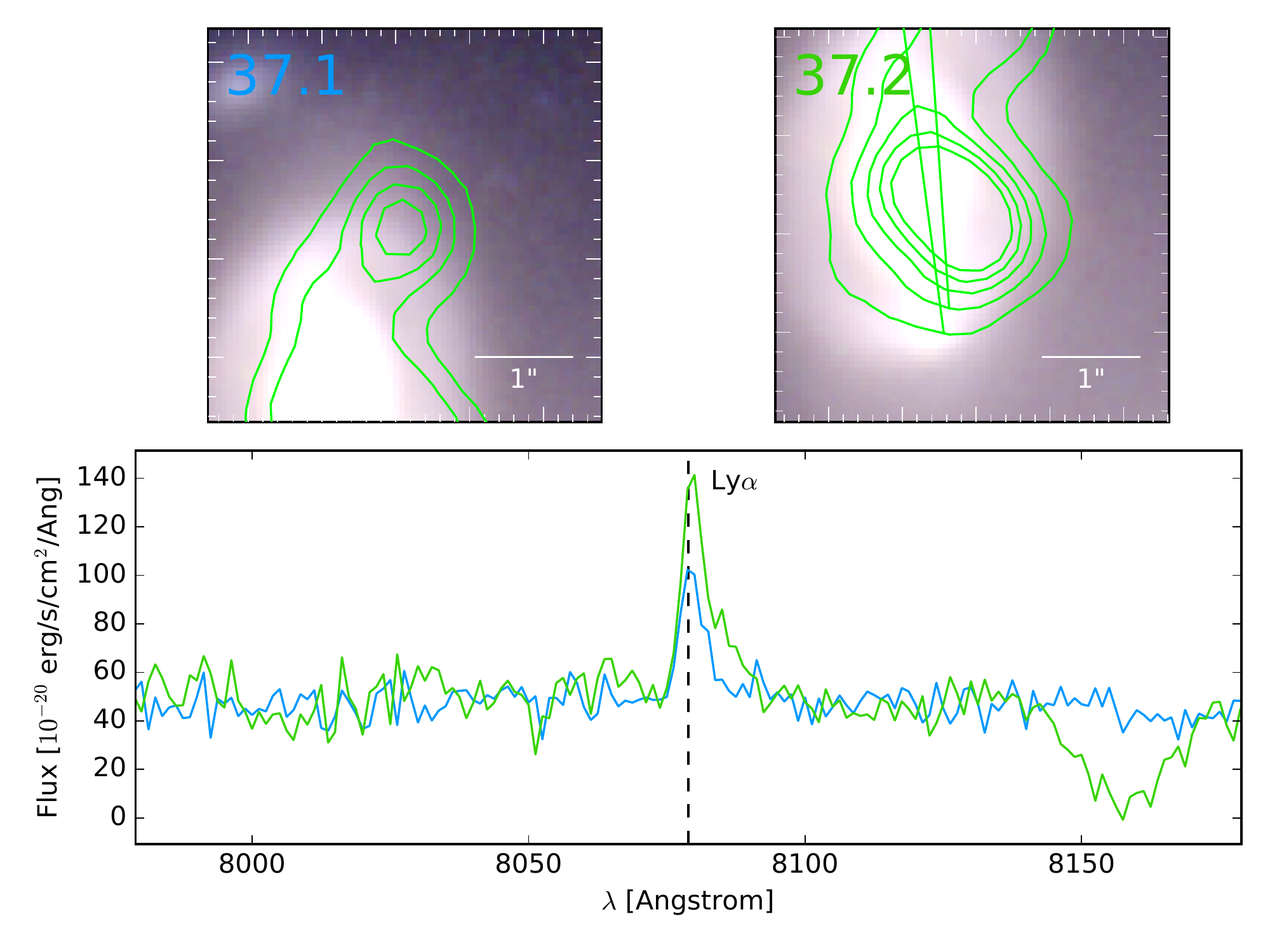}
\includegraphics[width=0.45\textwidth]{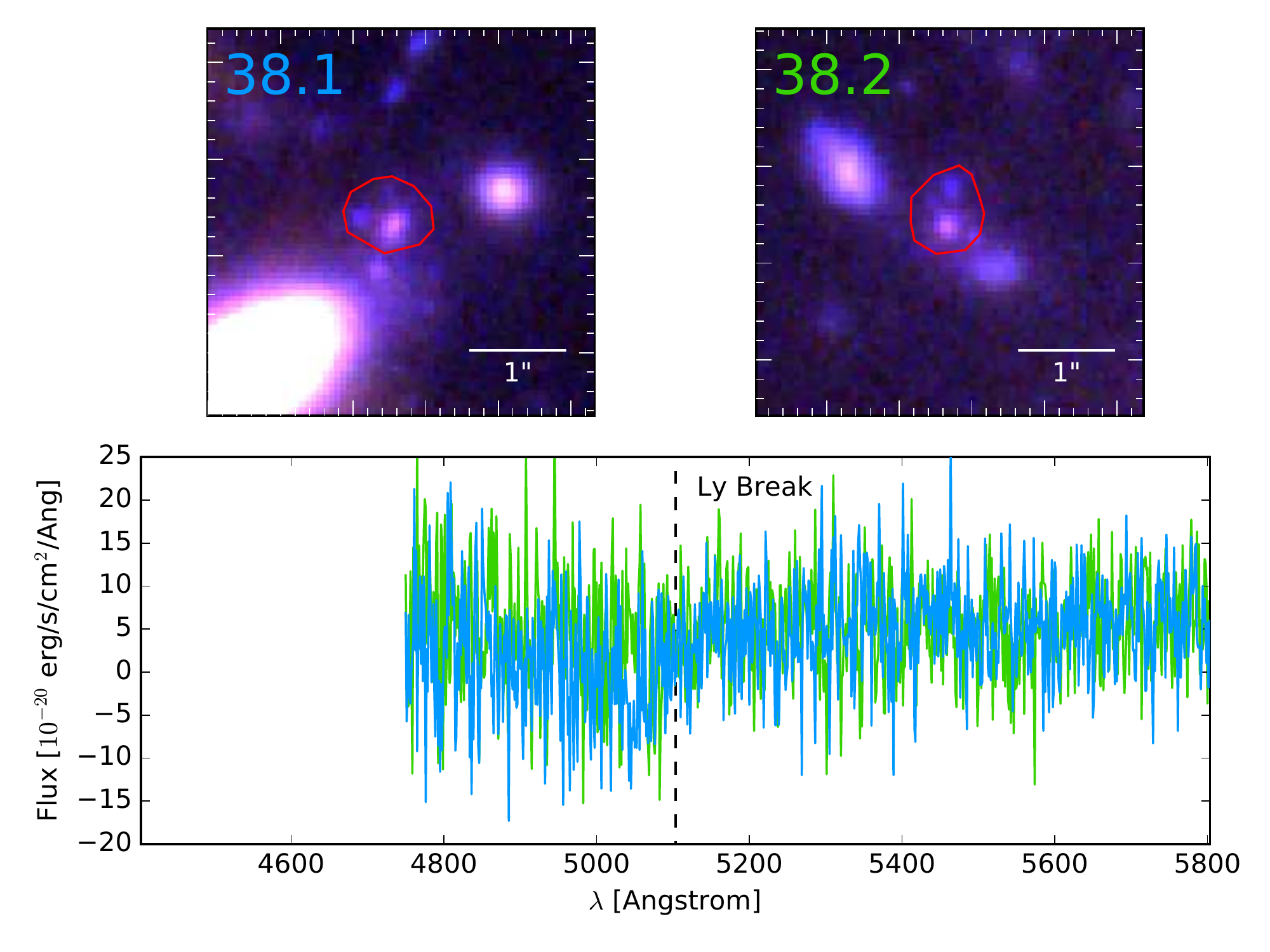}
}
\centering{
\includegraphics[width=0.45\textwidth]{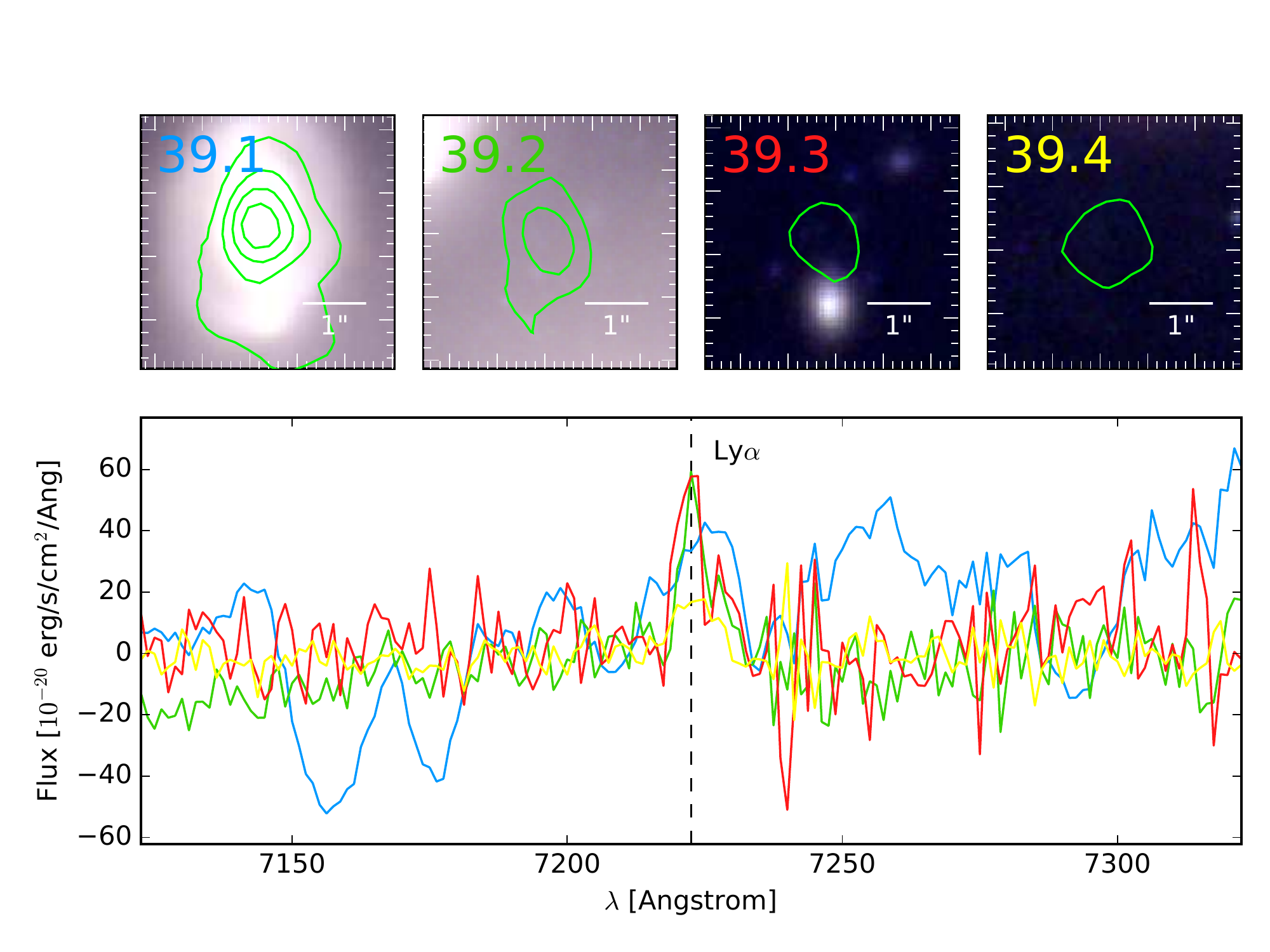}
\includegraphics[width=0.45\textwidth]{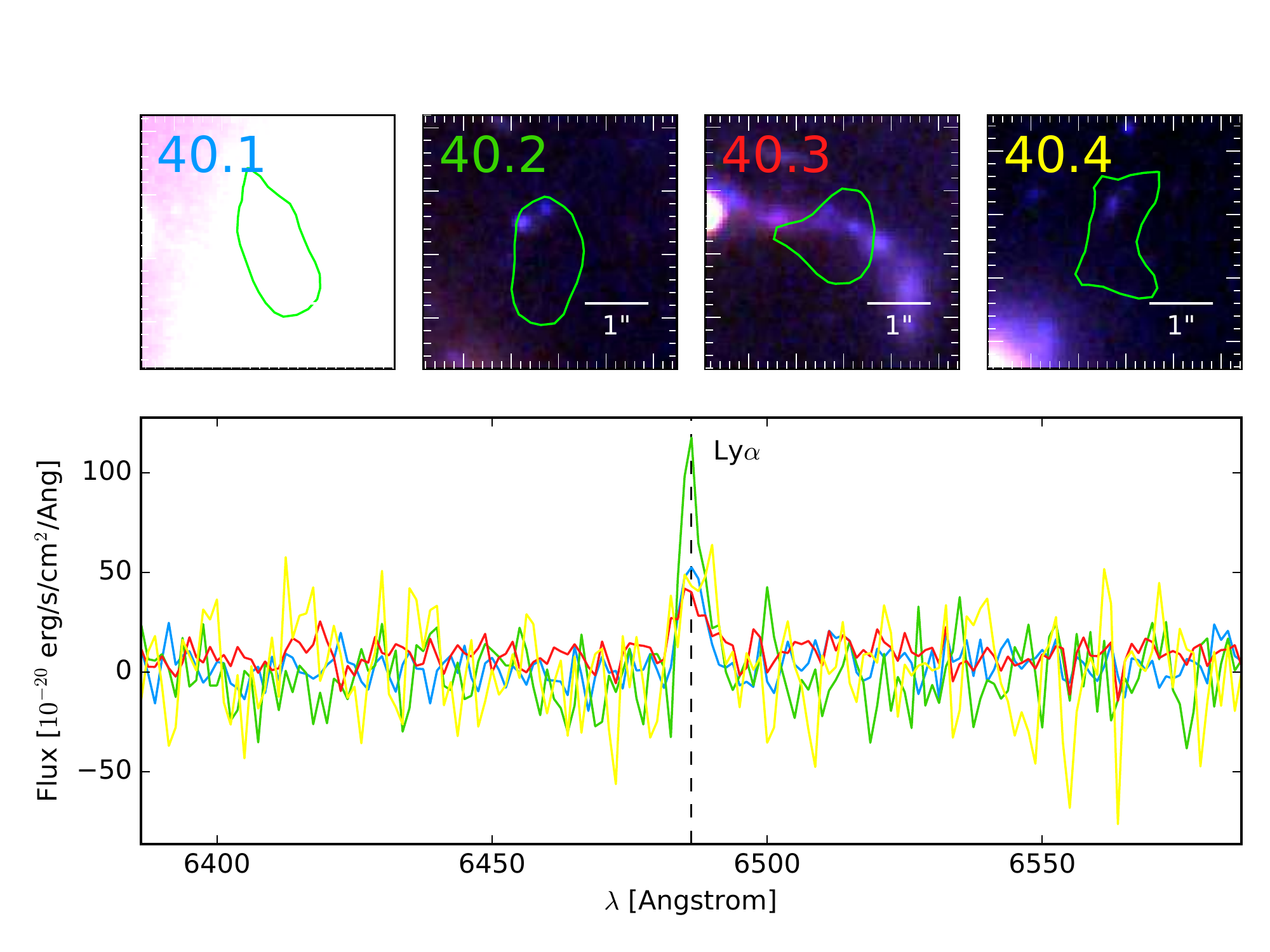}
}
\caption{Multiple-Images with spectroscopic redshifts (continued)}
\end{figure*}

\begin{figure*} \ContinuedFloat
\centering{
\includegraphics[width=0.45\textwidth]{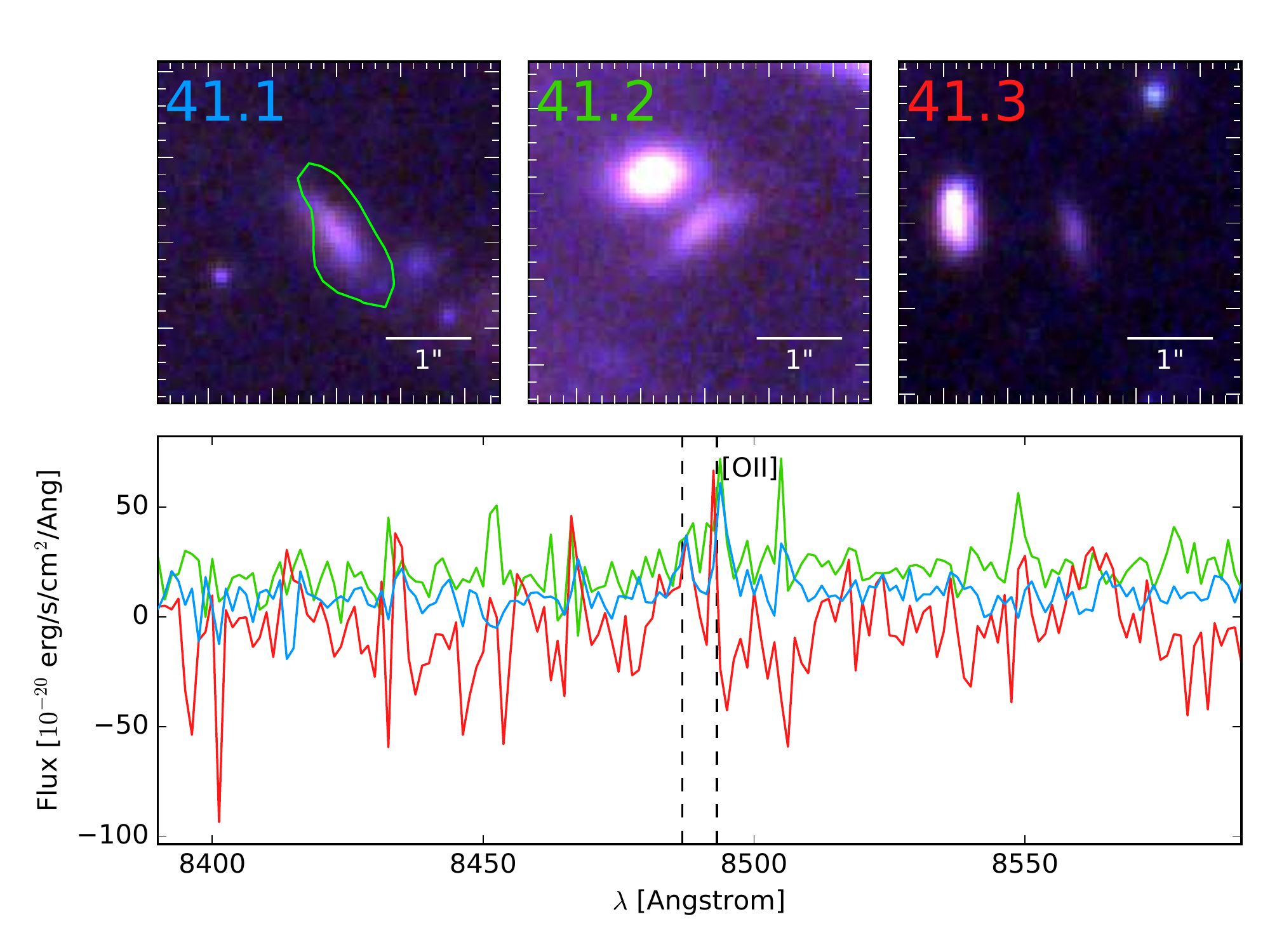}
\includegraphics[width=0.45\textwidth]{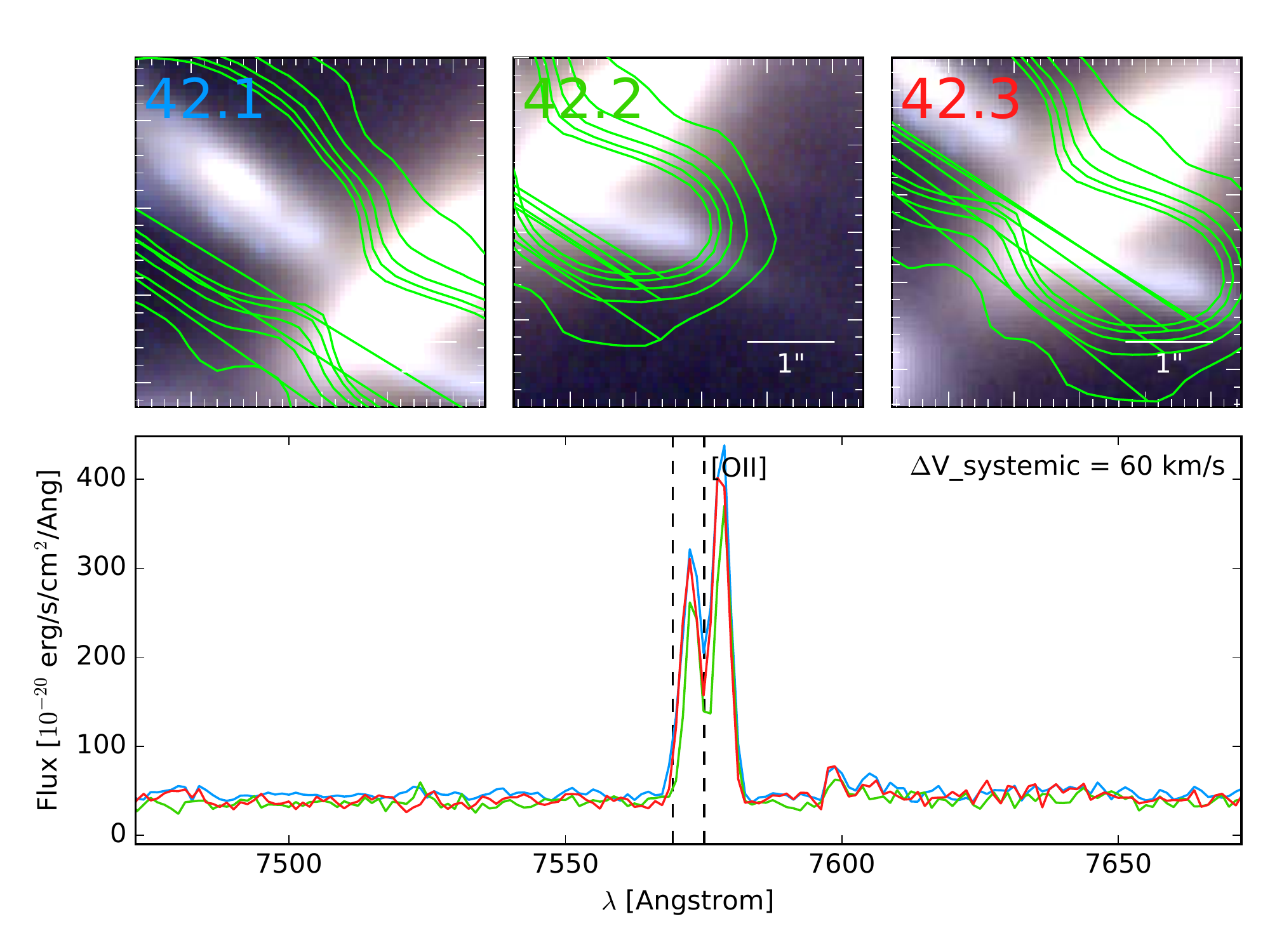}
}
\caption{Multiple-Images with spectroscopic redshifts (continued)}
\end{figure*}

\begin{figure*}
  \centering{
\includegraphics[width=0.45\textwidth]{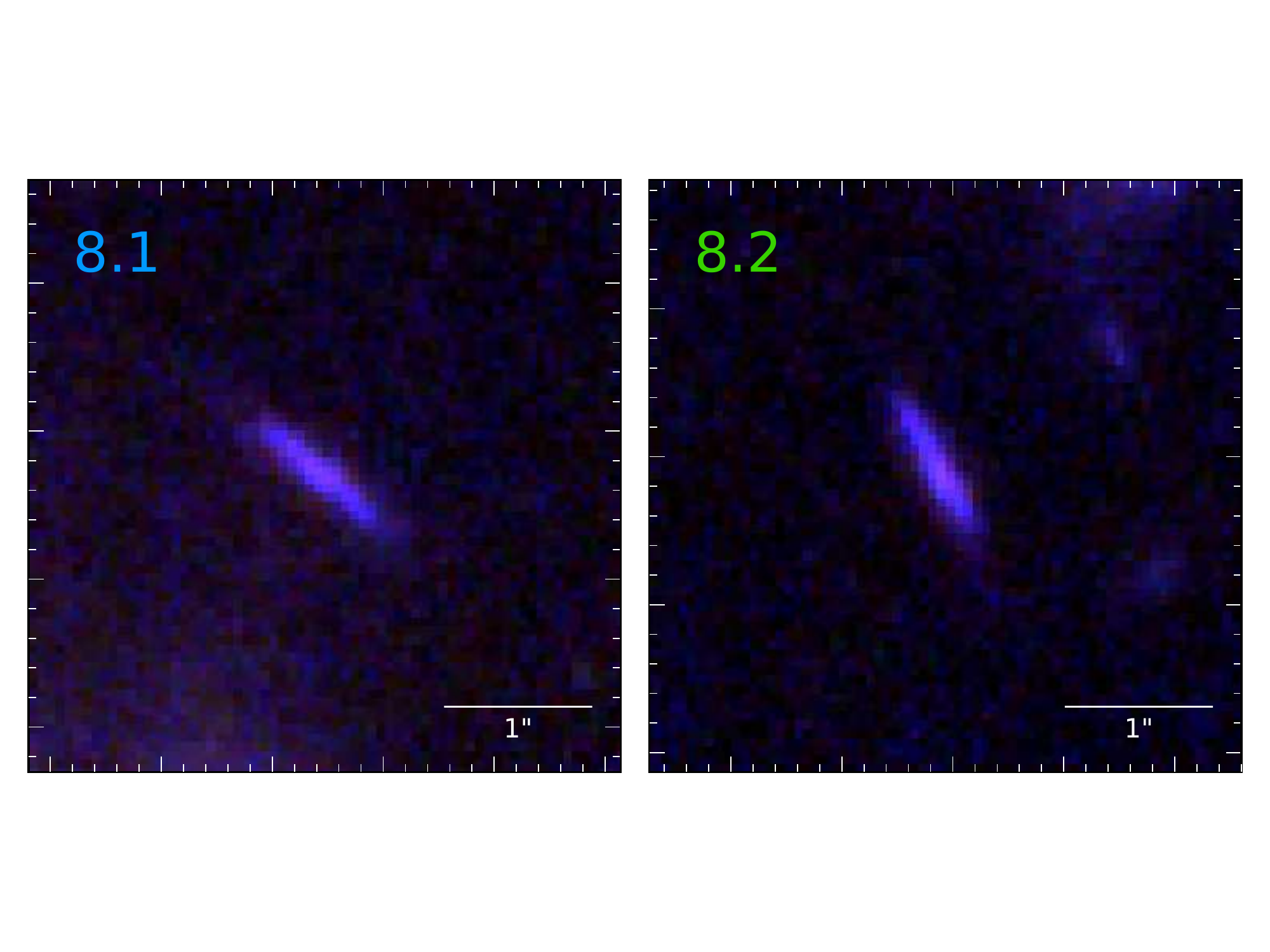}\\
}
\centering{
\includegraphics[width=0.45\textwidth]{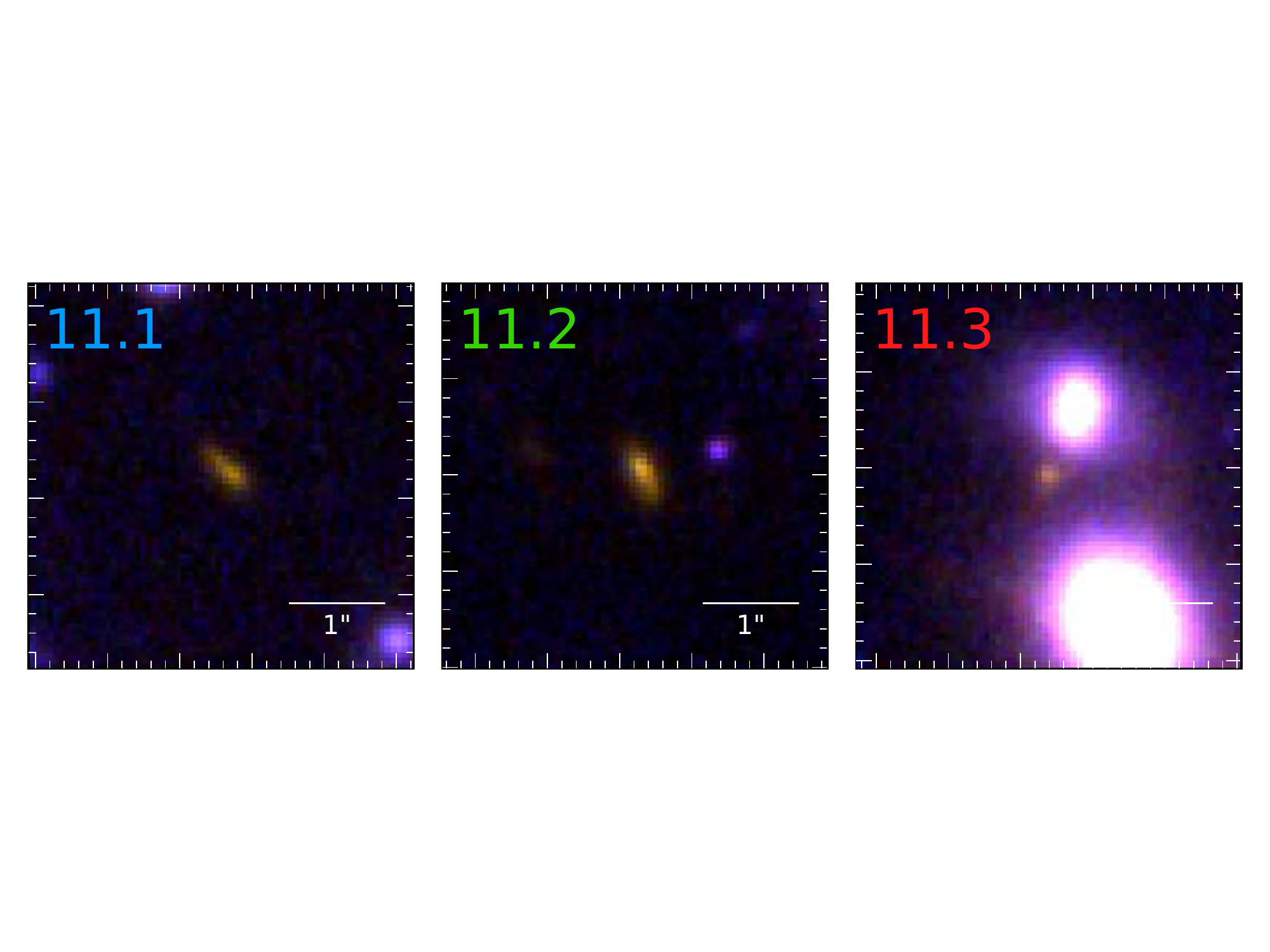}
\includegraphics[width=0.45\textwidth]{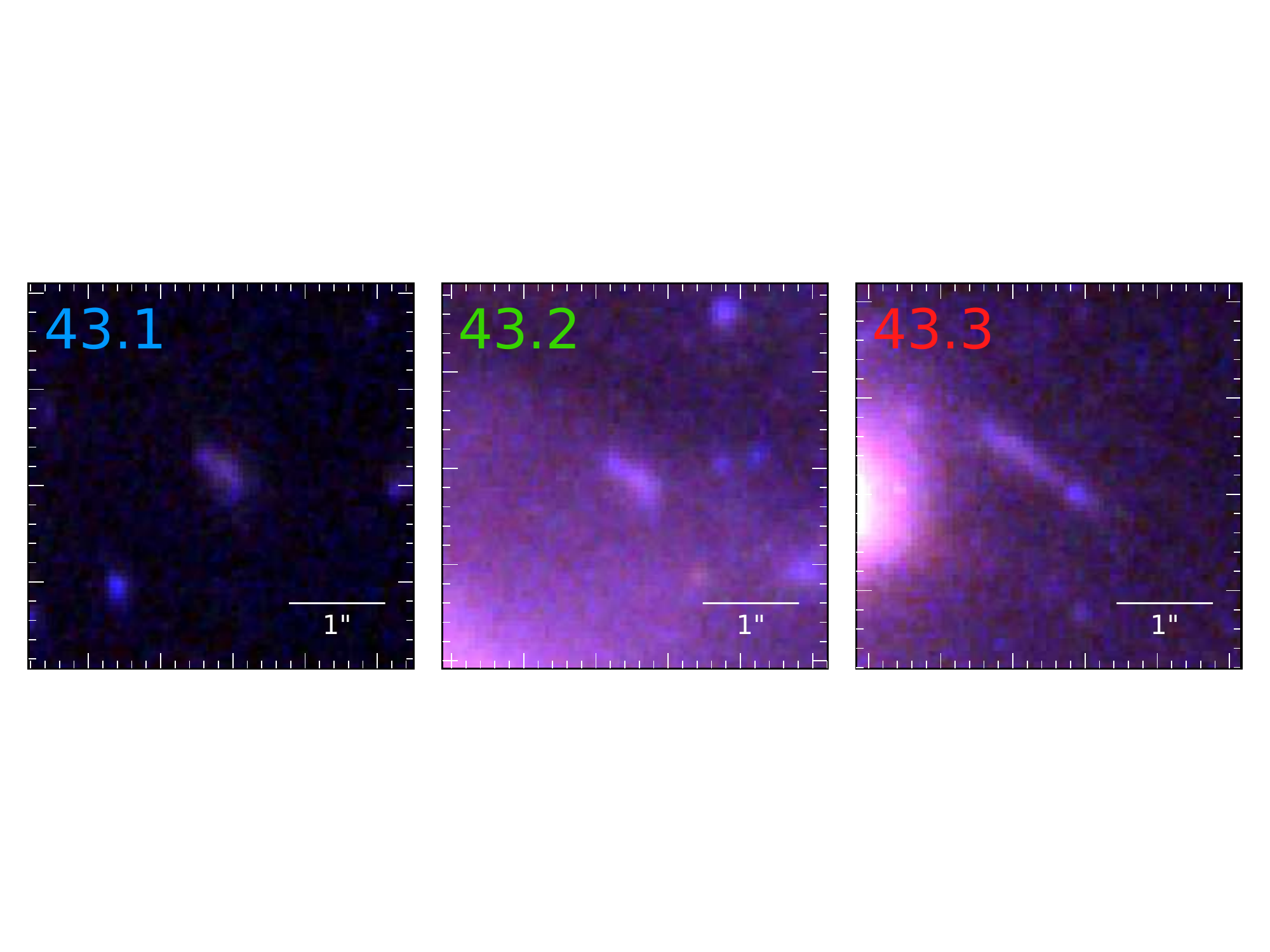}
}
\centering{
\includegraphics[width=0.45\textwidth]{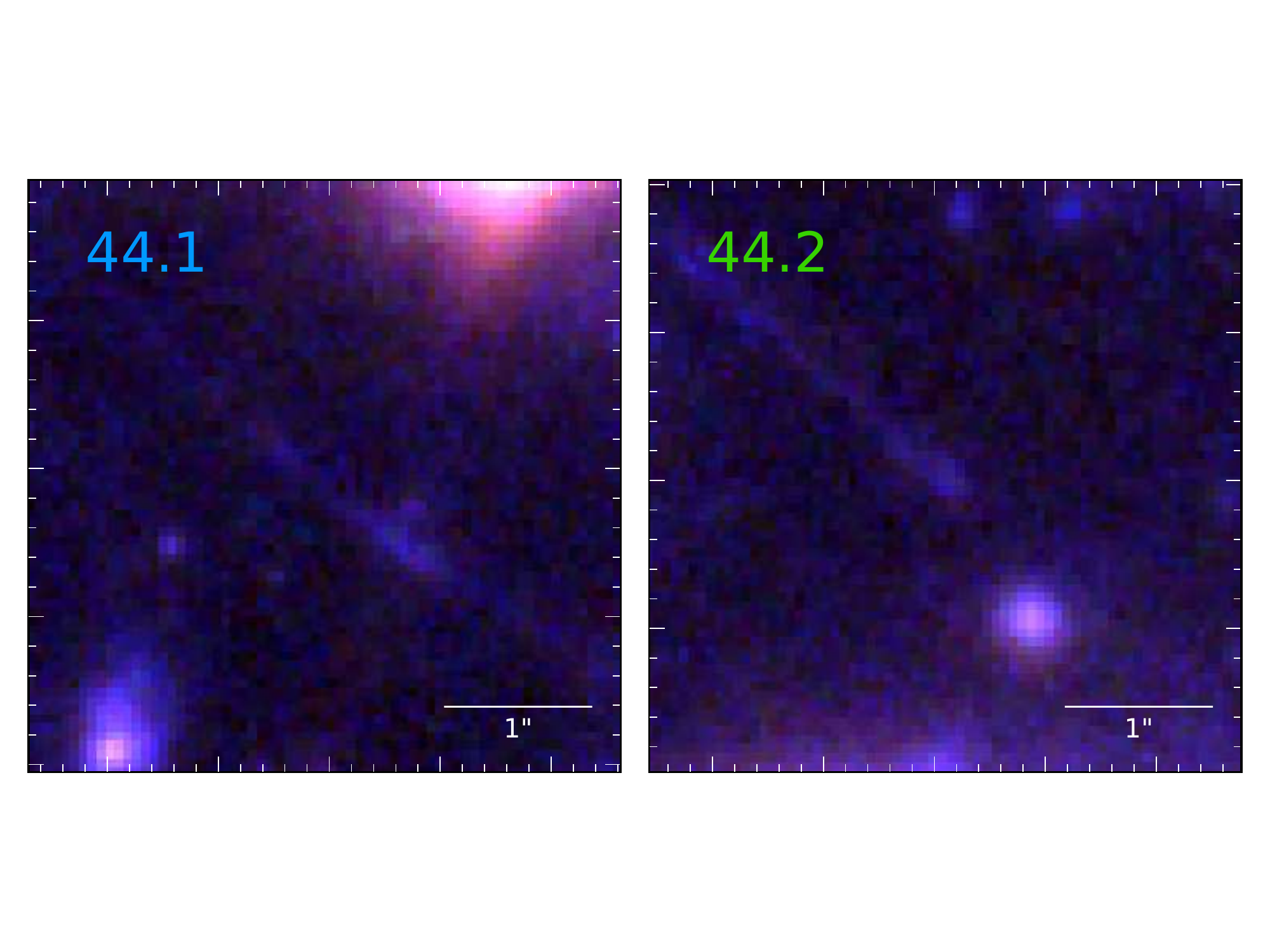}
\includegraphics[width=0.45\textwidth]{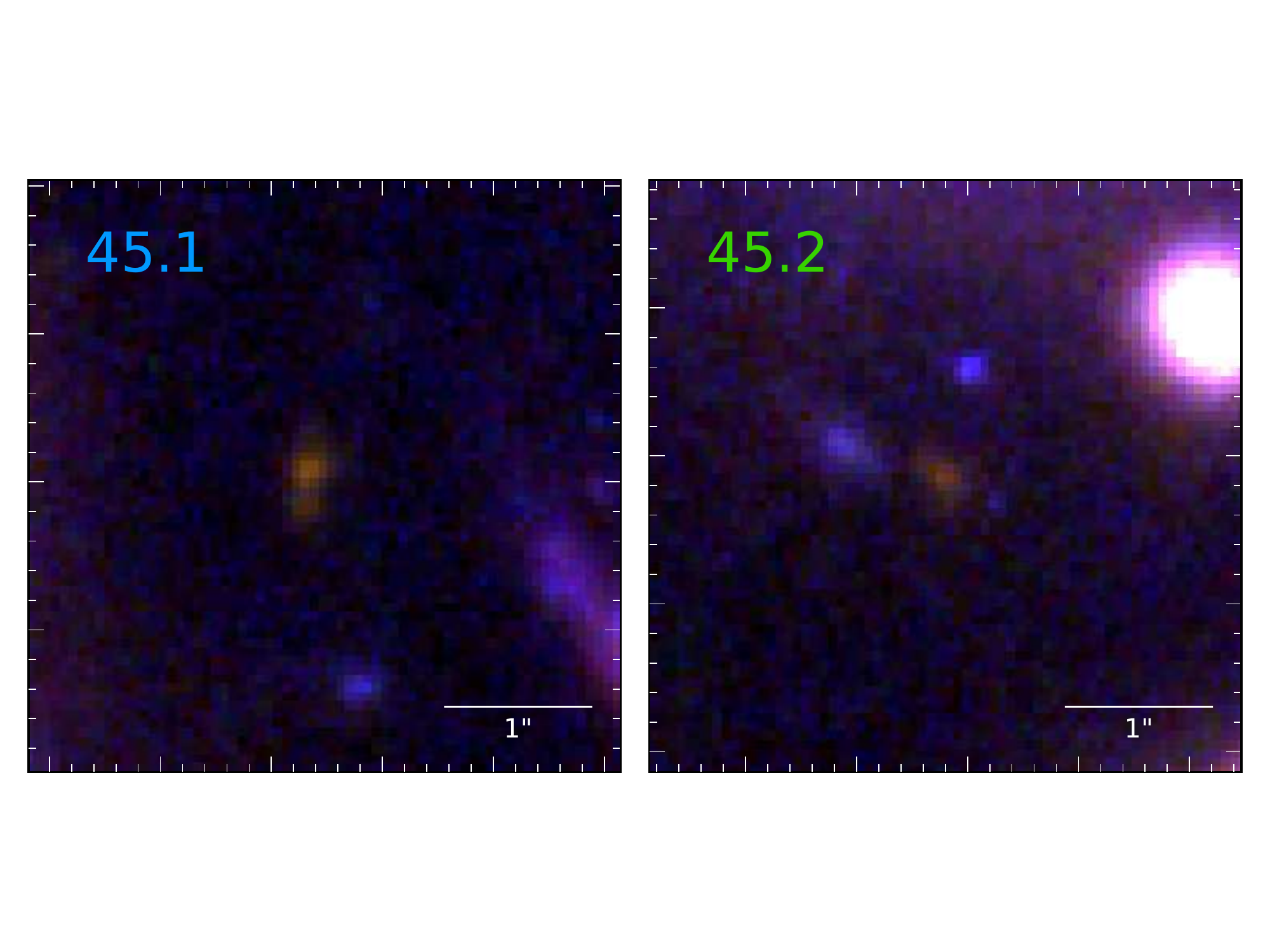}
}
\caption{Multiply-imaged systems without spectroscopic redshifts.  The combination of colours, morphologies, and relative positions in the image plane strongly suggest that each system is a real multiple image set.  Without spectroscopic information, we allow the final redshift to vary as a free parameter in the model (Table \ref{tbl:Multi-Images}).}
\label{fig:NoSpecFigs}
\end{figure*}

\section{Master Redshift Table}
\label{app:Redshifts}
Table \ref{tbl:Redshifts} shows a portion of the master redshift catalog compiled in this work. This is only a small subset of the full catalog, shown for demonstration purposes.  The entire table is provided as an online supplement to this manuscript.  For clarity, only the $F814W$ photometry is shown in this sample.  However, in the full table magnitudes for all seven HFF bands are provided.  The ``Type'' column classifies objects into categories based on prominent emission or absorption features seen in the spectrum, using a scheme similar to the MUSE Ultra Deep Field catalog \citep{ina17}.  Our type codes are as follows:
\begin{itemize}
  \item Type 0: Stars
  \item Type 1: H$\alpha$ emitters [$z < 0.4$]
  \item Type 2: [\ion{O}{II}] emitters [$0.4 < z < 1.5$]
  \item[] (H$\alpha$ has redshifted out of the MUSE window)
  \item Type 3: Absorption line galaxies
  \item[] (no emission features; typically cluster members)
  \item Type 4: \ion{C}{III}] emitters [$1.5 < z < 3$]
  \item Type 5: Lyman $\alpha$ emitters [$z > 3$]
  \item Type 6: [\ion{O}{III}] emitters
  \item[] (only if H$\alpha$ and [\ion{O}{II}] are not present; very rare)
\end{itemize}
The ``Confidence'' column presents the confidence of the redshift measurement, from low (confidence 1) to high (confidence 3).  We use the following criteria to judge the redshift confidence:
\begin{itemize}
\item Confidence 3: Secure redshift; multiple prominent spectral features or one strong feature (such as Lyman $\alpha$) that is unambiguous to identify.
\item Confidence 2: Probable redshift; features are less strong but still clearly identifiable.  They may be overly broad or noisy, leading to a less precise peak (or trough) estimate.  The redshift precision is still less than $\delta z = 0.001$
\item Confidence 1: Possible redshift; identified by one (ambiguous) line feature, or a few, very low S/N lines.  Often faint absorption-line galaxies.  These redshifts should be used with caution. 
\end{itemize}

\begin{table*}
  \centering
  \caption{Master Redshift Catalog (Sample)}
  \label{tbl:Redshifts}
  \begin{tabular}[t]{lllllllllrl}
    \hline
    ID & RA & Dec & $z$  & Type & Conf & $m_{F814W}$ & e\_$m_{F814W}$ & $\mu$   & Multi-Image & Lines\\
    \hline
2085 & 39.9773329 & -1.5995876 & 0.8047 & 2 & 2 &  25.186 & 0.009 & 1.606  &  & [OII], [OIII]\\ 
2159 & 39.9747501 & -1.5991900 & 3.2797 & 5 & 3 &  25.814 & 0.018 & 2.813  &  & Ly-a, CIII] (weak)\\
2510 & 39.9734309 & -1.5988101 & 0.0000 & 0 & 3 &  -2.000 & 0.001 & 1.000  &  & Star\\
2526 & 39.9731475 & -1.5980583 & 1.0693 & 2 & 1 &  23.780 & 0.006 & 2.293  &  & [OII]3730, Mg2799 (abs)\\
2577 & 39.9769563 & -1.5986203 & 0.3702 & 3 & 3 &  21.841 & 0.001 & 1.000  &  & Ca-K, Ca-H, Mg-b, Na-D\\
2615 & 39.9756001 & -1.5977416 & 0.3692 & 3 & 3 &  21.785 & 0.001 & 1.000  &  & Ca-K, Ca-H, H-b\\
2653 & 39.9718952 & -1.5974092 & 0.3239 & 1 & 3 &  23.302 & 0.003 & 1.000  &  & Ha, Hb, OII, OIII\\
2655 & 39.9751731 & -1.5973908 & 0.3564 & 3 & 1 &  24.770 & 0.009 & 1.000  &  & H+K? (v. weak), D4000?\\ 
2714 & 39.9732177 & -1.5991269 & 0.3747 & 3 & 3 &  20.482 & 0.001 & 1.000  &  & Ca H+K, Mgb, NaD, Ha, Hb (abs)\\
2914 & 39.9706141 & -1.5972957 & 0.0000 & 0 & 3 &  -2.000 & 0.000 & 1.000  &  & Star\\
8620 & 39.9671636 & -1.5768729 & 0.8041 & 2 & 3 &  23.178 & 0.003 & 14.143 &  1.1 & [OII], Hb, [OIII], Balmer series\\
\hline
    \end{tabular}
\end{table*}

\section{GLASS Redshift Comparison}
\label{app:GLASS}

In Table \ref{tbl:GlassCompare} we compare the MUSE and GLASS redshift catalogs of A370.  There are approximately 300 objects in common between the two catalogs, matched within a 1$\arcsec$ radius.  A majority of the matches (165 galaxies) have only a small redshift discrepancy ($\delta z < 0.025$), agreeing well with each other given measurement uncertainty and the lower resolution of the \emph{HST} grism.  Additionally, we find $\sim$ 80 galaxies which have a MUSE-measured redshift but a value of -1 (no measurement) in the GLASS catalog.  These are almost entirely cluster-member galaxies which have strong optical absorption features that are too blue to be seen in the GLASS wavelength range.  While other matches have a more substantial disagreement in their redshift values, in many cases we can compare the features seen in both the GLASS and MUSE data sets and correct mis-identified features in a single catalog.  This provides a better, more accurate redshift measurement in both catalogs, and can increase the redshift confidence of a measurement made from a single, low-significance feature.  In the following table, we summarize the corrections and improvements made by these comparisons, highlighting the benefits of extended spectroscopic coverage.

\begin{table*}
  \centering
  \caption{Comparisons between MUSE and GLASS redshifts (Sample)}
  \label{tbl:GlassCompare}
  \begin{tabular}{lcccccp{90mm}}
    \hline
    ID$_{\rm MUSE}$ & $z_{\rm MUSE}$ & C$_{\rm MUSE}$ & ID$_{\rm GLASS}$ & $z_{\rm GLASS}$ & C$_{\rm GLASS}$ & Description\\
	\hline
    3421  & 1.3399 & 3 & 3613 & 1.450 $\rightarrow$ 1.340 & 2 & Noise peak in GLASS G102 misidentified as [\ion{O}{II}].  MUSE redshift is based on a clear, bright [\ion{O}{II}] feature.  Possible [\ion{O}{III}] and H$\alpha$ lines seen at the MUSE redshift in GLASS G141 data, though they are noisy.\\
    4506  & 1.4500 & 3 & 3596 & 1.423 $\rightarrow$ 1.450 & 4 & [\ion{O}{II}], [\ion{O}{III}] and H$\alpha$ identified in GLASS, but very broad.  Misalignment in the wavelength solution due to variance in the grism PA makes the redshift measurement uncertain.  Better resolution in MUSE removes this uncertainty.\\
    4991  & 0.8048 & 3 & 3439 & 0.360 $\rightarrow$ 0.805 & 2 $\rightarrow$ 3 & Emission line at 9000\AA\ originally identified as  H$\alpha$ in GLASS G102, but MUSE data reveals it to be [\ion{O}{III}].  This is verified by additional emission lines seen at shorter wavelengths.\\
    5109  & 0.3464 & 3 & 3524 & 0.370 $\rightarrow$ 0.346 & 3 & Faint [\ion{S}{II}] emission seen in GLASS G102 Band misidentified as H$\alpha$.  The actual H$\alpha$ and [\ion{S}{II}] emission match those seen in MUSE.\\
    5263  & 0.5802 & 3 & 3345 & 1.067 $\rightarrow$ 0.580 & 2.5 $\rightarrow$ 3 & H$\alpha$ feature at 10400\AA\ in GLASS G102 misidentified as [\ion{O}{III}].  [\ion{O}{II}], [\ion{O}{III}], and Balmer-series emission lines in MUSE confirm this redshift.\\
    5398  & 0.3050 & 3 & 3603 & 0.380 $\rightarrow$ 0.305 & 2 & Continuum feature in G102 misidentified as H$\alpha$ + [\ion{N}{II}].  The actual feature is seen in data but very broad.  Better resolution in MUSE identifies the lines unambiguously.\\
    6253  & 1.0616 & 3 & 3069 & 0.570 $\rightarrow$ 1.062 & 2.5 $\rightarrow$ 4 & [\ion{O}{III}] feature at 10200\AA\ in GLASS G102 misidentified as H$\alpha$.  Strong [\ion{O}{II}] doublet in MUSE confirms this redshift.  There is also a faint trace of H$\alpha$ in the G141 data.\\
    6841  & 0.7338 & 3 & 3000 & 0.703 $\rightarrow$ 0.734 & 2 & Noise peak in GLASS originally identified as H$\alpha$.  The real (but faint) H$\alpha$ feature is seen in GLASS G102/G141 overlap region where sensitivity is low.  This matches the MUSE H$\alpha$ line and is consistent with several other optical lines.\\ 
    6898 & 0.3746 & 1 $\rightarrow$ 2 & 2582 & 0.380 & 3 & MUSE redshift is measured from several strong absorption features, including Ca H+K, Mg-b, and Na-D.  However, the galaxy is extremely close to the southern BCG ($z = 0.3733$) and possibly contaminated.  In the 2D MUSE narrow-band image, however, we find a distinct ``emission'' feature (really the continuum flux between Ca H and Ca K) that matches the appearance of the galaxy in the HFF image, distinct from the BCG.  The GLASS spectrum also shows a probable H$\alpha$ absorption feature consistent with the MUSE redshift, strengthening the result.  Therefore, we increase the MUSE confidence value from 1 to 2.\\
    7390  & 0.0000 & 3 & 2711 & N/A $\rightarrow$ 0.000 & 0 $\rightarrow$ 4 & This object does not have a redshift in the GLASS catalog, but a broad TiO feature matching MUSE data at 8600\AA\ confirms it is a star.\\    
    7656  & 0.3664 & 3 & 2702 & 1.610 $\rightarrow$ 0.366 & 2 &  GLASS redshift based on a noise peak at 13000\AA\ misidentified as [\ion{O}{III}].  Noisy H$\alpha$ and [\ion{S}{II}] matching the MUSE redshift are seen in G102.  The MUSE redshift is determined from several strong optical lines.\\
    7815  & 0.2561 & 3 & 2597 & N/A $\rightarrow$ 0.256 & 0 $\rightarrow$ 2 & This object does not have a redshift in the GLASS catalog, but H$\alpha$ is seen at the blue edge of G102, matching the MUSE redshift.  This feature was not originally considered because it is in the noisy, low-transmission region of the band.\\ 
    8079  & 1.5182 & 2 $\rightarrow$ 3 & 2555 & 1.520 & 3 $\rightarrow$ 4 & Multiple Image 9.1.  MUSE redshift is measured from faint \ion{C}{III}] at 4800\AA\, located in the low-transmission region of the data.  The GLASS data show clear H$\alpha$ and [\ion{O}{III}] emission at the same redshift, confirming our result.  Therefore, we increase the MUSE confidence value from 2 to 3.\\
    8121 & 0.3803 & 3 & 2603 & 1.755 $\rightarrow$ 0.380 & 3 & Probable noise peak at 13700\AA\ in GLASS G141 data misidentified as [\ion{O}{III}].   Faint traces of [\ion{S}{III}] seen in G141 at the MUSE redshift.  MUSE redshift is confirmed by the presence of several strong optical emission features, including H$\alpha$, [\ion{O}{II}], and [\ion{O}{III}].\\
    8552  & 1.5182 & 2 $\rightarrow$ 3 & 2430 & 1.520 & 3 $\rightarrow$ 4 & Multiple Image 9.3.  MUSE redshift is measured from faint \ion{C}{III}] at 4800\AA\, located in the low-transmission region of the data.  The GLASS data show clear H$\alpha$ and [\ion{O}{III}] emission at the same redshift, confirming our result.  Therefore, we increase the MUSE confidence value from 2 to 3.\\
    8606  & 1.5182 & 2 $\rightarrow$ 3 & 2242 & 1.520 & 3 $\rightarrow$ 4 & Multiple Image 9.2.  MUSE redshift is measured from faint \ion{C}{III}] at 4800\AA\, located in the low-transmission region of the data.  The GLASS data show clear H$\alpha$ and [\ion{O}{III}] emission at the same redshift, confirming our result.  Therefore, we increase the MUSE confidence value from 2 to 3.\\      
    \hline
    \end{tabular}
\end{table*}
\begin{table*}\ContinuedFloat
  \centering
  \caption{Comparisons between MUSE and GLASS redshifts (continued)}
  \begin{tabular}{lcccccp{90mm}}
    \hline
    ID$_{\rm MUSE}$ & $z_{\rm MUSE}$ & C$_{\rm MUSE}$ & ID$_{\rm GLASS}$ & $z_{\rm GLASS}$ & C$_{\rm GLASS}$ & Description\\
	\hline
    9141  & 0.3644 & 3 & 2275 & 0.790 $\rightarrow$ 0.364 & 3 $\rightarrow$ 4 & Emission line at 8900\AA\ originally identified as [\ion{O}{III}] in GLASS, but it is actually H$\alpha$.  This is confirmed in MUSE, along with several other optical lines.\\ 
    9197 & 1.1956 & 3 & 2390 & 1.930 $\rightarrow$ 1.196 & 4 & H$\alpha$ feature at 14600\AA\ in GLASS data misidentified as the [\ion{O}{III}] doublet.  The broad shape of the feature is due to a large rotation curve.  The H$\alpha$ solution is consistent with the MUSE redshift, measured from a moderate [\ion{O}{II}] emission feature.\\         
    9336  & 0.3923 & 3 & 2269 & 1.030 $\rightarrow$ 0.392 & 3 & H$\alpha$ feature at 9100\AA\ misidentified as [\ion{O}{III}].   [\ion{O}{II}], [\ion{O}{III}], and Balmer-series emission lines in MUSE confirm this redshift.\\
    9914 & 0.2071 & 3 & 2060 & 1.080 $\rightarrow$ 0.207 & 3 & Probable noise peak at 13600\AA\ in GLASS G141 data misidentified as H$\alpha$.  There are no features in GLASS data at the MUSE redshift.  However, the MUSE redshift is based on several strong emission features, including H$\alpha$, H $\beta$, and [\ion{O}{III}], so we are confident of its accuracy.\\
    10288 & 0.4649 & 3 & 2366 & 0.380 $\rightarrow$ 0.465 & 4 $\rightarrow$ 2 & Several absorption features seen in MUSE data, conclusively identifying the redshift.  Origin of the GLASS redshift is unclear.  Possible H$\alpha$ feature seen in G102 at the MUSE redshift, though the galaxy is clearly elliptical and other Balmer-series lines appear as absorption features.\\
    10544 & 0.2560 & 3 & 1979 & 1.065 $\rightarrow$ 0.256 & 1 $\rightarrow$ 3 & GLASS redshift based on a likely noise peak, identified as H$\alpha$ in G141 at 13500\AA.\  A noisy H$\alpha$ feature can be seen in G102 at the MUSE redshift, as well as faint [\ion{S}{III}] features in G141.  Several other emission lines are also seen in MUSE. \\
    10793 & 2.3830 & 2 $\rightarrow$ 3 & 1908 & 1.525 $\rightarrow$ 2.383 & 3 $\rightarrow$ 4 & [\ion{O}{II}] feature at 12500\AA\ in GLASS G102 misidentified as [\ion{O}{III}].  \ion{Mg}{II} emission at the MUSE redshift is also detected in G102 data.  The MUSE redshift is determined from a faint \ion{C}{III}] feature, but combined with the GLASS data we increase the confidence of the redshift from 2 to 3. \\
    11269 & 0.3813 & 3 & 1658 & 1.250 $\rightarrow$ 0.381 & 2 & GLASS redshift based on a noise peak at 8500\AA\ misidentified as [\ion{O}{II}].  This feature is not seen in the MUSE data.  Instead, the MUSE redshift is based on clear Ca H+K absorption features.  The GLASS data potentially identifies an H$\alpha$ feature at the MUSE redshift, though the galaxy is elliptical and does not show any other Balmer emission features. \\
    13253 & 0.7156 & 3 & 1278 & 1.260 $\rightarrow$ 0.716 & 3 & H$\alpha$ feature seen in the G102/G141 overlap misidentified as [\ion{O}{III}].  Additional lines in the MUSE data confirm this redshift.\\
    13464 & 1.0348 & 3 & 1274 & 1.230 $\rightarrow$ 1.035 & 3 & MUSE redshift determined from strong Ca H+K features.  GLASS redshift misidentifies G-band and H $\gamma$ absorption as the 4000\AA\ break.  Ca H+K and 4000\AA\ features are seen in G102 at the MUSE redshift, but they are noisy.\\
    13495 & 1.2656 & 2 $\rightarrow$ 3 & 1185 & 1.270 & 4 & MUSE redshift is measured from faint [\ion{O}{II}] emission, along with noisy [\ion{Ne}{II}], H$\zeta$, and H$\delta$ which fall in the skyline-contaminated region of the spectrum.  The GLASS data clearly show H$\alpha$, H$\beta$, and [\ion{O}{III}] at the same redshift, confirming our result.  Therefore, we increase the MUSE confidence value from 2 to 3.\\
    15035 & 0.3277 & 3 &  855 & N/A $\rightarrow$ 0.328 & 0 $\rightarrow$ 3 & This object does not have a redshift in the GLASS catalog, but H$\alpha$ is seen in G102, confirming the MUSE redshift. The H$\alpha$ feature was judged to be of too low significance at the G102 edge, with only one position angle of grism data available, to provide an original GLASS redshift.\\
    15715 & 0.3735 & 3 & 1024 & 0.820 $\rightarrow$ 0.374 & 3 $\rightarrow$ 4 & A large rotation curve makes the H$\alpha$ feature at 9000\AA\ appear as a doublet in GLASS, which is misidentified as [\ion{O}{III} + H$\beta$].  IFU spectroscopy from MUSE reveals the rotation curve and confirms a single H$\alpha$ feature along with several other optical lines.\\
    16354 & 0.2557 & 3 & 1191 & N/A $\rightarrow$ 0.256 & 0 $\rightarrow$ 3 & This object does not have a redshift in the GLASS catalog, but H$\alpha$ is seen at the blue edge of G102, matching the MUSE redshift.  This feature was not originally considered because it is in the noisy, low-transmission region of the band.\\  
 
    \hline
    \end{tabular}
\end{table*}


\bsp	
\label{lastpage}

\begin{thebibliography}{99}
\bibitem[\protect\citeauthoryear{Acebron et al.}{2017}]{ace17} Acebron A., Jullo E., Limousin M., Tilquin A., Giocoli C., Jauzac M., Mahler G., Richard J., 2017, MNRAS, 470, 1809 
\bibitem[\protect\citeauthoryear{Bacon et al.}{2010}]{bac10} Bacon R., et al., 2010, SPIE, 7735, 773508 
\bibitem[\protect\citeauthoryear{Bacon et al.}{2015}]{bac15} Bacon R., et al., 2015, A\&A, 575, A75
\bibitem[\protect\citeauthoryear{Bahcall \& Kulier}{2014}]{bah14} Bahcall N.~A., Kulier A., 2014, MNRAS, 439, 2505 
\bibitem[\protect\citeauthoryear{Baldry et al.}{2014}]{bal14} Baldry I.~K., et al., 2014, MNRAS, 441, 2440
\bibitem[\protect\citeauthoryear{Bertin \& Arnouts}{1996}]{ber96} Bertin E., Arnouts S., 1996, A\&AS, 117, 393 
\bibitem[\protect\citeauthoryear{Bina et al.}{2016}]{bin16} Bina D., et al., 2016, A\&A, 590, A14 
\bibitem[\protect\citeauthoryear{Blake, James, \& Poole}{2014}]{bla14} Blake C., James J.~B., Poole G.~B., 2014, MNRAS, 437, 2488 
\bibitem[\protect\citeauthoryear{Bond, Kofman, \& Pogosyan}{1996}]{bon96} Bond J.~R., Kofman L., Pogosyan D., 1996, Natur, 380, 603 
\bibitem[\protect\citeauthoryear{Bozek et al.}{2016}]{boz16} Bozek B., Boylan-Kolchin M., Horiuchi S., Garrison-Kimmel S., Abazajian K., Bullock J.~S., 2016, MNRAS, 459, 1489 
\bibitem[\protect\citeauthoryear{Bullock \& Boylan-Kolchin}{2017}]{bul17} Bullock J.~S., Boylan-Kolchin M., 2017, ARA\&A, 55, 343 
\bibitem[\protect\citeauthoryear{Caminha et al.}{2017a}]{cam17a} Caminha G.~B., et al., 2017, A\&A, 600, A90
\bibitem[\protect\citeauthoryear{Caminha et al.}{2017b}]{cam17b} Caminha G.~B., et al., 2017, A\&A, 607, A93
\bibitem[\protect\citeauthoryear{Chiriv{\`i} et al.}{2018}]{chi18} Chiriv{\`i} G., Suyu S.~H., Grillo C., Halkola A., Balestra I., Caminha G.~B., Mercurio A., Rosati P., 2018, A\&A, 614, A8 
\bibitem[\protect\citeauthoryear{Conselice}{2014}]{con14} Conselice C.~J., 2014, ARA\&A, 52, 291
\bibitem[\protect\citeauthoryear{Diego et al.}{2007}]{die07} Diego J.~M., Tegmark M., Protopapas P., Sandvik H.~B., 2007, MNRAS, 375, 958 
\bibitem[\protect\citeauthoryear{Diego et al.}{2015}]{die15} Diego J.~M., Broadhurst T., Molnar S.~M., Lam D., Lim J., 2015, MNRAS, 447, 3130
\bibitem[\protect\citeauthoryear{Diego et al.}{2018}]{die18} Diego J.~M., et al., 2018, MNRAS, 473, 4279
\bibitem[\protect\citeauthoryear{Ebeling, Stephenson, \& Edge}{2014}]{ebe14} Ebeling H., Stephenson L.~N., Edge A.~C., 2014, ApJ, 781, L40 
\bibitem[\protect\citeauthoryear{El{\'{\i}}asd{\'o}ttir et al.}{2007}]{eli07} El{\'{\i}}asd{\'o}ttir {\'A}., et al., 2007, arXiv, arXiv:0710.5636 
\bibitem[\protect\citeauthoryear{Ettori et al.}{2013}]{ett13} Ettori S., Donnarumma A., Pointecouteau E., Reiprich T.~H., Giodini S., Lovisari L., Schmidt R.~W., 2013, SSRv, 177, 119
\bibitem[\protect\citeauthoryear{Finley et al.}{2017}]{fin17} Finley H., et al., 2017, A\&A, 608, A7 
\bibitem[\protect\citeauthoryear{Fort et al.}{1986}]{for86} Fort B., Mellier Y., Picat J.~P., Rio Y., Lelievre G., 1986, SPIE, 627, 321
\bibitem[\protect\citeauthoryear{Franx et al.}{2008}]{fra08} Franx M., van Dokkum P.~G., F{\"o}rster Schreiber N.~M., Wuyts S., Labb{\'e} I., Toft S., 2008, ApJ, 688, 770-788 
\bibitem[\protect\citeauthoryear{Gavazzi et al.}{2004}]{gav04} Gavazzi R., Mellier Y., Fort B., Cuillandre J.-C., Dantel-Fort M., 2004, A\&A, 422, 407 
\bibitem[\protect\citeauthoryear{Girardi et al.}{2015}]{gir15} Girardi M., et al., 2015, A\&A, 579, A4
\bibitem[\protect\citeauthoryear{Griffiths et al.}{2018}]{gri18} Griffiths A., et al., 2018, MNRAS, 475, 2853
\bibitem[\protect\citeauthoryear{Grillo et al.}{2016}]{gri16} Grillo C., et al., 2016, ApJ, 822, 78 
\bibitem[\protect\citeauthoryear{Halkola, Seitz, \& Pannella}{2007}]{hal07} Halkola A., Seitz S., Pannella M., 2007, ApJ, 656, 739 
\bibitem[\protect\citeauthoryear{Hammer}{1987}]{ham87} Hammer F., 1987, hrpg.work, 467 
\bibitem[Hern{\'a}n-Caballero et al.(2017)]{her17} Hern{\'a}n-Caballero, A., P{\'e}rez-Gonz{\'a}lez, P.~G., Diego, J.~M., et al.\ 2017, \apj, 849, 82 
\bibitem[\protect\citeauthoryear{Hoag et al.}{2016}]{hoa16} Hoag A., et al., 2016, ApJ, 831, 182
\bibitem[\protect\citeauthoryear{Hoekstra}{2007}]{hoe07} Hoekstra H., 2007, MNRAS, 379, 317 
\bibitem[\protect\citeauthoryear{Inami et al.}{2017}]{ina17} Inami H., et al., 2017, A\&A, 608, A2 
\bibitem[\protect\citeauthoryear{Ishigaki et al.}{2018}]{ish18} Ishigaki M., Kawamata R., Ouchi M., Oguri M., Shimasaku K., Ono Y., 2018, ApJ, 854, 73 
\bibitem[\protect\citeauthoryear{Jauzac et al.}{2016}]{jau16} Jauzac M., et al., 2016, MNRAS, 463, 3876
\bibitem[\protect\citeauthoryear{Jauzac et al.}{2018}]{jau18} Jauzac M., et al., 2018, MNRAS, 481, 2901    
\bibitem[\protect\citeauthoryear{Jullo et al.}{2007}]{jul07} Jullo E., Kneib J.-P., Limousin M., El{\'{\i}}asd{\'o}ttir {\'A}., Marshall P.~J., Verdugo T., 2007, NJPh, 9, 447 
\bibitem[\protect\citeauthoryear{Jullo \& Kneib}{2009}]{jul09} Jullo E., Kneib J.-P., 2009, MNRAS, 395, 1319 
\bibitem[\protect\citeauthoryear{Jullo et al.}{2010}]{jul10} Jullo E., Natarajan P., Kneib J.-P., D'Aloisio A., Limousin M., Richard J., Schimd C., 2010, Sci, 329, 924 
\bibitem[\protect\citeauthoryear{Karman et al.}{2015}]{kar15} Karman W., et al., 2015, A\&A, 574, A11 
\bibitem[\protect\citeauthoryear{Kawamata et al.}{2018}]{kaw18} Kawamata R., Ishigaki M., Shimasaku K., Oguri M., Ouchi M., Tanigawa S., 2018, ApJ, 855, 4
\bibitem[\protect\citeauthoryear{Kneib et al.}{1996}]{kne96} Kneib J.-P., Ellis R.~S., Smail I., Couch W.~J., Sharples R.~M., 1996, ApJ, 471, 643 
\bibitem[\protect\citeauthoryear{Koopmans et al.}{2006}]{koo06} Koopmans L.~V.~E., Treu T., Bolton A.~S., Burles S., Moustakas L.~A., 2006, ApJ, 649, 599
\bibitem[\protect\citeauthoryear{Koyama}{2016}]{koy16} Koyama K., 2016, RPPh, 79, 046902 
\bibitem[\protect\citeauthoryear{Kravtsov \& Borgani}{2012}]{kra12} Kravtsov A.~V., Borgani S., 2012, ARA\&A, 50, 353 
\bibitem[\protect\citeauthoryear{Lagattuta et al.}{2017}]{lag17} Lagattuta D.~J., et al., 2017, MNRAS, 469, 3946
\bibitem[\protect\citeauthoryear{Laporte \& White}{2015}]{lap15} Laporte C.~F.~P., White S.~D.~M., 2015, MNRAS, 451, 1177 
\bibitem[\protect\citeauthoryear{Leclercq et al.}{2017}]{lec17} Leclercq F., et al., 2017, A\&A, 608, A8 
\bibitem[\protect\citeauthoryear{Li et al.}{2016}]{li16} Li R., Frenk C.~S., Cole S., Gao L., Bose S., Hellwing W.~A., 2016, MNRAS, 460, 363
\bibitem[\protect\citeauthoryear{Limousin et al.}{2016}]{lim16} Limousin M., et al., 2016, A\&A, 588, A99
\bibitem[\protect\citeauthoryear{Lin et al.}{2006}]{lin06} Lin Y.-T., Mohr J.~J., Gonzalez A.~H., Stanford S.~A., 2006, ApJ, 650, L99 
\bibitem[\protect\citeauthoryear{L{\'o}pez-Corredoira, Guti{\'e}rrez, \& G{\'e}nova-Santos}{2017}]{lop17} L{\'o}pez-Corredoira M., Guti{\'e}rrez C.~M., G{\'e}nova-Santos R.~T., 2017, ApJ, 840, 62 
\bibitem[\protect\citeauthoryear{Lotz et al.}{2017}]{lot17} Lotz J.~M., et al., 2017, ApJ, 837, 97 
\bibitem[\protect\citeauthoryear{Mahler et al.}{2018}]{mah18} Mahler G., et al., 2018, MNRAS, 473, 663 
\bibitem[\protect\citeauthoryear{Marrone et al.}{2012}]{mar12} Marrone D.~P., et al., 2012, ApJ, 754, 119 
\bibitem[\protect\citeauthoryear{Maseda et al.}{2018}]{mas18} Maseda M.~V., et al., 2018, ApJ, 865, L1 
\bibitem[\protect\citeauthoryear{McPartland et al.}{2016}]{mcp16} McPartland C., Ebeling H., Roediger E., Blumenthal K., 2016, MNRAS, 455, 2994 
\bibitem[\protect\citeauthoryear{Mellier et al.}{1988}]{mel88} Mellier Y., Soucail G., Fort B., Mathez G., 1988, A\&A, 199, 13 
\bibitem[\protect\citeauthoryear{Meneghetti et al.}{2017}]{men17} Meneghetti M., et al., 2017, MNRAS, 472, 3177 
\bibitem[\protect\citeauthoryear{Merten et al.}{2015}]{mer15} Merten J., et al., 2015, ApJ, 806, 4
\bibitem[\protect\citeauthoryear{Morandi, Ettori, \& Moscardini}{2007}]{mor07} Morandi A., Ettori S., Moscardini L., 2007, MNRAS, 379, 518 
\bibitem[\protect\citeauthoryear{Natarajan et al.}{2017}]{nat17} Natarajan P., et al., 2017, MNRAS, 468, 1962 
\bibitem[\protect\citeauthoryear{Newman et al.}{2013a}]{new13a} Newman A.~B., Treu T., Ellis R.~S., Sand D.~J., Nipoti C., Richard J., Jullo E., 2013, ApJ, 765, 24
\bibitem[\protect\citeauthoryear{Newman et al.}{2013b}]{new13b} Newman A.~B., Treu T., Ellis R.~S., Sand D.~J., 2013, ApJ, 765, 25 
\bibitem[\protect\citeauthoryear{Nierenberg et al.}{2013}]{nie13} Nierenberg A.~M., Treu T., Menci N., Lu Y., Wang W., 2013, ApJ, 772, 146 
\bibitem[\protect\citeauthoryear{Ogrean et al.}{2016}]{ogr16} Ogrean G.~A., et al., 2016, ApJ, 819, 113 
\bibitem[\protect\citeauthoryear{Patr{\'{\i}}cio et al.}{2016}]{pat16} Patr{\'{\i}}cio V., et al., 2016, MNRAS, 456, 4191 
\bibitem[\protect\citeauthoryear{Patr{\'{\i}}cio et al.}{2018}]{pat18} Patr{\'{\i}}cio V., et al., 2018, MNRAS, 477, 18
\bibitem[\protect\citeauthoryear{Percival \& White}{2009}]{per09} Percival W.~J., White M., 2009, MNRAS, 393, 297
\bibitem[\protect\citeauthoryear{Richard et al.}{2010}]{ric10} Richard J., Kneib J.-P., Limousin M., Edge A., Jullo E., 2010, MNRAS, 402, L44 
\bibitem[\protect\citeauthoryear{Richard et al.}{2014}]{ric14} Richard J., et al., 2014, MNRAS, 444, 268 
\bibitem[\protect\citeauthoryear{Richard et al.}{2015}]{ric15} Richard J., et al., 2015, MNRAS, 446, L16 
\bibitem[\protect\citeauthoryear{Schaye et al.}{2015}]{sch15} Schaye J., et al., 2015, MNRAS, 446, 521
\bibitem[\protect\citeauthoryear{Schmidt et al.}{2014}]{sch14} Schmidt K.~B., et al., 2014, ApJ, 782, L36 
\bibitem[\protect\citeauthoryear{Schwinn et al.}{2017}]{sch17} Schwinn J., Jauzac M., Baugh C.~M., Bartelmann M., Eckert D., Harvey D., Natarajan P., Massey R., 2017, MNRAS, 467, 2913 
\bibitem[\protect\citeauthoryear{Sebesta et al.}{2016}]{seb16} Sebesta K., Williams L.~L.~R., Mohammed I., Saha P., Liesenborgs J., 2016, MNRAS, 461, 2126
\bibitem[\protect\citeauthoryear{Shirasaki, Nagai, \& Lau}{2016}]{shi16} Shirasaki M., Nagai D., Lau E.~T., 2016, MNRAS, 460, 3913 
\bibitem[\protect\citeauthoryear{Springel, Frenk, \& White}{2006}]{spr06} Springel V., Frenk C.~S., White S.~D.~M., 2006, Natur, 440, 1137 
\bibitem[\protect\citeauthoryear{Stanford et al.}{2006}]{sta06} Stanford S.~A., et al., 2006, ApJ, 646, L13 
\bibitem[\protect\citeauthoryear{Strait et al.}{2018}]{str18} Strait V., et al., 2018, ApJ, 868, 129 
\bibitem[\protect\citeauthoryear{Soto et al.}{2016}]{sot16} Soto K.~T., Lilly S.~J., Bacon R., Richard J., Conseil S., 2016, MNRAS, 458, 3210 
\bibitem[\protect\citeauthoryear{Soucail et al.}{1987}]{sou87} Soucail G., Mellier Y., Fort B., Hammer F., Mathez G., 1987, A\&A, 184, L7
\bibitem[\protect\citeauthoryear{Soucail et al.}{1999}]{sou99} Soucail G., Kneib J.~P., B{\'e}zecourt J., Metcalfe L., Altieri B., Le Borgne J.~F., 1999, A\&A, 343, L70 
\bibitem[\protect\citeauthoryear{Tasitsiomi et al.}{2004}]{tas04} Tasitsiomi A., Kravtsov A.~V., Gottl{\"o}ber S., Klypin A.~A., 2004, ApJ, 607, 125 
\bibitem[\protect\citeauthoryear{Treu et al.}{2015}]{tre15} Treu T., et al., 2015, ApJ, 812, 114 
\bibitem[\protect\citeauthoryear{Vogelsberger et al.}{2014}]{vog14} Vogelsberger M., et al., 2014, MNRAS, 444, 1518 
\bibitem[\protect\citeauthoryear{Wang et al.}{2015}]{wan15} Wang X., et al., 2015, ApJ, 811, 29 
\bibitem[\protect\citeauthoryear{Weilbacher, Streicher, \& Palsa}{2016}]{wei16} Weilbacher P.~M., Streicher O., Palsa R., 2016, ascl.soft, ascl:1610.004 
\bibitem[\protect\citeauthoryear{Wetzel et al.}{2013}]{wet13} Wetzel A.~R., Tinker J.~L., Conroy C., van den Bosch F.~C., 2013, MNRAS, 432, 336 
\bibitem[\protect\citeauthoryear{Williams, Sebesta, \& Liesenborgs}{2018}]{wil18} Williams L.~L.~R., Sebesta K., Liesenborgs J., 2018, MNRAS, 480, 3140   
\bibitem[\protect\citeauthoryear{Wisotzki et al.}{2016}]{wis16} Wisotzki L., et al., 2016, A\&A, 587, A98
\bibitem[\protect\citeauthoryear{Wisotzki et al.}{2018}]{wis18} Wisotzki L., et al., 2018, Natur, 562, 229 
\bibitem[\protect\citeauthoryear{Wen \& Han}{2013}]{wen13} Wen Z.~L., Han J.~L., 2013, MNRAS, 436, 275 
\bibitem[\protect\citeauthoryear{Wong et al.}{2012}]{won12} Wong K.~C., Ammons S.~M., Keeton C.~R., Zabludoff A.~I., 2012, ApJ, 752, 104 
\bibitem[\protect\citeauthoryear{Wong et al.}{2013}]{won13} Wong K.~C., Zabludoff A.~I., Ammons S.~M., Keeton C.~R., Hogg D.~W., Gonzalez A.~H., 2013, ApJ, 769, 52 

\end{thebibliography}
\end{document}